\documentclass{aa}  

\usepackage{amsmath, amssymb, amsfonts, amscd}
\usepackage{natbib}
\usepackage{graphicx}
\usepackage{txfonts}
\usepackage[final]{hyperref}
\usepackage{mathrsfs}
\usepackage{color}
\usepackage{mathabx}
\usepackage{supertabular}
\usepackage{multirow}
\usepackage{ragged2e}
\usepackage[switch]{lineno}
\usepackage{tikz}

\newcommand{\twolabelsat}[4]{%
\begin{tikzpicture}
\path (0,0) node {\vphantom{.}};
\path (#1,0) node[inner sep=0cm, outer sep=0cm] {#2};
\path (#3,0) node[inner sep=0cm, outer sep=0cm] {#4};
\end{tikzpicture}
}

\newcommand*\patchAmsMathEnvironmentForLineno[1]{%
  \expandafter\let\csname old#1\expandafter\endcsname\csname #1\endcsname
  \expandafter\let\csname oldend#1\expandafter\endcsname\csname end#1\endcsname
  \renewenvironment{#1}%
     {\linenomath\csname old#1\endcsname}%
     {\csname oldend#1\endcsname\endlinenomath}}%
\newcommand*\patchBothAmsMathEnvironmentsForLineno[1]{%
  \patchAmsMathEnvironmentForLineno{#1}%
  \patchAmsMathEnvironmentForLineno{#1*}}%
\AtBeginDocument{%
\patchBothAmsMathEnvironmentsForLineno{equation}%
\patchBothAmsMathEnvironmentsForLineno{align}%
\patchBothAmsMathEnvironmentsForLineno{flalign}%
\patchBothAmsMathEnvironmentsForLineno{alignat}%
\patchBothAmsMathEnvironmentsForLineno{gather}%
\patchBothAmsMathEnvironmentsForLineno{multline}%
}

\hypersetup{
   colorlinks=true,
   citecolor=blue,
   linkcolor=blue,
   urlcolor=black
   }

\definecolor{emerald}{rgb}{0.0,0.5,0.0}
\definecolor{smcolor}{rgb}{0.7,0.3,0.0}
\definecolor{blue-violet}{rgb}{0.54, 0.17, 0.89}

\newcommand{\rec}[1]{#1}

\newcommand{\LEt}[1]{}

\def\inumber{i}                                 
\def\GammaF{\Gamma}                     
\def\Nset{\mathbb{N}}                   
\def\Zset{\mathbb{Z}}                   
\def\Rset{\mathbb{R}}                   
\def\Cset{\mathbb{C}    }                       
\def\define{\equiv}                             
\def\scale{\, \propto \,}                       

\def\cot{{\rm cot}}

\def\argmax{{\rm argmax}}

\newcommand{\notation}[4]{#1_{#2 ; #3}^{#4}}            
\newcommand{\ddroit}{{\rm d}}                           
\newcommand{\vect}[1]{\boldsymbol{#1}}          
\newcommand{\evect}[1]{\vect{{\rm e}}_{#1}}     
\newcommand{\Cvar}[1]{\tilde{#1}}                       
\newcommand{\normvar}[1]{\tilde{#1}}  
\newcommand{\dd}[2]{\partial_{#2} #1}           
\newcommand{\ddd}[3]{\partial_{#2 #3} #1}       
\newcommand{\DD}[2]{\dfrac{\ddroit #1}{\ddroit #2}}                     
\newcommand{\DDn}[3]{\dfrac{\ddroit^{#3} #1}{\ddroit #2^{#3}}} 
\newcommand{\scal}[2]{\langle #1 , #2 \rangle}  
\newcommand{\abs}[1]{\left| #1 \right|}         
\newcommand{\conj}[1]{\overline{#1}}
\newcommand{\matx}[1]{\boldsymbol{#1}}
\newcommand{\transp}[1]{#1^{\rm T}}             
\newcommand{\infvar}[1]{\ddroit #1}

\newcommand{\kron}[2]{\delta_{#1, #2}}
\newcommand{\mean}[1]{\overline{#1}}

\def\dotp{\cdot }
\def\crossp{\times}

\def\idmat{\matx{I}}
\newcommand{\idmati}[1]{\idmat_{#1}}
\newcommand{\normi}[2]{\lVert  #1 \rVert_{#2}}
\newcommand{\normtwo}[1]{\normi{#1}{2}}
\newcommand{\timeav}[1]{\left\langle #1 \right\rangle}
\def\standarddev{\sigma}
\def\standarddevn{\hat{\standarddev}}

\newcommand{\lnormi}[1]{\ell^{#1}}
\def\ltwonorm{\lnormi{2}}

\def\ffunc{f}
\def\gfunc{g}
\def\hfunc{h}

\def\nab{\nabla}                                        
\def\grad{\nab}                                 
\def\lap{\nab^2}                                        
\def\div{\grad \dotp}                           
\def\XX{X}      

\newcommand{\cosi}[1]{C_{#1}}
\newcommand{\sini}[1]{S_{#1}}

\def\coordsystem{S}

\newcommand{\domainbd}[1]{\partial #1}
\def\surface{S}
\def\volume{V}
\def\surfacev{\vect{\surface}}

\def\length{\ell}
\def\sphere{\mathcal{S}}
\def\sphcap{\mathcal{C}}
\def\sphcapbd{\domainbd{\sphcap}}
\def\oceancap{\mathcal{O}}
\def\domainstar{\domain_{*}}
\def\rdstar{\rr_{*}}
\def\domainstarbd{\domainbd{\domainstar}}
\def\oceancapbd{\domainbd{\oceancap}}
\newcommand{\sphcapi}[1]{\sphcap_{#1}}

\def\llat{l}                                            
\def\lmax{L}
\newcommand{\lmaxi}[1]{\lmax_{#1}}
\def\mm{m}                                      
\def\kk{k}                                              
\def\nk{n}
\def\jj{j}
\def\qq{q}
\def\pp{p}
\def\Legendre{P}                                
\def\SPH{Y}     
\def\SPHref{\hat{\SPH}}
\def\hypergtwoFone{{}_2F_{1}}
\def\hypergtwoFonereg{{}_2\tilde{F}_{1}}
\newcommand{\Ylm}[2]{\SPH_{#1}^{#2}}
\newcommand{\Ylmref}[2]{\SPHref_{#1}^{#2}}
\newcommand{\hyperci}[1]{a_{#1}}
\newcommand{\llati}[1]{\llat_{#1}}
\newcommand{\Ylmi}[1]{\SPH_{#1}}
\newcommand{\Ylmrefi}[1]{\SPHref_{#1}}

\def\SCH{\Theta}
\newcommand{\Hlm}[2]{\SCH_{#1}^{#2}}
\newcommand{\alm}[2]{\alpha_{#1, #2}}

\newcommand{\wignerD}[3]{D_{#1,#2}^{#3}}
\newcommand{\wignerd}[3]{d_{#1,#2}^{#3}}
\def\wignermat{\matx{D}}
\newcommand{\wignerDmati}[3]{\wignermat_{#1,#2}^{#3}}
\def\vectSPHrot{\matx{Y}}
\def\vectSPHini{\hat{\matx{Y}}}
\newcommand{\wignermati}[1]{\wignermat_{#1}}

\def\rotmat{\matx{R}}
\def\irotation{{\rm rot}}
\newcommand{\rotmati}[2]{\rotmat_{#1, #2}}
\newcommand{\rmati}[1]{\rotmat_{#1}}
\def\eulermat{\matx{R}}
\def\angalp{\alpha}
\def\angbet{\beta}
\def\anggam{\gamma}
\def\coordsini{\coordsystem}

\def\nbasis{K}
\newcommand{\nbasisi}[1]{\nbasis_{#1}}
\newcommand{\basis}[1]{\mathcal{#1}}
\newcommand{\basisref}[1]{\hat{\basis{#1}}}
\def\anybasis{\basis{H}}
\def\transitionmat{\matx{P}}

\newcommand{\transmati}[2]{\transitionmat_{#1,#2}}

\def\Abasis{\basis{A}}
\def\Bbasis{\basis{B}}

\def\Abasisref{\basisref{A}}
\def\Bbasisref{\basisref{B}}
\def\basisfunc{Y}
\def\basiscoor{y}
\newcommand{\vcoori}[1]{\matx{\basiscoor}_{#1}}
 
\newcommand{\coori}[2]{\basiscoor_{#1,#2}}

\def\vcoorA{\vcoori{\Abasis}}
\def\vcoorB{\vcoori{\Bbasis}}

\newcommand{\coorAi}[1]{\coori{\Abasis}{#1}}
\newcommand{\coorBi}[1]{\coori{\Bbasis}{#1}}

\def\rescoor{\epsilon}

\newcommand{\resci}[2]{\rescoor_{#1,#2}}

\newcommand{\vfunci}[1]{\matx{\basisfunc}_{#1}}

\newcommand{\funci}[2]{\basisfunc_{#1,#2}}

\def\vfuncAref{\vfunci{\Abasisref}}
\def\vfuncBref{\vfunci{\Bbasisref}}
\def\vfuncA{\vfunci{\Abasis}}
\def\vfuncB{\vfunci{\Bbasis}}

\def\vfuncA{\vfunci{\Abasis}}
\def\vfuncB{\vfunci{\Bbasis}}
\def\wignermatA{\wignermati{\Abasis}}
\def\wignermatB{\wignermati{\Bbasis}}

\newcommand{\funcArefi}[1]{\funci{\Abasisref}{#1}}
\newcommand{\funcBrefi}[1]{\funci{\Bbasisref}{#1}}
\def\nbasisA{M}
\def\nbasisB{N}
\def\sphcapAref{\sphcapi{\Abasisref}}
\def\sphcapBref{\sphcapi{\Bbasisref}}
\def\resfunc{\varepsilon}
\def\resmat{\matx{\resfunc}}

\newcommand{\resfi}[2]{\resfunc_{#1,#2}}

\newcommand{\resmi}[1]{\resmat_{#1}}

\def\resmAref{\resmi{\Abasisref}}
\def\resmBref{\resmi{\Bbasisref}}

\def\gyrocoeff{\beta}
\def\gyromat{\matx{B}}
\newcommand{\gyromati}[2]{\gyromat_{#1,#2}}
\newcommand{\gyroci}[2]{\gyrocoeff_{#1,#2}}
\def\gyrompp{\gyromati{\potfunc}{\potfunc}}
\def\gyromps{\gyromati{\potfunc}{\streamfunc}}
\def\gyromsp{\gyromati{\streamfunc}{\potfunc}}
\def\gyromss{\gyromati{\streamfunc}{\streamfunc}}

\def\kkb{\jj}
\def\jjb{\kk}
\def\tidalmat{\matx{H}}

\def\nn{n}                                              

\newcommand{\LegP}[1]{\Legendre_{#1}}                     
\newcommand{\LegF}[2]{\Legendre_{#1}^{#2}}                
\newcommand{\expo}[1]{{\rm e}^{#1}}                             
\newcommand{\integ}[4]{\int_{#3}^{#4} #1 \infvar{#2} }   
\newcommand{\ointeg}[4]{\oint_{#3}^{#4} #1 \infvar{#2} }
\newcommand{\pochham}[2]{\left( #1 \right)_{#2}}

\def\Ggrav{G}                                   

\def\ipla{{\rm p}}                              
\def\imoon{\Moon}

\def\ipert{{\rm s}}
\def\iscont{{\rm S}}
\def\iocean{{\rm oc}}                           
\def\isol{{\rm sol}}                            
\def\idrag{{\rm R}}                             
\def\icore{{\rm core}}                             
\def\icont{{\rm c}}
\def\ieq{{\rm eq}}
\def\iwater{{\rm w}}

\def\xx{x}
\def\yy{y}
\def\zz{z}
\def\XX{X}
\def\YY{Y}
\def\ZZ{Z}
\def\rr{r}                                           
\def\col{\theta}                                
\def\lon{\varphi}                             
\def\time{t}                                    

\def\angrot{\phi}
\def\colpla{\hat{\theta}}
\def\lonpla{\hat{\varphi}}
\def\rrpla{\hat{r}}

\def\lonscont{\lonpla_{\iscont}}
\def\colscont{\colpla_{\iscont}}
\def\lonoc{\lonpla_{\iocean}}
\def\coloc{\colpla_{\iocean}}
\def\colbd{\col_0}
\def\colcont{\col_{\icont}}
\newcommand{\loni}[1]{\lon_{#1}}
\def\lonm{\loni{\mm}}
\def\inorth{{\rm NP}}
\def\colnp{\col_{\inorth}}
\def\lonnp{\lon_{\inorth}}

\def\er{\evect{\rr}}                            
\def\etheta{\evect{\col}}                       
\def\ephi{\evect{\lon}}                 
\def\nvect{\vect{n}}

\def\ex{\evect{\xx}}
\def\ey{\evect{\yy}}
\def\ez{\evect{\zz}}
\def\eX{\evect{\XX}}
\def\eY{\evect{\YY}}
\def\eZ{\evect{\ZZ}}
\def\Uvect{\vect{U}}
\def\Uvectrot{\Uvect_{\irotation}}

\def\framesymb{\mathcal{R}}                     

\newcommand{\framei}[1]{\framesymb_{#1}}
\def\framestd{\framei{}}
\def\framerotation{\framei{\irotation}}
\def\Xvect{X}                                   
\def\framepla{\framestd}
\def\frameoc{\framei{\iocean}}

\newcommand{\rframe}[5]{\framesymb_{#1} {:} \left( #2, #3, #4, #5 \right) }

\def\smaxis{a}                                  
\def\meanmotion{n}
\def\norb{\meanmotion_\ipert}                             
\def\spinrate{\Omega}                   

\def\Mbody{M}                                   
\def\Rbody{R}                                   
\def\Mpla{\Mbody_\ipla}                 
\def\Mmoon{\Mbody_{\imoon}}
\def\Mpert{\Mbody_{\ipert}}
\def\npert{\meanmotion_{\ipert}}
\def\nmoon{\meanmotion_{\imoon}}
\def\rpert{\rr_{\ipert}}

\def\rpertvect{\vect{\rr}_{\ipert}}
\def\Rpla{\Rbody_\ipla}                 

\def\iearth{{\rm E}}                             
\def\Mearth{\Mbody_{\iearth}}           
\def\Rearth{\Rbody_{\iearth}}           
\def\srearth{\spinrate_{\iearth}}

\def\Hlayer{H}                                  
\def\density{\rho}                              
\def\chartime{\tau}                             
\def\freq{\sigma}                               
\def\ggravi{g}                                  
\def\Hoc{\Hlayer}                       
\def\fdrag{\freq_\idrag}                        

\def\rhowater{\density_{\iwater}}
\def\Rcore{\Rbody_\icore}                       
\def\Mcore{\Mbody_\icore}                       
\def\force{F}
\def\forcev{\vect{\force}}
\def\rhopla{\density}           
\def\fcorio{\vect{f}}
\def\fcorion{\normvar{\fcorio}}

\def\mupla{\mu}                                 
\def\tauA{\chartime_{\rm A}}            
\def\tauM{\chartime_{\rm M}}            
\def\alphaA{\alpha}                             
\def\rhocore{\mean{\density}}           
\def\Cmu{\Cvar{\mu}}                    

\def\ftide{\freq}                                       
\def\period{P}                                  
\def\Prot{\period_{\rm rot}}
\def\msigma{\mm,\ftide}                 

\def\Cmul{\Cmu_\llat^{\ftide}}          
\def\coeffAmu{A}
\def\Al{\coeffAmu_{\llat}}
\def\gravpot{U}                                 
\def\gravpotn{\normvar{\gravpot}}
\def\vel{V}                                             
\def\dep{\xi}                                   
\def\iforcing{{\rm T}}                          
\def\iocdyn{{\rm D}}
\def\Ftide{\Xi}                                        
\def\Ftideno{\normvar{\Ftide}}
\def\Utide{\gravpot_\iforcing}          
\def\zetaoc{\zeta}
\def\zetaeq{\zetaoc_{\ieq}}
\def\Vvect{\vect{\vel}}                 
\def\Vtheta{\vel_\col}                          
\def\Vphi{\vel_\lon}                            

\newcommand{\vartide}[1]{\delta #1}
\def\rhotide{\vartide{\density}}

\def\zetaocn{\normvar{\zetaoc}}
\def\zetaeqn{\zetaocn_{\ieq}}

\newcommand{\zetaeqni}[1]{\zetaocn_{\ieq; #1}}

\newcommand{\zetaoclmi}[3]{\zetaocn_{#1}^{#2,#3}}

\def\fdragn{\normvar{\freq}_{\idrag}}
\def\Rowave{\normvar{\freq}_{\rm G}}
\def\ftidequadn{\chi}
\def\ftiden{\normvar{\ftide}}

\def\timen{\normvar{\time}}
\def\depvect{\vect{\dep}}
\def\depn{\normvar{\depvect}}
\def\hdep{\vect{\dep}}
\def\hdepn{\depn}
\def\gradn{\normvar{\grad}}
\def\lapn{\gradn^2}
\def\divn{\gradn \dotp}
\def\intftide{\Delta \ftide}
\def\intftidemean{\mean{\intftide}}
\def\intftiden{\hat{\intftide}}
\newcommand{\intftidei}[1]{\intftide_{#1}}
\newcommand{\intftideni}[1]{\intftiden_{#1}}
\def\CDF{{\rm CDF}}
\def\Nint{N}
\newcommand{\ftidei}[1]{\ftide_{#1}}

\def\iref{0}
\def\timeref{\time_{\iref}}
\def\speedref{\vel_{\iref}}

\def\Ulmsig{\notation{\gravpotn}{\iforcing}{\llat}{\mm,\ftide}}  
\def\Uplmsig{\notation{\gravpotn}{\iresp}{\llat}{\mm,\ftide}}     
\newcommand{\Ulmsigi}[1]{\notation{\gravpot}{\iforcing}{\llat_{#1}}{\mm_{#1},\ftide}}
\newcommand{\Ulmnsiglmi}[2]{\notation{\gravpotn}{\iforcing}{#1}{#2 , \ftide}}

\def\iresp{{\rm D}}
\def\Utiden{\gravpotn_{\iforcing}}
\def\Uresp{\gravpot_{\iresp}}
\def\Urespn{\gravpotn_{\iresp}}
\newcommand{\Utideni}[1]{\gravpotn^{\ftide}_{{\iforcing} ; #1}}

\def\Utidensig{\Utiden^{\ftide}}
\def\Urespnsig{\Urespn^{\ftide}}
\def\Ftidenosig{\Ftideno^{\ftide}}
\newcommand{\Ftidenoi}[1]{\Ftideno^{\ftide}_{#1}}

\def\potfunc{\Phi}
\def\streamfunc{\Psi}
\newcommand{\fpoti}[1]{\potfunc_{#1}}
\newcommand{\fstreami}[1]{\streamfunc_{#1}}
\def\fpotcoeff{\alpha}
\def\fstreamcoeff{\gamma}
\def\sphcoeff{\eta}
\newcommand{\fpotci}[1]{\fpotcoeff_{#1}}
\newcommand{\fstreamci}[1]{\fstreamcoeff_{#1}}
\newcommand{\sphci}[1]{\sphcoeff_{#1}}

\newcommand{\phicolfi}[1]{F_{#1}}
\newcommand{\psicolfi}[1]{G_{#1}}
\newcommand{\sphcolfi}[1]{H_{#1}}
\newcommand{\mmi}[1]{\mm_{#1}}

\def\pcoeff{p}
\newcommand{\pfunci}[1]{\pcoeff_{#1}}
\newcommand{\ppoti}[1]{\pfunci{#1}}
\newcommand{\pstreami}[1]{\pfunci{- #1}}

\def\lovegrav{k}                                        
\def\lovedep{h}                                 
\def\torque{\mathcal{T}}                        
\def\iload{{\rm L}}
\def\kl{\lovegrav_\llat^\ftide}                 
\def\hl{\lovedep_\llat^\ftide}                  
\def\kloadl{\lovegrav_{\iload; \llat}^\ftide}        
\def\hloadl{\lovedep_{\iload ; \llat}^\ftide} 

\def\kquad{\lovegrav_{2}^{\ftide}}

\def\kp{\notation{\lovegrav}{\iresp}{\llat}{\msigma}}    
\def\kpquad{\notation{\lovegrav}{\iresp}{2}{2,\ftide}}

\def\tiltfactor{\gamma}                 
\newcommand{\tiltdi}[1]{\notation{\tiltfactor}{\iocdyn}{\llati{#1}}{\ftide}}                
\newcommand{\tiltgi}[1]{\notation{\tiltfactor}{\iforcing}{\llati{#1}}{\ftide}}     
\def\tiltoperator{\Gamma}

\def\tiltopd{\tiltoperator_{\iocdyn}}
\def\tiltopg{\tiltoperator_{\iforcing}}

\def\tiltmat{\matx{\tiltoperator}}
\def\tiltmatd{\tiltmat_{\iocdyn}}
\def\tiltmatg{\tiltmat_{\iforcing}}
\def\unitvect{\matx{E}}
\newcommand{\unitvi}[2]{\unitvect_{#1,#2}}

\def\gravpotnv{\matx{\gravpotn}}
\def\Utidenv{\gravpotnv_{\iforcing}}
\def\nsph{K}
\def\npot{M}
\def\nstream{N}

\def\cstreamsph{\upsilon}

\newcommand{\cstreamsphi}[2]{\cstreamsph_{#1,#2}}
\def\couplingmat{\matx{A}}

\def\forcingmat{\matx{R}}
\newcommand{\cmati}[1]{\couplingmat_{#1}}

\newcommand{\fmati}[1]{\forcingmat_{#1}}
\def\cmatpot{\cmati{\potfunc}}

\def\fmatpot{\fmati{\potfunc}}

\def\eigenvalpot{\lambda}
\def\eigenvalstr{\nu}
\def\eigenval{\lambda}
\def\eigenvmatpot{\matx{\Lambda}}
\def\eigenvmatstr{\matx{N}}
\def\potfuncv{\matx{\potfunc}}
\def\streamfuncv{\matx{\streamfunc}}
\def\Xvect{\matx{X}}
\newcommand{\eigenvpoti}[1]{\eigenvalpot_{#1}}
\newcommand{\eigenvstri}[1]{\eigenvalstr_{#1}}
\newcommand{\eigenvali}[1]{\eigenval_{#1}}

\newcommand{\torquei}[1]{\torque_{#1}}

\def\torquez{\torquei{\zz}}
\def\torquezquad{\torque_{\zz;2}}
\def\domain{\mathcal{V}}

\def\idiss{{\rm diss}}
\def\power{\mathcal{P}}
\def\powertide{\power_{\iforcing}}
\def\powertideoc{\power_{\iforcing ; \iocean}}
\def\powertidesol{\power_{\iforcing ; \isol}}
\def\powerdiss{\power_{\idiss}}
\def\powerdissoc{\power_{\idiss ; \iocean}}
\def\powerdisssol{\power_{\idiss ; \isol}}

\def\irefcase{*}
\def\Hocrc{\Hoc^{\irefcase}}
\def\fdragrc{\fdrag^{\irefcase}}
\def\colscontrc{\colscont^{\irefcase}}
\def\colcontrc{\colcont^{\irefcase}}
\def\muplarc{\mupla^{\irefcase}}
\def\tauArc{\tauA^{\irefcase}}

\def\Pl{\LegP{\llat}}                                           
\def\Plm{\LegF{\llat}{\mm}}                             

\newcommand{\eq}[1]{Eq.~(\ref{#1})}
\newcommand{\eqs}[2]{Eqs.~(\ref{#1}) and~(\ref{#2})}
\newcommand{\eqsto}[2]{Eqs.~(\ref{#1}-\ref{#2})}

\newcommand{\eqsfour}[4]{Eqs.~(\ref{#1}), (\ref{#2}), (\ref{#3}), and~(\ref{#4})}
\newcommand{\append}[1]{Appendix~\ref{#1}}

\newcommand{\fig}[1]{Fig.~\ref{#1}}

\newcommand{\sect}[1]{Sect.~\ref{#1}}

\newcommand{\comments}[1]{}

\usepackage{etoolbox}
\usepackage[refpage]{nomencl}

\providetoggle{nomsort}
\settoggle{nomsort}{true} 

\makeatletter
\iftoggle{nomsort}{%
    \let\old@@@nomenclature=\@@@nomenclature        
        \newcounter{@nomcount} \setcounter{@nomcount}{0}%
        \newcommand{\threedigits}[1]{\ifnum#1<100 0\two@digits{#1} \else \number#1\fi}
        \renewcommand\the@nomcount{\threedigits{\value{@nomcount}}}
        \def\@@@nomenclature[#1]#2#3{
          \addtocounter{@nomcount}{1}%
        \def\@tempa{#2}\def\@tempb{#3}%
          \protected@write\@nomenclaturefile{}%
          {\string\nomenclatureentry{\the@nomcount\nom@verb\@tempa @[{\nom@verb\@tempa}]%
          \begingroup\nom@verb\@tempb\protect\nomeqref{\theequation}%
          |nompageref}{\thepage}}%
          \endgroup
          \@esphack}%
      }{}
\makeatother

\newcommand{\mynomone}[3][section]{%
  \begingroup\edef\x{\endgroup
  \unexpanded{\nomenclature{#2}}%
    {\unexpanded{#3} \hspace*{\fill}  (\csname the#1\endcsname)}}\x}

\newcommand{\mynomtwo}[4][section]{%
  \begingroup\edef\x{\endgroup
  \unexpanded{\nomenclature[#2]{#3}}%
    {\unexpanded{#4} \hspace*{\fill}  (\csname the#1\endcsname)}}\x}

\renewcommand\nomgroup[1]{%
  \item[\bfseries
  \ifstrequal{#1}{A}{Acronyms}{%
  \ifstrequal{#1}{S}{Symbols}{%
  \ifstrequal{#1}{C}{Other Symbols}{}}}%
]}

\newcounter{logglabel}

\newcommand{\mynom}[3][S]{\nomenclature[#1]{#2}{#3~}}

\makenomenclature

\begin{document}

  \title{Can one hear supercontinents in the tides of ocean planets?}


  \author{Pierre Auclair-Desrotour
  \and Mohammad Farhat
  \and Gwenaël Boué
  \and Mickaël Gastineau
  \and Jacques Laskar
          }

  \institute{IMCCE, Observatoire de Paris, Université PSL, CNRS, Sorbonne Université, 77 Avenue Denfert-Rochereau, 75014 Paris, France \\
    \email{pierre.auclair-desrotour@obspm.fr} 
  }

  \date{Received ...; accepted ...}

  \abstract
{Recent observations and theoretical progress made about the history of the Earth-Moon system suggest that tidal dissipation in oceans primarily drives the long term evolution of orbital systems hosting ocean planets. Particularly, they emphasise the key role played by the geometry of land-ocean distributions in this mechanism. However, the complex way continents affect oceanic tides still remains to be elucidated.}
   {In the present study, we investigate the impact of a single supercontinent on the tidal response of an ocean planet and the induced tidally dissipated energy.}
   {The adopted approach is based on the linear tidal theory. By simplifying the continent to a spherical cap of given angular radius and position on the globe, we proceed to a harmonic analysis of the whole planet's tidal response including the coupling with the solid part due to ocean loading and self-attraction variations. In this framework, tidal flows are formulated analytically in terms of explicitly defined oceanic eigenmodes, as well as the resulting tidal Love numbers, dissipated power, and torque.}
   {The analysis highlights the symmetry breaking effect of the continent, which makes the dependence of tidal quantities on the tidal frequency become highly irregular. The metric introduced to quantify this continentality effect reveals abrupt transitions between polar and non-polar configurations, and between small-sized and medium-sized continents. Additionally, it predicts that a continent similar to South America or smaller (${\sim}30^\degree$-angular radius) does not alter qualitatively the tidal response of a global ocean whatever its position on the planet.   }
   {}


\maketitle



\section{Introduction}
\label{sec:intro}

`Can one hear the shape of a drum?' This question, raised by \cite{Kac1966}, emphasises the link connecting the geometry of a vibrating system to the frequencies at which it can vibrate. Mathematically, it defines a broad class of problems referred to as spectral geometry, which aims to establish relationships between the eigenmodes of Riemannian manifolds and their geometric features. In the aforementioned study, Kac investigates whether it is possible or not to infer some information about the shape of a vibrating drumhead from the sound it makes, which amounts to recovering the drumhead's geometry from its acoustic signature in an inverse problem approach. Unfortunately, the answer to this question was shown to be negative in the general case, as two different shapes are able to produce the very same sound \citep[e.g.][]{GWW1992}. Nevertheless, it remains possible to infer some specific geometric features of the system to a certain extent. 

Interestingly, this statement is also applicable to tidally forced ocean planets, which may be regarded as giant celestial vibrating drumheads. In the general case, one calls `ocean planets' rocky bodies that are partly or totally covered by a surface liquid layer\footnote{The extended definition of ocean planets -- or ocean worlds -- may also refer to bodies containing a substantial amount of water in the form of subsurface oceans, such as Jupiter's moon Europa \citep[e.g.][]{Kivelson2000}.} \citep[e.g.][]{Leger2004}. With more than 70 percent of its surface covered by oceans \citep[e.g.][]{ES2010}, the Earth thus appears as a typical ocean planet. It is actually the only planet of this kind that can be found presently in the Solar system, although observational evidences indicate that Mars possibly had large paleo-oceans too before it dried \citep[][]{CH2003,Dohm2009,Scheller2021}.

This uniqueness of the Earth, however, diminishes if one considers extrasolar systems. The remarkable exoplanetary diversity established thus far suggests that ocean planets are rather ordinary, if not widespread. Particularly, a substantial amount of rocky planets were found to orbit in the habitable zone of their host stars, which is the region where temperature conditions can sustain liquid water on the planets' surfaces \citep[][]{Kasting1993}. This is the case of many rocky planets orbiting red and brown dwarfs \citep[e.g.][]{PL2007,Raymond2007,Kopparapu2017} such as the super-Earths GJ~1214~b \citep[e.g.][]{Charbonneau2009}, LHS~1140~b \citep[e.g.][]{Lillo-Box2020}, and TOI-1452~b \citep[][]{Cadieux2022}, or the Earth-sized planets TRAPPIST-1 e, f, and g \citep[e.g.][]{Grimm2018}, the latter being suspected to harbour ocean-scale volumes of liquid water \citep[e.g.][]{Bolmont2017,Bourrier2017,Turbet2018}. Improving our understanding of the surface conditions, climate, and fate of these planets requires to better constrain their long-term orbital evolution \citep[e.g.][]{Pierrehumbert2010}.

Tides are the main mechanism that controls the evolution of planetary systems over timescales ranging from millions to billions of years \citep[e.g.][]{Hut1981,CL2001,Levrard2009}. Apart from the atmospheric thermal tides that are induced by insolation variations \citep[e.g.][]{LC1969,Leconte2015,ADL2019}, they result from the mutual gravitational interactions between celestial bodies, which undergo the differential attraction of their neighbours. Combined with dissipative processes, this tidal forcing generates a delayed mass redistribution, thereby leading to exchanges of angular momentum between the orbit of the tidal perturber and the spin of the tidally forced body \citep[e.g.][]{Correia2014}. Additionally, tides are accompanied with energy dissipation where the mechanical energy lost by the orbital system is converted into heat. While it is negligible on Earth, this tidal heating is sometimes able to partly melt the body and to generate surface volcanism, as observed on the Jovian moon Io \citep[e.g.][]{Peale1979}. 

Oceanic tides take the form of frequency-resonant surface gravity modes distorted by the planet's rotation \citep[e.g.][]{ADLML2018}, which strongly differs from the smooth viscoelastic elongation of solid bodies \citep[e.g.][]{Correia2014}. As a consequence, tidal dissipation in oceans may be increased by several orders of magnitude while crossing a resonance associated with a predominant mode \citep[e.g.][]{AG2010}. This resonant behaviour notably explains why, for present day Earth, the oceanic contribution (${\sim}2.5~$TW) to the semidiurnal component ($M_2$) of dissipated tidal energy is an order of magnitude greater than the solid counterpart \citep[][]{Lambeck1977,LPL1997,ER2001,Tyler2021}. Analogously, oceanic tidal flows presumably warm up the icy moons of the Solar system outer planets that are suspected to harbour subsurface oceans, such as Europa, Callisto, or Titan \citep[e.g.][]{Tyler2008,Tyler2014,Kamata2015,Beuthe2016,Matsuyama2018}.

Starting with \cite{MacDonald1964}, several authors investigated the role played by oceanic tides in the evolution of the Earth-Moon system, using either semi-analytical approaches with simplified geometries \citep[e.g.][]{Webb1980,Webb1982,BR1999,Farhat2022b} -- including global ocean models \citep[e.g.][]{ADLML2018,Motoyama2020,Tyler2021} --, or numerical methods based upon realistic land-ocean distributions \citep[e.g.][]{LeProvost1994,LeProvost1998,Arbic2010,Kodaira2016,Green2017,Blackledge2020,Daher2021}. For a didactical review of the pioneering studies of the field, the reader is referred to \cite{Lambeck1977} and \cite{BR1999}. This long series of works particularly emphasises the strong interplay between the resonant oceanic tidal flows and continents, which shapes the four billion year-history of the Earth's length of the day (LOD)\mynom[A]{LOD}{Length of the day} and Earth-Moon distance \citep[see e.g.][]{Daher2021,Farhat2022b}. Moreover, \cite{Green2017} show that the topography significantly increases tidal dissipation, and \cite{Blackledge2020} underline the sensitivity of the latter to coastlines' fractality. However, the way the large-scale geometry of the ocean basin alters the planet's tidal response cannot be easily disentangled from other effects in realistic models due to its complexity. 

The present work is an attempt to address this issue by generalising the semi-analytical tidal theory of hemispherical oceans \citep[][]{LHP1970,Webb1980,Webb1982,Farhat2022b} to ocean basins of arbitrary sizes. In this approach, the oceanic depth is assumed to be uniform and thin compared to the planet radius. Adopting the so-called 'shallow water' approximation \citep[e.g.][]{Vallis2006}, we ignore both the ocean stratification and the associated baroclinic component of tidal flows -- that is the contribution of internal gravity waves \citep[e.g.][]{GZ2008} -- so that the described tide is purely barotropic. Also, the geometry of the continent is simplified to a spherical cap of specified angular radius in order to avoid mathematical complications. Finally, bottom friction is modelled by a standard Rayleigh drag, and the coupling effect between the ocean and the deformable solid surface is taken into account. Analogously with the examples of the global and hemispherical oceans, this formalism allows the planet's tidal response and the resulting dissipated energy to be formulated in terms of explicit eigenmodes, each of these eigenmodes being resonant for a specific tidal frequency. Additionally, the ocean dynamics is controlled by a small number of dimensionless parameters, which provides an appropriate framework for probing the parameter space and characterising the continentality effect. 

In \sect{sec:oceanic_response}, we detail the theory by successively introducing the Laplace's Tidal Equations (LTEs)\mynom[A]{LTEs}{Laplace's Tidal Equations} that govern the oceanic tidal response, the eigenmodes of the ocean basin, and the expressions of the tidal Love numbers, torque, and power. In \sect{sec:reference_case}, we examine the frequency-behaviour of an Earth-sized planet with an equatorial hemispherical ocean. This reference case is used in \sect{sec:parametric_study} to characterise the sensitivity of the planet's tidal response to the position of the continent on the globe, its size, the oceanic depth, the dissipation timescale, the elasticity of the solid part, and its anelasticity timescale. We elaborate on the continentality effect in \sect{sec:metric_continent} by introducing a metric that quantifies the sensitivity of the tidal torque to the position and size of the supercontinent. Finally, in \sect{sec:conclusions}, the conclusions of the study are summarised. We stress here that \sect{sec:oceanic_response} details technical aspects of the developed model. Thus we invite the reader that might be only interested in the application of the theory to skip this section and to jump directly to \sect{sec:reference_case}. It is also noteworthy that all the notations introduced in the main text can be retrieved in the nomenclature made in \append{app:nomenclature}.

\section{Oceanic tidal response}
\label{sec:oceanic_response}

\begin{figure}[htb]
   \centering
   \includegraphics[width=0.48\textwidth,trim = 0.cm 0.cm 490pt 366pt,clip]{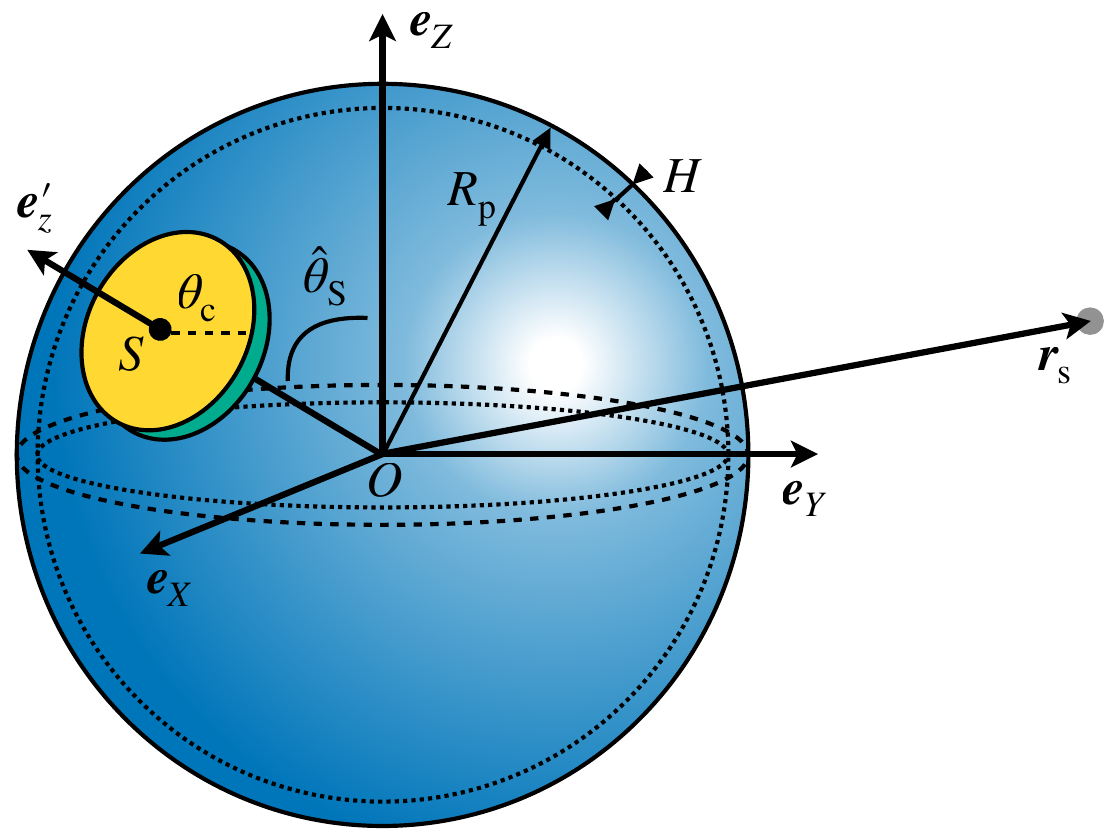}
      \caption{Geometry of the studied system and associated parameters. The unit vector $\ez^\prime$, which indicates the position of the continental centre, is defined as $\ez^\prime \define - \ez$, with $\ez$ pointing towards the centre of the ocean basin. In the diagram, the longitude of the continental centre is set to $\lonscont = 0$.}
       \label{fig:geometry_planet}%
\end{figure}

We establish here the equations governing the tidal dynamics of the ocean basin harboured by a rocky planet of radius $\Rpla$\mynom{$\Rpla$}{Planet radius}. These equations are based on the commonly used shallow water approach, where the ocean is considered as a thin liquid layer of uniform density and depth $\Hoc \ll \Rpla$ \citep[e.g.][]{Vallis2006}\mynom{$\Hoc$}{Oceanic depth}. In this approach, the fluid is supposed to be incompressible. The tidal response thus described is said to be barotropic because it does not depend on the vertical structure of the ocean \citep[][]{Vallis2006}. The continental geometry is simplified to a spherical cap of angular radius $\colcont$\mynom{$\colcont$}{Angular radius of the supercontinent}, as shown by \fig{fig:geometry_planet}. Accordingly, the angular radius of the ocean basin is defined as $\colbd \define \pi - \colcont$\mynom{$\colbd$}{Angular radius of the ocean basin}. 

\subsection{Geometry of the ocean basin}
\label{ssec:geometry_ocean}

To write down the equations that describe the oceanic tidal waves, we shall preliminarily introduce two frames of references and their associated systems of coordinates. First, we denote by $\rframe{}{O}{\eX}{\eY}{\eZ} $\mynom{$\framepla$}{Geocentric frame of reference rotating with the planet}\mynom{$\eX$}{Cartesian unit vector associated with $\framepla$}\mynom{$\eY$}{Cartesian unit vector associated with $\framepla$}\mynom{$\eZ$}{Cartesian unit vector associated with $\framepla$} the frame of reference rotating with the planet and having its centre of gravity, O, as origin. The Cartesian basis of unit vectors $\left( \eX, \eY, \eZ \right)$ is such that $\eX$ and $\eY$ correspond to two orthogonal directions in the planet's equatorial plane, and $\eZ$ to the direction of the spin vector. The unit vector $\eZ$ thus designates the position of the North pole on the unit sphere. The geocentric frame of reference $\framepla$ is associated with the usual spherical coordinates $\left( \rrpla, \colpla, \lonpla \right)$, where $\rrpla$\mynom{$\rrpla$}{Radial coordinate in $\framepla$}, $\colpla$\mynom{$\colpla$}{Colatitude in $\framepla$}, and $\lonpla$\mynom{$\lonpla$}{Longitude in $\framepla$} designate the radial, colatitudinal and longitudinal coordinates, respectively. The position of the circular supercontinent on the globe is defined by the coordinates of its centre, $ \left( \colscont, \lonscont \right)$\mynom{$\colscont$}{Colatitude of the continental centre in $\framepla$}\mynom{$\lonscont$}{Longitude of the continental centre in $\framepla$}. As the centre of the oceanic basin corresponds to the antipodal point, its coordinates on the globe are given by $\left( \coloc, \lonoc \right) = \left( \pi - \colscont , \pi + \lonscont  \right)$\mynom{$\coloc$}{Colatitude of the oceanic centre in $\framepla$}\mynom{$\lonoc$}{Longitude of the continental centre in $\framepla$}. 

The oceanic frame of reference, $\rframe{\iocean}{O}{\ex}{\ey}{\ez} \mynom{$O$}{Planet's centre of mass}$\mynom{$\frameoc$}{Frame of reference rotating with the planet and oriented by the centre of the ocean basin}\mynom{$\ex$}{Cartesian unit vector associated with $\frameoc$}\mynom{$\ey$}{Cartesian unit vector associated with $\frameoc$}\mynom{$\ez$}{Cartesian unit vector associated with $\frameoc$}, is such that $\ez$ points towards the centre of the ocean basin, while $\ex$ and $\ey$ define two orthogonal axes in the plane normal to $\ez$. This frame of reference is associated with the spherical coordinates $\left( \rr , \col , \lon \right)$, where $\rr = \rrpla$\mynom{$\rr$}{Radial coordinate in $\frameoc$} is the radial coordinate, $\col$\mynom{$\col$}{Colatitude in $\frameoc$} the colatitude ($\col=0$ at the centre of the ocean basin), and $\lon$\mynom{$\lon$}{Longitude in $\frameoc$} the longitude. The change of basis vectors $\left( \eX, \eY, \eZ \right) \rightarrow \left( \ex , \ey , \ez \right) $ is expressed as a function of the coordinates of the continental centre as 
\begin{equation}
\begin{bmatrix}
\ex \\ \ey \\ \ez
\end{bmatrix}
= 
\begin{bmatrix}
\cos \lonscont \cos \colscont & \sin \lonscont \cos \colscont & - \sin \colscont \\
 \sin \lonscont & - \cos \lonscont  & 0 \\ 
- \cos \lonscont \sin \colscont & - \sin \lonscont \sin \colscont & - \cos \colscont
\end{bmatrix}
\begin{bmatrix}
\eX \\ \eY \\ \eZ
\end{bmatrix}.
\label{change_basis_vectors_1}
\end{equation}
We note that the longitude of the continental centre does not affect the tidal response since the tidal forcing is periodic in longitude. Therefore, we set this longitude to $\lonscont = 0$ in the following, similar to the example shown by \fig{fig:geometry_planet}, which yields
\begin{equation}
\begin{bmatrix}
\ex \\ \ey \\ \ez
\end{bmatrix}
= 
\begin{bmatrix}
 \cos \colscont & 0 & - \sin \colscont \\
 0 & - 1 & 0 \\ 
-  \sin \colscont & 0 & - \cos \colscont 
\end{bmatrix}
\begin{bmatrix}
\eX \\ \eY \\ \eZ
\end{bmatrix}.
\label{change_basis_vectors}
\end{equation}
Finally, we introduce the set of unit vectors $\left( \er , \etheta, \ephi \right)$\mynom{$\er$}{Radial unit vector associated with $\frameoc$}\mynom{$\etheta$}{Colatitudinal unit vector associated with $\frameoc$}\mynom{$\ephi$}{Longitudinal unit vector associated with $\frameoc$} associated with the spherical coordinates $\left( \rr, \col , \lon \right)$, respectively. The change of basis vectors $\left( \ex , \ey , \ez \right) \rightarrow \left( \er , \etheta, \ephi \right)$ is given by the standard relation,
\begin{equation}
\begin{bmatrix}
\er  \\ \etheta \\ \ephi 
\end{bmatrix}
= 
\begin{bmatrix}
\sin \col \cos \lon & \sin \col \sin \lon & \cos \col \\
\cos \col \cos \lon & \cos \col \cos \lon & - \sin  \col \\
- \sin \lon & \cos \lon & 0
\end{bmatrix}
\begin{bmatrix}
\ex \\ \ey \\ \ez
\end{bmatrix}.
\end{equation}

\subsection{Laplace's tidal equations}
\label{ssec:laplace_equations}

The planet is subject to the effects of the tidal gravitational potential generated by the perturber (the Moon or the Sun in the Earth case), which is expressed, in the frame of reference $\framestd$, as \mynom{$\Utide$}{Tidal gravitational potential generated by the perturber}
\begin{equation}
\Utide \define \frac{\Ggrav \Mpert}{\abs{\Rpla \er - \rpertvect}} - \frac{\Ggrav \Mpert}{\rpert^2} \Rpla \cos \colpla - \frac{\Ggrav \Mpert}{\rpert},
\label{tidalpot}
\end{equation}
where the constant component and the component responsible for the Keplerian dynamics of the two-body system are removed (second and third terms in the right-hand member of \eq{tidalpot}). In this equation, $\Ggrav$\mynom{$\Ggrav$}{Universal gravitational constant} is the universal gravitational constant, $\Mpert $\mynom{$\Mpert$}{Mass of the perturber} the mass of the perturber, $\rpertvect$\mynom{$\rpertvect$}{Position vector of the perturber with respect to $O$} its position vector, and $\rpert \define \abs{\rpertvect}$\mynom{$\rpert$}{Planet-perturber distance} the planet-perturber distance. We note that the symbol $\define$\mynom{$\define$}{Symbol used for definitions in equations} is used throughout the text to distinguish between definitions and equalities. \rec{The tidal force per unit mass exerted by the perturber on the planet is given by $\forcev \define \grad \Utide$\mynom{$\forcev$}{Tidal force per unit mass induced by $\Utide$}. }Following the formalism introduced in earlier works \citep[e.g.][]{Tyler2011,Matsuyama2014,ADLML2018,Auclair2019,Motoyama2020,Farhat2022b}, we write down the momentum and mass conservation equations, respectively, as
\begin{align}
\label{momentum}
\dd{\Vvect}{\time} + \fdrag \Vvect + \fcorio \crossp \Vvect + \ggravi \grad \left( \tiltopd \zetaoc - \tiltopg \zetaeq \right)  &= 0, \\
\label{continuity}
\dd{\zetaoc}{\time} + \div  \left( \Hoc \Vvect \right) &= 0, 
\end{align}
with $\time$\mynom{$\time$}{Time} designating the time, $\ggravi$\mynom{$\ggravi$}{Surface gravity at rest} the surface gravity at rest, $\fdrag $\mynom{$\fdrag$}{Rayleigh drag frequency} the Rayleigh drag frequency used to describe the action of dissipative mechanisms, $\Vvect$\mynom{$\Vvect$}{Horizontal velocity vector} the horizontal velocity vector -- defined from the horizontal displacement vector $\hdep$\mynom{$\hdep$}{Horizontal displacement vector} as $\Vvect  \define \dd{\hdep}{\time} $ --, $\fcorio$\mynom{$\fcorio$}{Coriolis parameter} the Coriolis parameter, $\zetaoc$\mynom{$\zetaoc$}{Vertical displacement of the ocean's surface with respect to the oceanic floor} the vertical displacement of the ocean's surface with respect to the oceanic floor, and $\zetaeq \define \Utide / \ggravi$\mynom{$\zetaeq$}{Equilibrium displacement corresponding to the equipotential surface} the equilibrium displacement corresponding to the equipotential surface induced by the tidal gravitational potential. 

In the momentum equation, the notations $\tiltopd$\mynom{$\tiltopd,\tiltopg$}{Solid deformation operators} and $\tiltopg$ refer to non-trivial linear operators accounting for the effects of ocean loading, self-attraction, and deformation of the solid regions of the planet \citep[e.g.][]{Hendershott1972}. These operators are called the `solid deformation operators' in the following. They encompass the complex coupling between the oceanic shell and the solid interior described by Poisson's equation, and the momentum and rheological equations governing the tidal dynamics of the solid part. The two operators simplify to $\tiltopd = \tiltopg =1$ if one neglects both the tidal deformation of the solid part (infinite-rigidity approximation) and the variation of self-attraction induced by the oceanic tidal response \citep[Cowling approximation; see][]{Cowling1941,Unno1989}. In the general case, they are determined by the rheological behaviour of the solid part in its response to gravitational and surface forcings. 

Equations~(\ref{momentum}) and~(\ref{continuity}) are known as the Laplace's tidal equations (hereafter, LTEs) in reference to Laplace's masterpiece \citep[][]{Laplace1798}. We note that $\zetaeq$ results from the coupled oceanic tidal response and tidal deformation of the solid part, which includes both gravitational and loading interactions, as discussed further. The Coriolis parameter is expressed as a function of the colatitude of the current point in $\framepla$, 
\begin{equation}
\fcorio \define 2 \spinrate \cos \colpla \, \er,
\end{equation}
where $\spinrate>0$\mynom{$\spinrate$}{Planet's spin angular velocity} is the planet's spin angular velocity. The horizontal gradient operator and divergence of the horizontal velocity are expressed in $\frameoc$, and the associated basis vectors, $\left( \etheta , \ephi \right)$, as
\begin{align}
\grad \define & \ \Rpla^{-1} \left[ \etheta \dd{}{\col} + \ephi \left( \sin \col \right)^{-1} \dd{}{\lon} \right], \\
\div \Vvect = & \left( \Rpla \sin \col \right)^{-1} \left[ \dd{}{\col} \left( \sin \col \Vtheta \right) + \dd{\Vphi}{\lon} \right].
\end{align}
The Rayleigh drag term $\fdrag \Vvect$ in \eq{momentum} accounts for the cumulated effects of dissipative mechanisms on tidal flows. The frequency $\fdrag$ is the inverse of the effective dissipation timescale associated with these dissipative mechanisms. It takes values around ${\sim}10^{-5}~{\rm s^{-1}}$ for the present day Earth \citep[e.g.][]{Webb1980,Farhat2022b}\footnote{Other values of $\fdrag$ may be found in the literature. For example, \cite{Wunsch1997} uses a much smaller value to study the dynamics of the long-period tides, $\fdrag = 2.5 \times 10^{-7} \ {\rm s^{-1}}$, while \cite{Motoyama2020} set this value to $\fdrag \sim 4 \times 10^{-6} \ {\rm s^{-1}}$ following the prescription given by \cite{Schwiderski1980}.}.

\begin{table*}[h]
\centering
\caption{\label{tab:control_param} Dimensionless control parameters determining the regime of the oceanic tidal response.   }
\begin{tabular}{lllll} 
 \hline 
 \hline 
\textsc{Parameter} & \textsc{Description} & \textsc{Limits} & \textsc{Asymptotic regimes} & \textsc{Reference}  \\ 
 \hline 
  \multicolumn{5}{c}{\textit{Geometry of the ocean basin}} \\[0.1cm]
   \multirow{2}{*}{$\colcont$} &  \multirow{2}{*}{Size of the supercontinent} & $\colcont = 0^\degree$ & Global ocean & \multirow{2}{*}{\fig{fig:geometry_planet}} \\
   & & $\colcont = 180^\degree$ & Dry planet & \\[0.1cm]
   \multirow{2}{*}{$\colscont$} &  \multirow{2}{*}{Position of the supercontinent on the globe} & $\colscont = 0^\degree$ & Polar continent & \multirow{2}{*}{\eq{change_basis_vectors_1}} \\
   & & $\colscont = 90^\degree$ & Equatorial continent & \\[0.1cm] 
  \multicolumn{5}{c}{\textit{Ocean dynamics}} \\[0.1cm]
\multirow{2}{*}{ $\ftiden \define \dfrac{\ftide}{2 \spinrate}$} & \multirow{2}{*}{Distortion of forced tidal waves by Coriolis forces} & $\ftiden \ll 1$ & Sub-inertial regime & \multirow{2}{*}{\eq{ftiden}}  \\
& & $\ftiden \gg 1$ & Super-inertial regime &  \\[0.3cm] 
 \multirow{2}{*}{$ \Rowave \define \dfrac{\sqrt{\ggravi \Hoc}}{2 \spinrate \Rpla}$} & \multirow{2}{*}{Deviation of free surface waves by Coriolis forces} & $ \Rowave \ll 1$ & Fast rotator regime & \multirow{2}{*}{\eq{Rowave_fdragn}} \\
 & & $ \Rowave \gg 1$ & Slow rotator regime & \\[0.3cm]
  \multirow{2}{*}{$\fdragn \define \dfrac{\fdrag}{2 \spinrate }$} &  \multirow{2}{*}{Damping of Coriolis effects by friction} & $\fdragn \ll 1$ & Quasi-adiabatic regime & \multirow{2}{*}{\eq{Rowave_fdragn}} \\
  & & $\fdragn \gg 1$ & Frictional regime & \\[0.3cm]
 \multicolumn{5}{c}{\textit{Solid deformation}} \\[0.1cm]
 \multirow{2}{*}{$\tiltdi{}, \tiltgi{} $} & \multirow{2}{*}{Visco-elastic response of the solid part} & $\tiltdi{}, \tiltgi{} = 1 $ & Rigid body (with CA) & \multirow{2}{*}{\eq{tiltd}} \\
 & & else & Solid-ocean coupling &  \\[0.1cm]
\hline
 \end{tabular}
 \tablefoot{The acronym `CA'\mynom[A]{CA}{Cowling approximation} refers to the Cowling approximation \citep[][]{Cowling1941}, mentioned in \sect{ssec:laplace_equations}.}
 \end{table*}

\subsection{Nondimensional tidal equations}

The nondimensional momentum and continuity equations are obtained by choosing as reference time and velocity scales the inertial time, $\timeref$\mynom{$\timeref$}{Reference time for normalisation}, and the typical velocity of long-wavelength surface gravity waves, $\speedref$\mynom{$\speedref$}{Reference velocity for normalisation}, which are defined, respectively, as
\begin{align}
\label{time_velocity_ref}
& \timeref \define \left( 2 \spinrate \right)^{-1}, & \speedref \define \sqrt{\ggravi \Hoc}.
\end{align}
Introducing the normalised time $\timen$\mynom{$\timen$}{Normalised time}, Coriolis parameter $\fcorion$\mynom{$\fcorion$}{Normalised Coriolis parameter}, horizontal gradient $\gradn$\mynom{$\gradn$}{Normalised gradient operator}, the complex horizontal displacement vector $\hdepn$\mynom{$\hdepn$}{Normalised horizontal displacement vector}, and vertical displacements $\zetaocn$\mynom{$\zetaocn$}{Normalised oceanic surface elevation} and $\zetaeqn$\mynom{$\zetaeqn$}{Normalised equilibrium surface elevation}, such that 
\begin{equation}
\begin{array}{lll}
\time = \timeref \timen, & \fcorio = \timeref^{-1} \fcorion & \grad = \Rpla^{-1} \gradn,  \\
 \hdep = \Re \left( \Rpla \hdepn \right) , & \zetaoc =  \Re \left( \Hoc \zetaocn \right), & \zetaeq = \Re \left( \Hoc \zetaeqn  \right), 
\end{array}
\end{equation}
with $\Re$ referring to the real part of a complex number, we end up with the nondimensional complex LTEs given by
\begin{align}
\label{momentum_norm}
\left[ \ddd{}{\timen}{\timen}  + \left( \fdragn + \fcorion \crossp \right) \dd{}{\timen} \right] \hdepn + \Rowave^2 \gradn \Ftideno& = 0 , \\
\label{continuity_norm}
\zetaocn + \divn \hdepn & = 0,
\end{align}
with $\Ftideno \define \tiltopd \zetaocn - \tiltopg \zetaeqn$\mynom{$\Ftideno$}{Normalised forcing term}. In the above equations, tidal dynamics are controlled by two dimensionless parameters,
\begin{align}
\label{Rowave_fdragn}
& \Rowave \define \frac{\speedref}{2 \spinrate \Rpla} =\frac{\sqrt{\ggravi \Hoc}}{2 \spinrate \Rpla}, &\fdragn \define \frac{\fdrag}{2 \spinrate }.
\end{align}

The first parameter, $\Rowave$\mynom{$\Rowave$}{Normalised Rossby deformation length}, can be considered as a normalised Rossby deformation length since it compares the typical propagation velocity of surface gravity waves ($\speedref$) with the Earth's rotation velocity. If $\Rowave \ll 1$ (fast rotator regime), the inertial forces resulting from Coriolis acceleration predominate with respect to the restoring forces of surface gravity waves (pressure forces and gravity). Conversely, if $\Rowave  \gg 1$ (slow rotator regime), Coriolis terms are not strong enough to significantly deviate free surface gravity waves. As it describes the ratio of drag forces to Coriolis forces, the second dimensionless parameter, $\fdragn$\mynom{$\fdragn$}{Normalised friction parameter}, may be regarded as an Ekman number. If $\fdragn \ll 1$, the drag does not alter the tidal response much. Conversely, $\fdragn \gg 1$ characterises a frictional (or viscous) regime where inertial effects are annihilated by the strong damping associated with the drag. 

In addition to $\Rowave $ and $\fdragn$, the time-derivative operators in \eq{momentum_norm} introduce a third dimensionless parameter describing the ratio of tidal forces to Coriolis forces,
\begin{equation}
\label{ftiden}
\ftiden \define \frac{\ftide}{2 \spinrate}.
\end{equation}
The notation $\ftide$\mynom{$\ftide$}{Tidal frequency} in the above equation designates the typical frequency of tidal flows, which will serve as the tidal frequency of the considered tidal force in the Fourier expansion of tidal quantities detailed further. If $\ftiden \ll 1$ (sub-inertial regime)\mynom{$\ftiden$}{Rossby number of forced waves}, the acceleration term of the momentum equation can be neglected with respect to Coriolis terms. Conversely, if $\ftiden \gg 1$ (super-inertial regime), the flow is strongly driven by the tidal forcing and it is thus hardly distorted by the planet's rotation. We note that $\ftiden$ is the inverse of the so-called spin parameter that is commonly used to characterise the oscillations of rotating fluids in planetary and stellar hydrodynamics \citep[e.g.][]{LS1997}. The dimensionless parameters that control the planet's tidal response are summarised in Table~\ref{tab:control_param}.

\subsection{Helmholtz decomposition}
\label{ssec:helmholtz_decomposition}

\cite{Proudman1920} demonstrated that Helmholtz's theorem \citep[e.g.][]{Arfken2005} can be used to decompose the horizontal displacement vector field into curl-free and divergence-free vector fields,
\begin{equation}
\label{helmholtz_xi}
\hdepn = \gradn \potfunc + \gradn \streamfunc \crossp \er .
\end{equation}
The curl-free ($\gradn \crossp \left( \gradn \potfunc \right) = 0$) and divergence-free ($\divn \left( \gradn \streamfunc \crossp \er \right) = 0 $) components of $\hdepn$ are defined from the divergent displacement potential $\potfunc$\mynom{$\potfunc$}{Divergent displacement potential} and the rotational displacement streamfunction $\streamfunc$\mynom{$\streamfunc$}{Rotational displacement streamfunction}, respectively \citep[][]{Webb1980,Tyler2011}. We emphasise that the Helmholtz decomposition is not unique for bounded domains such as the considered ocean basin. This results from the fact that additional physical constraints on the boundary condition are necessary to define $\potfunc$ and $\streamfunc$ \citep[e.g.][]{FoxKemper2003}. However choosing boundary conditions for the two functions is not straightforward given that the two components of \eq{helmholtz_xi} cannot be disentangled in the total flux. For instance, the impermeability condition at the coastline is formulated as $ \hdepn \dotp \nvect = 0 $, where $\nvect$\mynom{$\nvect$}{Outward pointing unit vector defining the normal to the coast} designates the outward pointing unit vector defining the normal to the coast. This theoretically requires to find another boundary condition and to solve it for~$\gradn \potfunc $ and~$\gradn \streamfunc $ together with the first condition, which may lead to significant mathematical complications. 

To circumvent this difficulty, it is convenient to adopt a commonly used trick \citep[e.g.][]{Webb1980,Webb1982,GMW1983,Watterson2001,HH2020,Farhat2022b}, which consists in applying the impermeability condition to both components of \eq{helmholtz_xi},
\begin{align}
\label{bcond_phi_psi}
& \nvect \dotp \gradn \potfunc = 0 , & \nvect \dotp \left( \gradn \streamfunc \crossp \er \right) =0. 
\end{align}
The first condition of \eq{bcond_phi_psi} means that the gradient of $\potfunc$ is zero in the direction normal to the coastline, which is equivalent to Neumann condition \citep[specified value of the derivative of the solution applied at the boundary of the domain; e.g.][Section~6.1]{MF1953}. The second condition may be reformulated as $\left( \er \crossp \nvect \right) \dotp \gradn \streamfunc = 0$, implying that the streamfunction is a constant along the coastline. This corresponds to a Dirichlet condition \citep[specified value of the solution itself;][Section~6.1]{MF1953}. Since both $\potfunc$ and $\streamfunc$ are defined to a constant, the value of $\streamfunc$ at the coastline is set to zero, which simplifies the second condition of \eq{bcond_phi_psi} to $\streamfunc = 0$. Interestingly, this condition induces the orthogonality of the curl-free and divergence-free components of the horizontal displacement \citep[e.g.][Appendix~E]{Farhat2022b}, formulated as 
\begin{equation}
\label{ortho_phi_psi}
\integ{ \conj{\gradn \potfunc}  \dotp \left( \gradn \streamfunc \crossp \er \right) }{\surface}{\oceancap}{} = 0, 
\end{equation}
with $\infvar{\surface} = \sin \col \infvar{\col} \infvar{\lon}$\mynom{$\infvar{\surface}$}{Infinitesimal surface element} being an infinitesimal surface element of the unit sphere, $\oceancap$\mynom{$\oceancap$}{Domain occupied by the ocean basin}\mynom{$\oceancapbd$}{Boundary of the ocean basin} the domain occupied by the ocean basin, and $\conj{\gradn \potfunc} $\mynom{$\conj{z}$}{Conjugate of a complex number ($z$)} the complex conjugate of $\gradn \potfunc$. This property leads to appreciable simplifications in the LTEs, as discussed further. 

The divergent potential function and the streamfunction are defined on a compact connected domain, the ocean basin ($\oceancap$), and they satisfy either Neumann or Dirichlet conditions at its boundary, $\oceancapbd$. As a consequence, the two functions can be expanded in terms of the complete sets of orthogonal eigenfunctions $\left\{ \fpoti{\jjb} \right\}_{1 \leq \jjb \leq \infty}$ or $\left\{ \fstreami{\jjb} \right\}_{1 \leq \jjb \leq \infty}$ that are the solutions of the wave equations given by
\begin{align}
\label{eqwave_fpot}
& \left( \lapn + \eigenvalpot \right) \potfunc = 0 \ \mbox{on} \ \oceancap,  & \nvect \dotp \gradn \potfunc = 0 \ \mbox{at} \ \oceancapbd , \\
\label{eqwave_fstream}
& \left( \lapn + \eigenvalstr \right) \streamfunc = 0 \ \mbox{on} \ \oceancap,  & \streamfunc = 0 \ \mbox{at} \ \oceancapbd ,
\end{align}
where $\lapn$\mynom{$\lapn$}{Normalised Laplacian operator} designates the normalised Laplacian operator defined, for any function $\ffunc$, as
\begin{equation}
\lapn \ffunc \define \left( \sin \col \right)^{-2} \left[ \sin \col \dd{}{\col} \left( \sin \col \dd{\ffunc}{\col} \right) + \ddd{\ffunc}{\lon}{\lon}   \right].
\end{equation}

\def\wbox{1cm}
\def\hraisebox{0.25\textwidth}
\def\verspacem{1.6cm}
\def\horspacen{1.6cm}
\def\horspacenneg{1.4cm}
\begin{figure*}[htb]
   \centering
  \raisebox{\hraisebox}[1cm][0pt]{%
   \begin{minipage}{\wbox}%
   $\fpoti{\jjb}$ 
\end{minipage}}
   \raisebox{\hraisebox}[1cm][0pt]{%
   \begin{minipage}{\wbox}%
 \vspace{1.5cm}  \textsc{$0$} \\[\verspacem] \textsc{$1$} \\[\verspacem] \textsc{$2$} \\[\verspacem] \textsc{$3$} 
\end{minipage}}
   \includegraphics[width=0.73\textwidth,trim = 0.cm 0.cm 0.cm 0cm,clip]{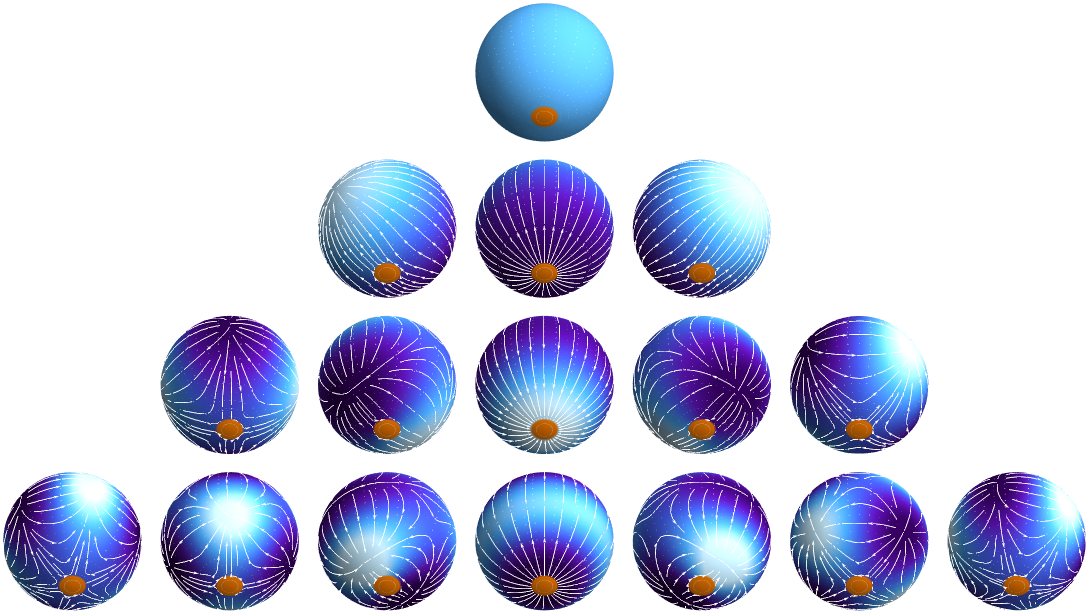} \\
\hspace{\horspacen}  \textsc{$\nk/ \mm$} \hspace{1.cm} $-3$ \hspace{\horspacenneg} $-2$ \hspace{\horspacenneg} $-1$ \hspace{\horspacenneg} $0$ \hspace{\horspacen} $1$ \hspace{\horspacen} $2$ \hspace{\horspacen} $3$  \hspace{1.3cm}~ \\
  \raisebox{\hraisebox}[1cm][0pt]{%
   \begin{minipage}{\wbox}%
   $\fstreami{\jjb}$ 
\end{minipage}}
 \raisebox{\hraisebox}[1cm][0pt]{%
   \begin{minipage}{\wbox}%
  \vspace{1.5cm}   \textsc{$0$} \\[\verspacem] \textsc{$1$} \\[\verspacem] \textsc{$2$} \\[\verspacem] \textsc{$3$} 
\end{minipage}}
   \includegraphics[width=0.73\textwidth,trim = 0.cm 0.cm 0.cm 0cm,clip]{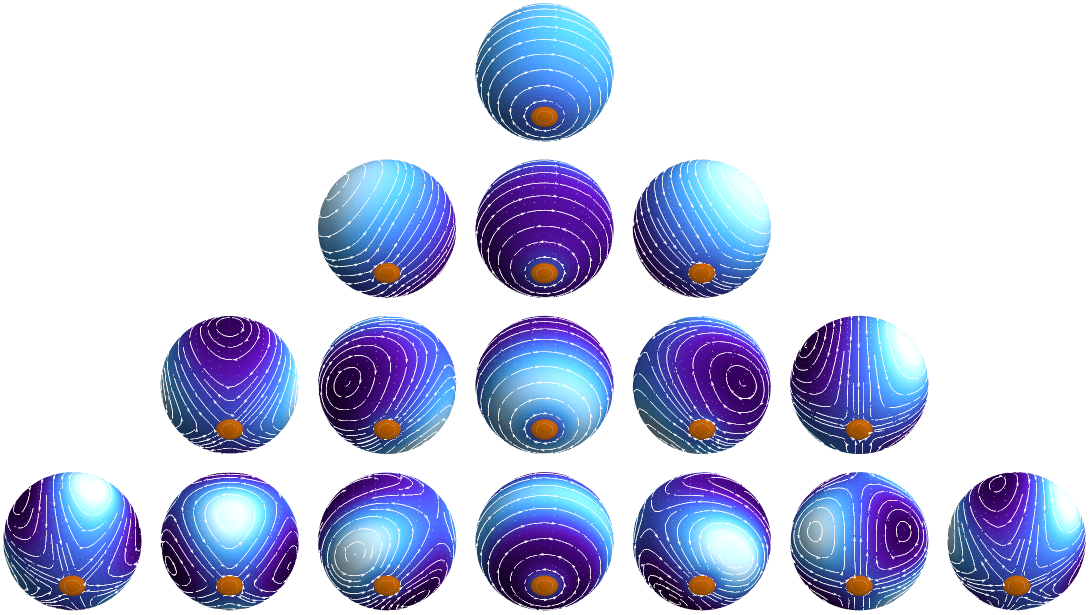} \\
 \hspace{\horspacen}   \textsc{$\nk/ \mm$} \hspace{1.cm} $-3$ \hspace{\horspacenneg} $-2$ \hspace{\horspacenneg} $-1$ \hspace{\horspacenneg} $0$ \hspace{\horspacen} $1$ \hspace{\horspacen} $2$ \hspace{\horspacen} $3$  \hspace{1.3cm}~
      \caption{Eigenfunctions describing the oceanic tidal response for a supercontinent of angular radius $\colcont = 10^\degree$ (i.e. $\colbd = 170^\degree$), and associated tidal flows. {\it Top:} set $\left\{ \fpoti{\jjb} \right\}$ (SCHNs). {\it Bottom:} set $\left\{ \fstreami{\jjb} \right\}$ (SCHDs). The orange disk designates the continent. The eigenfunctions are computed from the expression given by \eq{schc}, and their real parts are plotted for $\llati{\nk}$ such that $\nk = 0 ,  \ldots , 3$ (vertical axis) and $- \nk \leq \mm \leq \nk$ (horizontal axis). Bright or dark colours designate positive or negative values of the eigenfunctions, respectively. Streamlines indicate the tidal flows corresponding to $\gradn \fpoti{\jjb}$ for the set $\left\{ \fpoti{\jjb} \right\}$ and to $\gradn \fstreami{\jjb} \crossp \er$ for the set $\left\{ \fstreami{\jjb} \right\}$.}
       \label{fig:sch}%
\end{figure*}

The solutions of the colatitudinal component of \eqs{eqwave_fpot}{eqwave_fstream} are the associated Legendre functions of the first kind (ALFs)\mynom[A]{ALFs}{Associated Legendre Functions of the first kind} of real degrees, which are detailed in \append{app:legendre_functions}.  The corresponding eigenfunctions $\fpoti{\jjb}$\mynom{$\fpoti{\jjb}$}{$\jjb$-th eigenfunction of the set of SCHNs} are associated with the eigenvalues $\eigenvpoti{\jjb} \in \Rset^+$\mynom{$\eigenvpoti{\jjb}$}{Eigenvalue associated with $\fpoti{\jjb}$}, and the eigenfunctions $\fstreami{\jjb}$\mynom{$\fstreami{\jjb}$}{$\jjb$-th eigenfunction of the set of SCHDs} with the eigenvalues $\eigenvstri{\jjb} \in \Rset^+$\mynom{$\eigenvstri{\jjb} $}{Eigenvalue associated with $\fstreami{\jjb}$}, the relationship between the real degrees of the ALFs and the eigenvalues being detailed in \append{app:sch} (see Table~\ref{tab:sch_eigenv}). These eigenfunctions are obtained by multiplying the ALFs to the solutions of the longitudinal component of \eqs{eqwave_fpot}{eqwave_fstream}, namely $\exp \left( \pm \inumber \mm \lon \right)$, with $\inumber$\mynom{$\inumber$}{Imaginary number} being the imaginary number. The outcome is a set of functions known as spherical cap harmonics \citep[][]{Haines1985,HC1997,Thebault2004,Thebault2006}, and denoted by `SCHs'\mynom[A]{SCHs}{Spherical cap harmonics} in the following (see \eq{schc}, in \append{app:sch}). Thus, the set $\left\{ \fpoti{\jjb} \right\}_{1 \leq \jjb \leq \infty}$ is refered to as the `SCHNs' (Neumann condition)\mynom[A]{SCHNs}{SCHs with Neumann conditions}, and the set $\left\{ \fstreami{\jjb} \right\}_{1 \leq \jjb \leq \infty}$ as the `SCHDs' (Dirichlet condition)\mynom[A]{SCHDs}{SCHs with Dirichlet conditions}. For example, Figure~\ref{fig:sch} shows the first basis functions of the two sets for a supercontinent of angular radius $\colcont = 10^\degree$ (i.e. $\colbd = 170^\degree$). 

One shall notice that the SCHs and their associated eigenvalues both depend on the boundary condition applied at the coastline, meaning that the sets $\left\{ \fpoti{\jjb} \right\}_{1 \leq \jjb \leq \infty}$ and  $\left\{ \fstreami{\jjb} \right\}_{1 \leq \jjb \leq \infty}$ are necessarily different by construction. In each set, the basis functions are orthogonal to each other and normalised so that
\begin{equation}
\label{ortho_eigenf}
\integ{\conj{\fpoti{\jjb}} \fpoti{\kkb}}{\surface}{\oceancap}{} = \integ{\conj{\fstreami{\jjb}}\fstreami{\kkb} }{\surface}{\oceancap}{} = \kron{\jjb}{\kkb},
\end{equation}
where $\kron{\jjb}{\kkb}$\mynom{$\kron{\jjb}{\kkb}$}{Kronecker delta function} designates the Kronecker delta function, such that $\kron{\jjb}{\kkb} = 1$ for $\jjb = \kkb$ and $\kron{\jjb}{\kkb} = 0$ otherwise. However, two functions belonging to different sets are not orthogonal in the general case, 
\begin{equation}
\integ{\conj{\fpoti{\jjb}} \fstreami{\kkb}}{\surface}{\oceancap}{}  \neq 0,
\end{equation}
since the sets of basis functions $\left\{ \fpoti{\jjb} \right\}_{1 \leq \jjb \leq \infty}$ and $\left\{ \fstreami{\jjb} \right\}_{1 \leq \jjb \leq \infty}$ are not the same. 

Introducing the time-dependent coefficients $\ppoti{\jjb}$\mynom{$\ppoti{\jjb}$}{Weighting coefficient of $\fpoti{\jjb} $} and $\pstreami{\jjb}$\mynom{$\pstreami{\jjb}$}{Weighting coefficient of $\fstreami{\jjb} $}, we write down the divergent potential function and the streamfunction as
\begin{align}
\label{potfunc_series}
 \potfunc \left( \col , \lon , \timen \right) = &  \sum_{\jjb=1}^{\infty} \ppoti{\jjb} \left( \timen \right) \fpoti{\jjb} \left( \col, \lon \right) , \\
 \label{streamfunc_series}
 \streamfunc \left( \col , \lon , \timen \right) = & \sum_{\jjb=1}^{\infty} \pstreami{\jjb} \left( \timen \right) \fstreami{\jjb} \left( \col, \lon \right) .
\end{align}
By substituting \eq{potfunc_series} into \eq{continuity_norm}, the continuity equation becomes 
\begin{equation}
\zetaocn +  \lapn \potfunc = 0,
\end{equation}
which implies that 
\begin{equation}
\label{zetaocn}
\zetaocn = \sum_{\jjb=1}^{\infty} \eigenvpoti{\jjb} \ppoti{\jjb} \fpoti{\jjb}. 
\end{equation}
However, such a simple expression cannot be obtained for the equilibrium displacement ($\zetaeqn$) in the general case since it is naturally expanded in series of spherical harmonics (hereafter, SPHs)\mynom[A]{SPHs}{Spherical harmonics} in the coordinate system associated with the planet's spin, which do not satisfy Dirichlet or Neumann conditions at coastlines. The form of the expression given by \eq{zetaocn} can be obtained for $\zetaeqn$ only in the particular case of the hemispheric ocean ($\colbd = 90^\degree$), where the set of eigenfunctions $\left\{ \fpoti{\jjb} \right\}_{1 \leq \jjb \leq \infty}$ actually corresponds to a subset of the SPHs for the hemispherical domain, as shown by \cite{Farhat2022b}.

\subsection{Temporal equations}

The method used to establish the temporal differential equations for the coefficients $\ppoti{\jjb}$ and $\pstreami{\jjb}$ introduced in \eqs{potfunc_series}{streamfunc_series} closely follows that used for the hemispheric ocean configuration \citep[e.g.][]{Webb1980,Webb1982,Farhat2022b}. As a first step, we substitute $\potfunc$ and $\streamfunc$ by their series expansions in \eq{helmholtz_xi} and in the momentum equation given by \eq{momentum_norm}. As a second step, we do the dot product of the equation thus obtained by $ \conj{\gradn \fpoti{\jjb}} $, and we integrate it over the domain of the ocean basin. It follows
\begin{align}
\label{momentum_pr_1}
& \sum_{\kkb=1}^{\infty}  \left\{ \left[ \ddd{ }{\timen}{\timen} + \fdragn \dd{ }{\timen}  \right] \ppoti{\kkb} \integ{\conj{\gradn \fpoti{\jjb}} \dotp \gradn \fpoti{\kkb} }{\surface}{\oceancap}{}   \right.   \\
& + \left(  \ddd{}{\timen}{\timen} + \fdragn \dd{}{\timen}  \right) \pstreami{\kkb} \integ{ \conj{\gradn \fpoti{\jjb}} \dotp  \left( \gradn \fstreami{\kkb} \crossp \er \right) }{\surface}{\oceancap}{} \nonumber \\
&  +  \dd{\pstreami{\kkb}}{\timen} \integ{\conj{\gradn \fpoti{\jjb}} \dotp \left[ \fcorion \crossp \left( \gradn \fstreami{\kkb} \crossp \er  \right) \right] }{\surface}{\oceancap}{}  \nonumber \\
& \left. +  \dd{\ppoti{\kkb}}{\timen} \integ{\! \! \!  \conj{\gradn \fpoti{\jjb}} \dotp \left( \fcorion \crossp \gradn \fpoti{\kkb} \right) }{\surface}{\oceancap}{} \! \right\} \nonumber  + \Rowave^2 \integ{ \! \conj{\gradn \fpoti{\jjb}} \dotp \gradn \Ftideno }{\surface}{\oceancap}{} = 0. \nonumber
\end{align}

The dot product of two eigenfunctions that appears in the first term of \eq{momentum_pr_1} is computed by invoking, successively, Green's first identity \citep[e.g.][Chapter~7]{Strauss2007},
\begin{equation}
\integ{\conj{\gradn \fpoti{\jjb}} \dotp \gradn \fpoti{\kkb} }{\surface}{\oceancap}{} = \integ{\fpoti{\kkb} \left( \conj{\gradn \fpoti{\jjb}} \dotp \nvect \right) }{\length}{\oceancapbd}{} - \integ{\fpoti{\kkb} \conj{\lapn \fpoti{\jjb}} }{\surface}{\oceancap}{},
\end{equation}
the impermeability boundary condition on $\fpoti{\jjb}$ given by \eq{bcond_phi_psi}, the fact that the $\fpoti{\jjb}$ are eigenfunctions of the Laplacian operator as described by \eq{eqwave_fpot}, and the orthogonality property given by \eq{ortho_eigenf}. This yields
\begin{equation}
\label{gradphi}
\integ{\conj{\gradn \fpoti{\jjb}} \dotp \gradn \fpoti{\kkb} }{\surface}{\oceancap}{} = \eigenvpoti{\jjb} \kron{\jjb}{\kkb},
\end{equation}
and for the forcing term, 
\begin{equation}
\integ{\conj{\gradn \fpoti{\jjb}} \dotp \gradn \Ftideno  }{\surface}{\oceancap}{} = \eigenvpoti{\jjb} \integ{\conj{\fpoti{\jjb}} \Ftideno}{\surface }{\oceancap}{}.
\end{equation}
Since the tidal gravitational potential is expressed in terms of the SPHs associated with the coordinates $\left( \colpla , \lonpla \right)$, the calculation of the above integral requires computing the transition matrix between these SPHs and the oceanic eigenfunctions. This calculation is achieved in two steps. First, one computes the rotated SPHs associated with the change of coordinates $\left(\colpla, \lonpla \right) \rightarrow \left( \col , \lon \right)$, as detailed in Appendices~\ref{app:euler_rotation} and~\ref{app:rotation_sph}. Second, the transition matrices between the non-rotated SPHs and the SCHs are evaluated following the method described in \append{app:transition_matrices}.

The integral of the second term in the left-hand member of \eq{momentum_pr_1} actually vanishes owing to the orthogonality property of the curl-free and divergence-free components of the horizontal displacement (\eq{ortho_phi_psi}), 
\begin{equation}
\label{ortho_eigenfunctions}
\integ{ \conj{\gradn \fpoti{\jjb}} \dotp  \left( \gradn \fstreami{\kkb} \crossp \er \right) }{\surface}{\oceancap}{} = 0.
\end{equation}
Finally, the integrals of the third and fourth terms of \eq{momentum_pr_1} are simplified with the help of vectorial identities, yielding
\begin{align}
& \integ{ \conj{\gradn \fpoti{\jjb}} \dotp \left[ \fcorion \crossp \left( \gradn \fstreami{\kkb} \crossp \er  \right) \right] }{\surface}{\oceancap}{}  = \integ{ \left( \fcorion \dotp \er \right)\conj{\gradn \fpoti{\jjb}} \dotp \gradn \fstreami{\kkb} }{\surface}{\oceancap}{}, \nonumber \\
& \integ{\! \! \conj{\gradn \fpoti{\jjb}} \dotp \left( \fcorion \crossp \gradn \fpoti{\kkb} \right) }{\surface}{\oceancap}{}  = - \integ{\! \! \fcorion \dotp \left( \conj{\gradn \fpoti{\jjb}} \crossp \gradn \fpoti{\kkb} \right) }{\surface}{\oceancap}{}. 
\end{align}
The above simplifications lead to the first set of temporal equations for the coefficients $\ppoti{\kkb}$ and $\pstreami{\kkb}$,
\begin{align}
 & \sum_{\kkb=1}^{\infty}  \left\{ \left[ \ddd{ }{\timen}{\timen} + \fdragn \dd{ }{\timen}   \right] \ppoti{\kkb} \eigenvpoti{\jjb} \kron{\jjb}{\kkb}   +  \dd{\pstreami{\kkb}}{\timen} \integ{ \! \! \left( \fcorion \dotp \er \right) \conj{\gradn \fpoti{\jjb}} \dotp \gradn \fstreami{\kkb} }{\surface}{\oceancap}{} \right.  \nonumber    \\
\label{eq_pfunci_1}
& \left. -  \dd{\ppoti{\kkb}}{\timen}  \integ{\! \! \fcorion \dotp \left( \conj{\gradn \fpoti{\jjb}} \crossp \gradn \fpoti{\kkb} \right) }{\surface}{\oceancap}{} \right\}  + \Rowave^2 \eigenvpoti{\jjb} \integ{\conj{\fpoti{\jjb}} \Ftideno  }{\surface }{\oceancap}{}  = 0.
\end{align}

The second set of equations is obtained by repeating the same steps with $\conj{\gradn \fstreami{\jjb}} \crossp \er $ instead of $\conj{\gradn \fpoti{\jjb}}$ in the dot product operation,
\begin{align}
& \sum_{\kkb=1}^{\infty} \left\{ \left( \ddd{}{\timen}{\timen} + \fdragn \dd{}{\timen}  \right) \ppoti{\kkb} \integ{\left(\conj{\gradn \fstreami{\jjb}} \crossp \er \right) \dotp \gradn \fpoti{\kkb} }{\surface}{\oceancap}{}  \right. \\
& + \dd{\ppoti{\kkb}}{\timen}  \integ{  \left(  \conj{\gradn \fstreami{\jjb}}  \crossp \er \right) \dotp \left( \fcorion \crossp \gradn \fpoti{\kkb} \right)}{\surface}{\oceancap}{} \nonumber \\
& + \dd{\pstreami{\kkb}}{\timen} \integ{ \left( \conj{\gradn \fstreami{\jjb}} \crossp \er \right) \dotp \left[ \fcorion \crossp \left( \gradn \fstreami{\kkb} \crossp \er \right) \right]}{\surface}{\oceancap}{} \nonumber \\
& \left. + \left( \ddd{}{\timen}{\timen} + \fdragn \right) \pstreami{\kkb} \integ{\left( \conj{\gradn \fstreami{\jjb}} \crossp \er \right) \dotp \left( \gradn \fstreami{\kkb} \crossp \er \right)}{\surface}{\oceancap}{} \right\} \nonumber \\
& - \Rowave^2 \integ{\er \dotp \left( \conj{\gradn \fstreami{\jjb}} \crossp \gradn \Ftideno  \right)}{\surface}{\oceancap}{} = 0, \nonumber
\end{align}
which, using the orthogonality properties of the eigenfunctions and the relations
\begin{equation}
\label{gradpsi}
\integ{\conj{\gradn \fstreami{\jjb}} \dotp \gradn \fstreami{\kkb} }{\surface}{\oceancap}{} = \eigenvstri{\jjb} \kron{\jjb}{\kkb},
\end{equation}
simplifies to 
\begin{align}
& \sum_{\kkb=1}^{\infty} \left\{ \left( \ddd{}{\timen}{\timen} + \fdragn \right) \pstreami{\kkb}  \eigenvstri{\jjb} \kron{\jjb}{\kkb} - \dd{\ppoti{\kkb}}{\timen} \integ{ \left( \fcorion \dotp \er \right) \left( \conj{\gradn \fstreami{ \jjb}} \dotp \gradn \fpoti{\kkb} \right)}{\surface}{\oceancap}{}   \right.  \nonumber  \\
\label{eq_pfunci_2}
& \left. - \dd{\pstreami{\kkb}}{\timen} \! \! \integ{ \! \! \fcorion \! \dotp \! \left( \conj{\gradn \fstreami{\jjb}} \crossp \gradn \fstreami{\kkb}  \right) }{\surface}{\oceancap}{} \! \right\} \! - \Rowave^2 \! \integ{ \! \! \er \! \dotp \! \left( \conj{\gradn \fstreami{\jjb}} \crossp \gradn \Ftideno  \right) \!}{\surface}{\oceancap}{}   = 0 .
\end{align}

Thus, following the formalism used in earlier studies \citep[][]{Webb1980,Webb1982,Farhat2022b}, \eqs{eq_pfunci_1}{eq_pfunci_2} form an infinite linear system in the coefficients $\ppoti{\jjb}$ and $\pstreami{\jjb}$, written as
\begin{equation}
\label{eq_fpot}
\left( \ddd{}{\timen}{\timen} + \fdragn \dd{}{\timen}  \right) \ppoti{\jjb} + \eigenvpoti{\jjb}^{-1} \! \! \! \sum_{\substack{\kkb = -\infty \\ \kkb \neq 0}}^{+ \infty} \! \! \gyroci{\jjb}{\kkb} \dd{\pfunci{\kkb}}{\timen} + \Rowave^2  \integ{\! \! \conj{\fpoti{\jjb}} \Ftideno}{\surface }{\oceancap}{} = 0, 
\end{equation}
\begin{align}
\label{eq_fstream}
\left( \ddd{}{\timen}{\timen} + \fdragn \dd{}{\timen} \right) \pstreami{\jjb} + \eigenvstri{\jjb}^{-1} \! \! \! \sum_{\substack{\kkb = -\infty \\ \kkb \neq 0}}^{+ \infty} \! \! \gyroci{-\jjb}{\kkb} \dd{\pfunci{\kkb}}{\timen} & \\ 
- \frac{\Rowave^2}{\eigenvstri{\jjb}} \integ{ \er \dotp \left(\conj{\gradn{\fstreami{\jjb}}}  \crossp \gradn \Ftideno  \right)  }{\surface}{\oceancap}{} & = 0. \nonumber
\end{align}
In the above equations, the symbols $\gyroci{\jjb}{\kkb} $\mynom{$\gyroci{\jjb}{\kkb} $}{Gyroscopic coefficient} designate the so-called `gyroscopic coefficients' \citep[e.g.][]{Proudman1920,LHP1970,Webb1980},
\begin{align}
\label{gyro1_tide}
\gyroci{\jjb}{\kkb} \define & - \integ{\cos \colpla \, \er \dotp \left( \conj{\grad \fpoti{\jjb} } \crossp \grad \fpoti{\kkb} \right) }{\surface}{\oceancap}{}, \\ 
\gyroci{\jjb}{- \kkb} \define & \ \ \ \ \integ{\cos \colpla  \left( \conj{\grad \fpoti{\jjb} } \dotp \grad \fstreami{\kkb} \right)}{\surface}{\oceancap}{}, \\
\gyroci{-\jjb}{\kkb} \define &  - \integ{\cos \colpla \left( \conj{\grad \fstreami{\jjb}} \dotp \grad \fpoti{\kkb} \right)}{\surface}{\oceancap}{}, \\
\label{gyro4_tide}
\gyroci{-\jjb}{-\kkb} \define & - \integ{\cos \colpla \, \er \dotp \left(  \conj{\grad \fstreami{\jjb}} \crossp \grad \fstreami{\kkb} \right)}{\surface}{\oceancap}{}.
\end{align}
These coefficients account for the coupling effect of the Coriolis terms and the geometry of the ocean basin, which affects how the gradients of the eigenfunctions overlap. Their behaviour and properties are examined in \append{app:gyro_coeff}. The expression of the forcing term in \eq{eq_fstream} is given here for generality. This term is actually zero, as discussed in the next section, meaning that \eq{eq_fstream} does not depend on $\Ftideno$. 

\subsection{Ocean loading and self-attraction variation}
\label{ssec:ocean_loading}

At this stage, we still have to express $\Ftideno$ as a function of $\ppoti{\kkb}$ and $\pstreami{\kkb}$. To do so, we consider the fact that the gravitational tidal force acts on the planet as a periodic perturbation oscillating in time and longitude. Therefore, the tidal gravitational potential given by \eq{tidalpot} can be expanded in Fourier series of time and series of SPHs. In the general case, $\Utide$ is expressed as
\begin{equation}
\label{Utide}
\Utide = \speedref^2 \Re \left\{ \sum_{\ftide} \sum_{\llat=2}^{+ \infty} \sum_{\mm = -\llat}^{\llat} \Ulmsig \Ylmref{\llat}{\mm} \left( \colpla , \lonpla \right) \expo{\inumber \ftide \time} \right\} ,
\end{equation}
where $\ftide$ is the tidal frequency, $\speedref$ the reference velocity introduced in \eq{time_velocity_ref}, $\Ylmref{\llat}{\mm}$\mynom{$\Ylmref{\llat}{\mm}$}{Complex SPH associated with the coordinates of $\framepla$} the complex SPH of degree $\llat$\mynom{$\llat$}{Degree (or latitudinal wavenumber) of the SPHs} and order $\mm$\mynom{$\mm$}{Order (or longitudinal wavenumber) of the SPHs} associated with the coordinate system $\left( \colpla , \lonpla \right)$, defined in \eq{sphc}, and $\Ulmsig \in \Cset$\mynom{$\Ulmsig$}{$\Ylmref{\llat}{\mm}$-component of the normalised tidal gravitational potential} the associated frequency-dependent normalised component of the tidal gravitational potential. In the following, these notations are shortened for convenience. Similarly as the eigenfunctions of the ocean basin, $\fpoti{\jjb}$ and $\fstreami{\jjb}$, the SPHs are simply denoted by $\Ylmrefi{\jjb} = \Ylmref{\llati{\jjb}}{\mmi{\jjb}}$\mynom{$\Ylmrefi{\jjb}$}{$\jjb$-th function of the set of SPHs associated with $\framepla$}, and the corresponding coefficients by $\Utideni{\jjb} = \Ulmsigi{\jjb}$\mynom{$\Utideni{\jjb}$}{$\jjb$-th component of the normalised tidal potential}, the index $\jjb$ referring to an element of the set of SPHs, $\left\{ \Ylmrefi{\jjb} \right\}_{1 \leq \jjb \leq \infty}$. Moreover, all the indices used from now on ($\jjb$, $\kkb$, $\nn$, $\qq$) are supposed to run from one to infinity when bounds are not specified. \rec{It is noteworthy that the Fourier series expansion in \eq{Utide} can always be written in term of positive tidal frequencies as long as $\mm$ runs from $-\llat$ to~$\llat$. Therefore, we assume that $\ftide \geq 0$, and that the frequencies are all different from each other (no resonance).}

Since the components associated with two different tidal frequencies are not correlated in the linear tidal theory, the resulting tidal responses can be treated separately. We thus consider the contribution of the component associated with a given tidal frequency, $\Utidensig$, and the associated forcing term, $\Ftidenosig$, expressed as \mynom{$\Utidensig$}{$\ftide$-component of the normalised tidal gravitational potential}\mynom{$ \Ftidenosig$}{$\ftide$-component of the forcing term $\Ftideno$}\mynom{$\Utideni{\kkb}$}{$\Ylmrefi{\kkb}$-component of $\Utidensig$}\mynom{$\Ftidenoi{\kkb}$}{$\Ylmrefi{\kkb}$-component of $\Ftideno$}
\begin{align}
\label{Utidensig}
& \Utidensig = \sum_{\kkb} \Utideni{\kkb} \Ylmrefi{\kkb} \left( \colpla , \lonpla \right),
& \Ftidenosig = \sum_{\kkb} \Ftidenoi{\kkb}  \Ylmrefi{\kkb} \left( \colpla , \lonpla \right).
\end{align}  
As highlighted in earlier studies \citep[e.g.][]{Matsuyama2014,Matsuyama2018,Auclair2019} the combined contribution of ocean loading, self-attraction variation, and deformation of solid regions may be formulated as
\begin{equation}
\label{Ftidenoi}
\Ftidenoi{\kkb} = \tiltdi{\kkb} \left( \sum_{\nn} \scal{\Ylmrefi{\kkb}}{\fpoti{\nn}}  \eigenvpoti{\nn} \ppoti{\nn} \right) - \tiltgi{\kkb} \Utideni{\kkb},
\end{equation}
where we have made use of the scalar product defined by \eq{sprodYlm} and introduced the solid deformation factors, $\tiltdi{}$ and $\tiltgi{}$ \citep[see e.g.][]{Hendershott1972}, \mynom{$\tiltdi{},\tiltgi{}$}{Solid deformation factors}
\begin{align}
\label{tiltd}
 \tiltgi{} & \define 1 + \kl - \hl, \\
 \label{tiltg}
\tiltdi{} & \define 1 - \left( 1 + \kloadl - \hloadl \right) \frac{3 }{\left( 2 \llat + 1 \right) } \frac{\rhowater}{\rhocore}.
\end{align}

The factors $\tiltdi{}$ and $\tiltgi{}$ are independent of the order $\mm$ since they are expressed as functions of the Love numbers of the solid part, which is commonly assumed to be spherically symmetric in the tidal theory. The parameters $\kl$\mynom{$\kl$}{Degree-$\llat$ solid tidal gravitational Love number} and $\hl$\mynom{$\hl$}{Degree-$\llat$ solid tidal displacement Love number} in the first expression are known as the tidal gravitational and displacement Love numbers, respectively. They describe the response of the solid body to a gravitational tidal forcing, which takes the form of a self-attraction variation and a surface displacement. By analogy, the loading Love numbers $\kloadl$\mynom{$\kloadl$}{Degree-$\llat$ solid loading tidal Love number} and $\hloadl$\mynom{$\hloadl$}{Degree-$\llat$ solid loading displacement Love number} account for the gravitational and mechanical responses of solid regions to the cumulated gravitational and pressure forces generated by the oceanic mass redistribution, respectively. The second solid deformation factor, $\tiltgi{}$, also depends on the ratio of seawater density to the mean density of the solid regions, $\rhocore$\mynom{$\rhocore$}{Mean density of the solid regions}. We remark that $\tiltdi{}$ and $\tiltgi{}$ are sometimes considered as real factors \citep[e.g.][]{Matsuyama2014,Motoyama2020}, which corresponds to an adiabatic elastic response of the solid part, where dissipative mechanisms are ignored. However, the correspondence principle established by \cite{Biot1954} makes it possible to treat dissipative anelastic cases similarly as long as the anelasticity is linear. As a consequence, the expressions given by \eqs{tiltd}{tiltg} can be extended to any rheological model including dissipative processes. In that case, the solid Love numbers are complex transfer functions describing the visco-elastic response of solid regions when subjected to a harmonic tidal force \citep[][]{Remus2012,Auclair2019,Farhat2022a}. 

Using the expression given by \eq{Ftidenoi} and proceeding to a change of basis functions between the oceanic eigenfunctions and the SPHs associated with the coordinate system $\left( \colpla , \lonpla \right)$, we expand the integrals depending on the forcing in \eqs{eq_fpot}{eq_fstream} in series of $\ppoti{\kkb}$ and components of the tidal gravitational potential,
\begin{align}
\label{forcing_eq1}
\integ{\conj{\fpoti{\jjb}} \Ftideno  }{\surface }{\oceancap}{} = & \sum_{\nn} \left[\sum_{\kkb} \scal{\fpoti{\jjb}}{\Ylmrefi{\kkb}}  \tiltdi{\kkb}  \scal{\Ylmrefi{\kkb}}{\fpoti{\nn}}     \right] \eigenvpoti{\nn} \ppoti{\nn} \\
&  - \sum_{\kkb} \scal{\fpoti{\jjb}}{\Ylmrefi{\kkb}}  \tiltgi{\kkb} \Utideni{\kkb}, \nonumber
\end{align}
\begin{align}
\integ{ \! \! \er \! \dotp \! \left( \conj{\gradn \fstreami{\jjb}} \crossp \gradn \Ftideno  \right) \!}{\surface}{\oceancap}{} =& \sum_{\nn} \left[ \sum_{\kkb, \qq} \cstreamsphi{\jjb}{\qq} \scal{\Ylmi{\qq}}{\Ylmrefi{\kkb}}  \tiltdi{\kkb}  \scal{\Ylmrefi{\kkb}}{\fpoti{\nn}}  \right] \eigenvpoti{\nn} \ppoti{\nn} \nonumber \\
\label{forcing_eq2_interm}
& - \sum_{\kkb,\qq}  \cstreamsphi{\jjb}{\qq} \scal{\Ylmi{\qq}}{\Ylmrefi{\kkb}} \tiltgi{\kkb} \Utideni{\kkb},
\end{align}
where $\cstreamsphi{\jjb}{\qq} $\mynom{$\cstreamsphi{\jjb}{\qq}$}{Coupling coefficients appearing in the forcing terms of the LTEs} designates the coefficients defined as
\begin{equation}
\label{cstreamkq}
\cstreamsphi{\jjb}{\qq} \define \integ{ \! \! \er \! \dotp \! \left( \conj{\gradn \fstreami{\jjb}} \crossp \gradn \Ylmi{\qq}  \right) \!}{\surface}{\oceancap}{}. 
\end{equation}
These coefficients are all zero as shown in \append{app:gyro_coeff}, which implies that 
\begin{equation}
\label{forcing_eq2}
\integ{ \! \! \er \! \dotp \! \left( \conj{\gradn \fstreami{\jjb}} \crossp \gradn \Ftideno  \right) \!}{\surface}{\oceancap}{} =  0.
\end{equation}
We remark that the quadrupolar component of the tidal gravitational potential (the component associated with the $\Ylmref{2}{2}$ SPH) is far greater than the components of higher degrees if the size of the planet is small compared with the planet-perturber distance. These components can thus be ignored in standard two-body systems. The series in the formulation of the tidal gravitational given by \eq{Utidensig} then reduces to one term only. This removes the summation over $\kkb$ in the second term of the right-hand member of \eq{forcing_eq1}.

\subsection{The tidal solution}
\label{ssec:tidal_response}

As a last step, the tidal equations given by \eqs{eq_fpot}{eq_fstream} are written down in the Fourier domain. To compute the solution numerically, all the sets of basis functions are truncated, meaning that the infinite spaces of functions they describe are approximated by finite spaces of functions. As the tidal gravitational tidal potential varies over planetary length scales, the tidal response is essentially described by the eigenmodes of largest wavelengths. As a consequence, the truncation does not alter much the solution as long as the number of eigenmodes of the set is sufficiently large. The numbers of basis functions for the sets $\left\{ \fpoti{\kkb} \right\}$, $\left\{ \fstreami{\kkb} \right\}$, and $\left\{ \Ylmrefi{\kkb} \right\}$ are denoted by $\npot$\mynom{$\npot$}{Number of basis functions or truncation degree of the set $\left\{ \fpoti{\kkb} \right\}$}, $\nstream$\mynom{$\nstream$}{Number of basis functions or truncation degree of the set $\left\{ \fstreami{\kkb} \right\}$}, and $\nsph$\mynom{$\nsph$}{Number of basis functions or truncation degree of the set $\left\{ \Ylmrefi{\kkb} \right\}$}, respectively, while the vectors describing the tidal response and the tidal gravitational potential in these sets are expressed as\mynom{$\potfuncv$}{Vector defining $\potfunc$ in terms of the $\fpoti{\jjb}$}\mynom{$\streamfuncv$}{Vector defining $\streamfunc$ in terms of the $\fstreami{\jjb}$}\mynom{$\Utidenv$}{Vector defining $\Utidensig$ in terms of the SPHs}
\begin{align}
\potfuncv  & = \transp{\left[ \ppoti{1}, \ldots, \ppoti{\kkb}, \ldots , \ppoti{\npot}  \right]}, \\
\streamfuncv  & =  \transp{\left[ \pstreami{1}, \ldots, \pstreami{\kkb}, \ldots , \pstreami{\nstream}  \right]}, \\
\Utidenv & = \transp{\left[\Utideni{1}, \ldots, \Utideni{\kkb} , \ldots , \Utideni{\nsph} \right]},
\end{align}
with $\transp{}$\mynom{$\transp{}$}{Transpose of a matrix} designating the transpose of a matrix. Converted into an algebraic form, the forcing term given by \eq{forcing_eq1} is written as 
\begin{equation}
\integ{\conj{\fpoti{\jjb}} \Ftideno  }{\surface }{\oceancap}{}  = \transp{\unitvi{\npot}{\jjb}} \left( \cmatpot  \potfuncv +  \fmatpot \Utidenv \right),
\end{equation}
where we have introduced the unit vectors $\unitvi{\npot}{\jjb}$\mynom{$\unitvi{\npot}{\jjb}$}{Unit vector of size $\npot$ made of Kronecker delta functions} defined as 
\begin{equation}
\unitvi{\npot}{\jjb} \define  \transp{\left[ \kron{1}{\jjb}, \ldots , \kron{\kkb}{\jjb} , \ldots , \kron{\npot}{\jjb} \right]}, 
\end{equation}
and the matrices $\cmatpot$, and $\fmatpot$\mynom{$\cmatpot,\fmatpot$}{Matrices accounting for the deformation of the solid regions},
\begin{align}
& \cmatpot  = \transp{\conj{\transmati{\SPHref}{\potfunc}}}  \tiltmatd  \transmati{\SPHref}{\potfunc} \eigenvmatpot, & \fmatpot =  -\transp{\conj{\transmati{\SPHref}{\potfunc}}} \tiltmatg .
\end{align}
In the above expressions, $\transmati{\SPHref}{\potfunc}$\mynom{$\transmati{\SPHref}{\potfunc}$}{Transition matrix from the SPHs $\Ylmrefi{\jjb}$ to the SCHs $\fpoti{\jjb}$} designates the transition matrix from the SPHs $\Ylmrefi{\jjb}$ to the eigenfunctions $\fpoti{\jjb}$. The other matrices are defined as\mynom{$\tiltmatd,\tiltmatg $}{Diagonal matrices of solid deformation factors}\mynom{$\eigenvmatpot,\eigenvmatstr $}{Diagonal matrices of eigenvalues} 
\begin{equation}
\tiltmatd \define
\begin{bmatrix}
\tiltdi{1} & 0 & \ldots & \ldots & 0 \\
0          & \ddots &  \ddots &         & \vdots \\
\vdots &      \ddots       & \tiltdi{\kkb} & \ddots & \vdots \\
\vdots &             &          \ddots         &\ddots & 0\\
0 & \ldots & \ldots & 0& \tiltdi{\nsph}
\end{bmatrix},
\end{equation}
\begin{equation}
\tiltmatg \define 
\begin{bmatrix}
\tiltgi{1} & 0 & \ldots & \ldots & 0 \\
0          & \ddots &  \ddots &         & \vdots \\
\vdots &      \ddots       & \tiltgi{\kkb} & \ddots & \vdots \\
\vdots &             &          \ddots         &\ddots & 0\\
0 & \ldots & \ldots & 0& \tiltgi{\nsph}
\end{bmatrix},
\end{equation}
\begin{equation}
\eigenvmatpot  \define
\begin{bmatrix}
\eigenvpoti{1} & 0 & \ldots & \ldots & 0 \\
0          & \ddots &  \ddots &         & \vdots \\
\vdots &       \ddots      & \eigenvpoti{\kkb} & \ddots & \vdots \\
\vdots &             &              \ddots     &\ddots & 0\\
0 & \ldots & \ldots &0  & \eigenvpoti{\npot}
\end{bmatrix},
\end{equation}
\begin{equation}
\eigenvmatstr \define
\begin{bmatrix}
\eigenvstri{1} & 0 & \ldots & \ldots & 0 \\
0          & \ddots &  \ddots &         & \vdots \\
\vdots &       \ddots      & \eigenvstri{\kkb} & \ddots & \vdots \\
\vdots &             &              \ddots     &\ddots & 0\\
0 & \ldots & \ldots &0  & \eigenvstri{\nstream}
\end{bmatrix},
\end{equation}
Also, we introduce the matrix of gyroscopic coefficients\mynom{$\gyromat$}{Matrix of gyroscopic coefficients} 
\begin{equation}
\gyromat \define
\begin{bmatrix}
 \eigenvmatpot^{-1}  \gyrompp &  \eigenvmatpot^{-1} \gyromps \\
  \eigenvmatstr^{-1}  \gyromsp & \eigenvmatstr^{-1} \gyromss
\end{bmatrix},
\end{equation}
where the block matrices $\gyrompp $\mynom{$\gyrompp$}{First block matrix of $\gyromat$}, $\gyromps $\mynom{$\gyromps$}{Second block matrix of $\gyromat$}, $\gyromsp$\mynom{$\gyromsp$}{Third block matrix of $\gyromat$} and $\gyromss$\mynom{$\gyromss$}{Fourth block matrix of $\gyromat$} are defined in terms of the $\gyroci{\jjb}{\kkb} $ defined in \eqsto{gyro1_tide}{gyro4_tide} as 
\begin{equation}
\gyrompp \define 
\begin{bmatrix}
\gyroci{1}{1} & \ldots & \gyroci{1}{\kkb}  & \ldots & \gyroci{1}{\npot}  \\
\vdots & & \vdots & & \vdots \\
\gyroci{\jjb}{1} & \ldots & \gyroci{\jjb}{\kkb}  & \ldots & \gyroci{\jjb}{\npot}  \\
\vdots & & \vdots & & \vdots \\
\gyroci{\npot}{1} & \ldots & \gyroci{\npot}{\kkb}  & \ldots & \gyroci{\npot}{\npot} 
\end{bmatrix},
\end{equation}
\begin{equation}
\gyromps \define 
\begin{bmatrix}
\gyroci{1}{-1} & \ldots & \gyroci{1}{-\kkb}  & \ldots & \gyroci{1}{-\nstream}  \\
\vdots & & \vdots & & \vdots \\
\gyroci{\jjb}{-1} & \ldots & \gyroci{\jjb}{-\kkb}  & \ldots & \gyroci{\jjb}{-\nstream}  \\
\vdots & & \vdots & & \vdots \\
\gyroci{\npot}{-1} & \ldots & \gyroci{\npot}{-\kkb}  & \ldots & \gyroci{\npot}{-\nstream} 
\end{bmatrix},
\end{equation}
\begin{equation}
\gyromsp \define 
\begin{bmatrix}
\gyroci{-1}{1} & \ldots & \gyroci{-1}{\kkb}  & \ldots & \gyroci{-1}{\npot}  \\
\vdots & & \vdots & & \vdots \\
\gyroci{-\jjb}{1} & \ldots & \gyroci{-\jjb}{\kkb}  & \ldots & \gyroci{-\jjb}{\npot}  \\
\vdots & & \vdots & & \vdots \\
\gyroci{-\nstream}{1} & \ldots & \gyroci{-\nstream}{\kkb}  & \ldots & \gyroci{-\nstream}{\npot} 
\end{bmatrix}
= -\transp{\conj{\gyromps}} ,
\end{equation}
\begin{equation}
\gyromss \define
\begin{bmatrix}
\gyroci{-1}{-1} & \ldots & \gyroci{-1}{-\kkb}  & \ldots & \gyroci{-1}{-\nstream}  \\
\vdots & & \vdots & & \vdots \\
\gyroci{-\jjb}{-1} & \ldots & \gyroci{-\jjb}{-\kkb}  & \ldots & \gyroci{-\jjb}{-\nstream}  \\
\vdots & & \vdots & & \vdots \\
\gyroci{-\nstream}{-1} & \ldots & \gyroci{-\nstream}{-\kkb}  & \ldots & \gyroci{-\nstream}{-\nstream} 
\end{bmatrix}.
\end{equation}

Finally, the matrix accounting for the gravitational and pressure effects of the oceanic surface displacement, $\couplingmat $\mynom{$\couplingmat$}{Matrix accounting for the gravitational and pressure effects}, the matrix describing the coupling between the forcing tidal gravitational potential and the oceanic eigenmodes, $\forcingmat$\mynom{$\forcingmat$}{Matrix describing the coupling between the forcing tidal gravitational potential and the oceanic eigenmodes}, and the solution vector, $\Xvect$\mynom{$\Xvect$}{Vector describing the horizontal displacement in terms of the oceanic SCHs}, are respectively expressed as
\begin{equation}
\begin{array}{lll}
\couplingmat \define 
\begin{bmatrix}
\cmatpot  & 0 \\
0 & 0
\end{bmatrix}, & 
\forcingmat \define 
\begin{bmatrix}
\fmatpot  \\
0
\end{bmatrix}, & 
\Xvect \define 
\begin{bmatrix}
\potfuncv \\
\streamfuncv
\end{bmatrix}. 
\end{array}
\end{equation}
The temporal tidal equations given by \eqs{eq_fpot}{eq_fstream} thus lead to the solution 
\begin{align}
\label{algebraic_LTE}
& \Xvect = \tidalmat \left( \ftiden , \fdragn, \Rowave, \colcont,\colscont;  \tiltdi{},  \tiltgi{} \right)  \Utidenv, \\
& \tidalmat \define \Rowave^2  \left[ \ftiden \left( \ftiden - \inumber \fdragn \right) \idmat - \inumber \ftiden \gyromat  - \Rowave^2 \couplingmat   \right]^{-1} \forcingmat ,
\end{align}
with $\idmat$\mynom{$\idmat$}{Identity matrix} being the identity matrix. In the above expression, the oceanic tidal response to the tidal perturbation in the Fourier domain is expressed as a complex frequency-dependent matrix ($\tidalmat$)\mynom{$\tidalmat$}{Tidal transfer function matrix}, which links the tidal gravitational potential ($\Utidenv$) to the potential and stream functions that describe the horizontal tidal displacement ($\Xvect $). 

\subsection{Tidal Love numbers, torque, and power}

To close this theoretical section, it remains to introduce the parameters that quantify the tidally dissipated energy and its effect on the evolution of planetary systems. The tidal mass redistribution itself is quantified by self-attraction variations, which is derived from the gravitational potential of the tidally distorted body, $\Uresp$\mynom{$\Uresp$}{Gravitational potential of the tidally distorted body}, defined at its surface as 
\begin{equation}
\Uresp = \speedref^2 \Re \left\{ \sum_{\ftide} \sum_{\llat=0}^{+ \infty} \sum_{\mm = -\llat}^{\llat} \Uplmsig \Ylmref{\llat}{\mm} \left( \colpla , \lonpla \right) \expo{\inumber \ftide \time} \right\} .
\end{equation}
The components $\Uplmsig$\mynom{$\Uplmsig$}{Complex $ \left( \ftide, \llat , \mm \right)$-component of $\Uresp$} of this gravitational potential are expressed as 
\begin{equation}
\label{Uplmsig}
\Uplmsig = \kl \Ulmsig + \left( 1 + \kloadl \right) \frac{3}{2 \llat +1} \frac{\rhowater}{\rhocore} \zetaoclmi{\llat}{\mm}{\ftide}, 
\end{equation}
where $\Ulmsig $ is the corresponding component of the forcing tidal potential introduced in \eq{Utide}, and $\zetaoclmi{\llat}{\mm}{\ftide}$\mynom{$\zetaoclmi{\llat}{\mm}{\ftide}$}{Complex $ \left( \ftide, \llat , \mm \right)$-component of the normalised surface elevation} the corresponding component of the normalised surface elevation, 
\begin{equation}
\zetaoclmi{\llat}{\mm}{\ftide} = \sum_{\jjb=1}^{\infty} \scal{\Ylmref{\llat}{\mm}}{\fpoti{\jjb}} \eigenvpoti{\jjb} \ppoti{\jjb} .
\end{equation}
The first term of \eq{Uplmsig} is the self-attraction variation due to the distortion of the solid part generated by the gravitational tidal force, while the second term is the contribution of the ocean loading, which includes both the potential generated by the oceanic mass redistribution and the potential resulting from the distortion of the solid part induced by ocean loading. 

The complex Love numbers\mynom{$\kp$}{Complex Love numbers} are the transfer functions accounting for the intrinsic response of an extended body undergoing a tidal gravitational forcing \citep[e.g.][]{Ogilvie2014}. They are defined, for each SPH, as the ratio of the gravitational potential of the tidal response to the forcing tidal potential evaluated at the body's surface, 
\begin{equation}
\label{planet_love_number_def}
\kp \left( \ftiden , \fdragn, \Rowave, \colcont,\colscont;  \tiltdi{},  \tiltgi{} \right) \define \frac{\Uplmsig}{\Ulmsig} ,
\end{equation}
which, using the expression of the tidal potential given by \eq{Uplmsig}, yields 
\begin{equation}
\kp = \kl  + \left( 1 + \kloadl \right) \frac{3}{2 \llat +1} \frac{\rhowater}{\rhocore} \frac{\zetaoclmi{\llat}{\mm}{\ftide}}{\Ulmsig }. 
\label{planet_love_number}
\end{equation}
Considering the above equation, we note that the Love number depends both on the order $\mm$ and the degree $\llat$ of the associated SPH in the general case contrary to the solid Love numbers, which only depend on the degrees (see \eqs{tiltd}{tiltg}). This is due to the fact that the oceanic tidal response is not spherically symmetric owing to the geometry of coastlines and Coriolis effects. 

The long-term effect of tides on the evolution of the planet-perturber system is quantified by the generated tidal torques. The main contributor to this evolution is the time-averaged tidal torque\mynom{$\torquez $}{Tidal torque exerted about the planet's spin axis} exerted about the spin axis of the planet \citep[][]{Zahn1966a}, 
\begin{equation}
\torquez \define \timeav{ \integ{  \dd{\Utide}{\lonpla} \rhotide }{\volume}{\domain}{}},
\end{equation}
which affects the long-term variation rate of the planet-perturber distance and the planet's spin angular velocity \citep[e.g.][]{Ogilvie2014}. In the above equation, $\domain$ designates the volume filled by the planet, $\infvar{\volume}$\mynom{$\infvar{\volume}$}{Infinitesimal volume parcel} an infinitesimal volume parcel, and $\rhotide$\mynom{$\rhotide$}{Local tidal density variations} the local density variations associated with the tidal response. These density variations are related through Poisson's equation to the tidal gravitational potential of the distorted planet, $\Uresp$, namely
\begin{equation}
\label{Poisson_eq}
\lap \Uresp = - 4 \pi \Ggrav \rhotide .
\end{equation}
This allows the tidal torque to be expressed in terms of the forcing and deformation tidal gravitational potentials, 
\begin{equation}
\label{torquez_gravpot}
\torquez = -   \frac{1}{4 \pi \Ggrav }  \timeav{  \integ{  \dd{\Utide}{\lonpla}  \lap \Uresp  }{\volume}{\domain}{} }. 
\end{equation}

The latter integral is defined over the volume filled by the tidally deformed planet ($\domain$), but it can actually be carried out over any region that includes this volume. By making use of this property, \cite{Ogilvie2013} demonstrates that both the tidal torque and the tidally dissipated power are straightforwardly related to the Fourier components of the two gravitational potentials. As a first step, one shall notice that $\lap \Utide = 0$ outside of the perturber, by virtue of Poisson's equation, which implies that 
\begin{equation}
 \lap \left(  \dd{\Utide}{\lonpla} \right)  = \dd{\left( \lap \Utide \right)}{\lonpla} = 0,
\end{equation}
since the Laplacian and $\dd{}{\lonpla}$ operators can be permuted. Let $\domainstar$\mynom{$\domainstar$}{Simply connected region that includes the planet but not the perturber} be any region that is simply connected and does not include the perturber. Then, the fact that $\lap \left(  \dd{\Utide}{\lonpla} \right) = 0$ in $\domainstar$ allows the integral of \eq{torquez_gravpot} to be rewritten as
\begin{equation}
\label{torque_integ}
\integ{ \! \! \!  \dd{\Utide}{\lon}  \lap \Uresp  }{\volume}{\domainstar}{} = \integ{ \! \! \!  \left[ \dd{\Utide}{\lonpla} \lap \Uresp - \Uresp \lap \left(  \dd{\Utide}{\lonpla} \right)  \right] }{\volume}{\domainstar}{} .
\end{equation}
By virtue of Green's second identity \citep[e.g.][Sect.~1.11]{Arfken2005}, \eq{torque_integ} yields
\begin{equation}
\integ{ \! \! \!  \dd{\Utide}{\lon}  \lap \Uresp  }{\volume}{\domainstar}{} = \ointeg{ \! \! \!  \left[ \dd{\Utide}{\lonpla} \grad \Uresp -  \Uresp \grad \left(  \dd{\Utide}{\lonpla} \right)  \right] \dotp }{\surfacev}{\domainstarbd}{} ,
\end{equation}
the notation $\domainstarbd$\mynom{$\domainstarbd$}{Boundary of $\domainstar$} referring to the boundary of the domain $\domainstar$, and $\infvar{\surfacev}$ to and outward pointing infinitesimal surface element vector. 

By applying the time-averaging operator to the integral we obtain
\begin{equation}
\torquez = \sum_{\ftide} \frac{\speedref^4}{8 \pi \Ggrav} \Re \left\{ \ointeg{ \! \! \!  \left[ \conj{ \Urespnsig}  \grad \left( \dd{\Utidensig }{\lonpla}  \right) - \conj{\dd{\Utidensig }{\lonpla}} \grad \Urespnsig   \right] \dotp }{ \surfacev}{\domainstarbd}{}  \right\},
\label{torquez_series}
\end{equation}
with $\Utidensig$ and $\Urespnsig$ defined as 
\begin{align}
\Utide \left(  \rr, \colpla, \lonpla , \time \right)  = \speedref^2 \Re \left\{ \sum_{\ftide}  \Utidensig \left( \rr, \colpla, \lonpla  \right) \expo{\inumber \ftide \time} \right\} , \\
\Uresp \left(  \rr, \colpla, \lonpla , \time \right)  = \speedref^2 \Re \left\{ \sum_{\ftide}  \Urespnsig \left( \rr, \colpla, \lonpla  \right) \expo{\inumber \ftide \time} \right\} .
\end{align}
In practice, $\domainstar$ can be chosen such that its boundary is a sphere of some radius $ \rdstar$\mynom{$\rdstar$}{Radius of the spherical volume $\domainstar$} centred on the planet. 

In the vicinity of $\domainstarbd$, the forcing tidal potential and the deformation tidal potential can be represented using interior and exterior multipole expansions in SPHs, respectively,
\begin{align}
\label{Utidensig_sph}
\Utidensig \left( \rr, \colpla, \lonpla \right) = & \sum_{\llat=2}^{+ \infty} \sum_{\mm = -\llat}^{\llat}  \Ulmsig \left( \frac{\rr}{\Rpla} \right)^{\llat} \Ylmref{\llat}{\mm} \left( \colpla , \lonpla \right) , \\
\label{Urespnsig_sph}
\Urespnsig \left( \rr , \colpla, \lonpla \right) = & \sum_{\llat=0}^{+ \infty} \sum_{\mm = - \llat}^{\llat}  \Uplmsig \left( \frac{\rr}{\Rpla} \right)^{-\left( \llat + 1 \right)} \Ylmref{\llat}{\mm} \left( \colpla , \lonpla \right) . 
\end{align}
By substituting \eqs{Utidensig_sph}{Urespnsig_sph} into \eq{torquez_series}, and for arbitrary radius $\rdstar$, we end up with the formulation of the torque as a function of the imaginary part of Love numbers,
\begin{equation}
\torquez = \sum_{\ftide}  \sum_{\llat = 2}^{+ \infty} \sum_{\mm = - \llat}^{\llat} \frac{\mm \left( 2 \llat + 1 \right) \Rpla \speedref^4}{8 \pi \Ggrav} \abs{ \Ulmsig}^2 \Im \left\{ \kp  \right\}.
\label{torquez}
\end{equation}
For the quadrupolar tidal gravitational potential \citep[e.g.][]{Ogilvie2014}, 
\begin{equation}
\Ulmnsiglmi{2}{2} = \sqrt{\frac{6 \pi}{5}} \left( \frac{\Ggrav \Mpert}{\smaxis \speedref^2}  \right) \left( \frac{\Rpla}{\smaxis} \right)^2,
\label{U22_norm}
\end{equation}
with $\smaxis$\mynom{$\smaxis$}{Semi-major axis of the perturber} being the semi-major axis of the perturber, we recover the well-known relationship between the tidal torque and the quadrupolar Love number \citep[e.g.][]{ME2013,Auclair2019},
\begin{equation}
\torquezquad = \frac{3}{2} \Ggrav \Mpert^2 \frac{\Rpla^5}{\smaxis^6} \Im \left\{ \kpquad \right\}.
\end{equation}

The time-averaged power input by the tidal force\mynom{$\powertide$}{Time-averaged power input by the tidal force} can be determined in a similar way. This power is defined as the work per unit time done by the tidal force on the tidal motions \citep[e.g.][]{Ogilvie2013},
\begin{equation}
\powertide \define \timeav{  \integ{\rhopla \Vvect  \dotp \grad \Utide }{\volume}{\domainstar}{} },
\end{equation}
where $\rhopla$\mynom{$\rhopla$}{Local density} is the local density of the planet. We note that $\Vvect$ designates here the velocity field of tidal motions in the whole planet and not only in the ocean basin. By proceeding to an integration by parts, we rewrite the above expression as
\begin{equation}
\label{powertide_2}
\powertide = \ointeg{ \Utide \rhopla \Vvect \dotp }{\surfacev}{\domainstarbd}{} - \integ{\Utide  \div \left( \rhopla \Vvect \right)  }{\volume}{\domainstar}{}.
\end{equation}
The first term of the right-hand member in \eq{powertide_2} is zero since the boundary $\domainstarbd$ is outside of the planet, where $\rhopla= 0$. By combining together the continuity equation,
\begin{equation}
\dd{\rhotide}{\time} + \div \left( \rhopla \Vvect \right) = 0,
\end{equation}
and Poisson's equation, given by \eq{Poisson_eq}, we express $\div \left( \rhopla \Vvect \right) $ as a function of the deformation tidal gravitational potential,
\begin{equation}
\div \left( \rhopla \Vvect \right)  = \frac{1}{4 \pi \Ggrav} \lap \left( \dd{\Uresp}{\time} \right),
\end{equation}
which yields
\begin{equation}
\powertide = - \frac{1}{4 \pi \Ggrav} \timeav{  \integ{\Utide \lap \left( \dd{\Uresp}{\time} \right) }{\volume}{\domainstar}{}}. 
\end{equation}
Using Green's second identity, the above expression becomes
\begin{equation}
\powertide =  \sum_{\ftide}  \frac{\speedref^4 \ftide}{8 \pi \Ggrav} \Im \left\{ \ointeg{  \left[  \conj{\Utidensig} \grad \Urespnsig -  \Urespnsig \conj{ \grad  \Utidensig }   \right] \dotp }{ \surfacev}{\domainstarbd}{}  \right\}.
\end{equation}
Finally we use the expressions of $\Utidensig$ and $\Urespnsig $ given by \eqs{Utidensig_sph}{Urespnsig_sph}\rec{, and the definition of the Love number given by \eq{planet_love_number_def}}. It follows 
\begin{equation}
\label{powertide_klove}
\powertide = - \sum_{\ftide}  \sum_{\llat = 2}^{+ \infty} \sum_{\mm = - \llat}^{\llat} \frac{   \left( 2 \llat + 1 \right) \Rpla \speedref^4 \ftide}{8 \pi \Ggrav} \abs{ \Ulmsig}^2 \Im \left\{ \kp  \right\}.
\end{equation}

The tidal input power in the ocean\mynom{$\powertideoc$}{Time-averaged tidal input power in the ocean} is expressed as a function of the harmonics of the surface tidal elevation only,
\begin{equation}
\label{powertideoc}
\powertideoc =  -\frac{3 \Rpla \speedref^4 }{8 \pi \Ggrav} \left(  \frac{\rhowater}{\rhocore} \right) \sum_{\ftide} \sum_{\llat=2}^{+ \infty}  \sum_{\mm = - \llat}^{\llat} \ftide   \Im \left\{ \conj{\Ulmsig}  \zetaoclmi{\llat}{\mm}{\ftide} \right\}.
\end{equation}
By noting that $\Utiden = \zetaeqn$ and expanding the component of the forcing tidal potential in series of oceanic eigenmodes, 
\begin{equation}
\Ulmsig = \sum_{\jjb=1}^{\infty} \scal{\Ylmref{\llat}{\mm}}{\fpoti{\jjb}} \zetaeqni{\jjb} ,
\end{equation}
we can rewrite the sum on degrees and orders in \eq{powertideoc} as
\begin{equation}
\sum_{\llat=0}^{+ \infty} \sum_{\mm = - \llat}^{\llat} \conj{\Ulmsig}  \zetaoclmi{\llat}{\mm}{\ftide} = \sum_{\jjb,\kkb} \left[  \sum_{\llat=0}^{+ \infty} \sum_{\mm = - \llat}^{\llat} \scal{\fpoti{\jjb}}{\Ylmref{\llat}{\mm}} \scal{\Ylmref{\llat}{\mm} }{\fpoti{\kkb}} \right]\conj{ \zetaeqni{\jjb}} \eigenvpoti{\kkb} \ppoti{\kkb} .
\end{equation}
Therefore, owing to the orthogonality of the $\fpoti{\jjb}$ eigenfunctions, \eq{powertideoc} simplifies to 
\begin{equation}
\label{powertideoc_2}
\powertideoc = \frac{1}{2} \Rpla^2 \Hoc^2 \ggravi \rhowater \sum_{\ftide} \sum_{\jjb=1}^{+\infty} \ftide \Im \left\{ \eigenvpoti{\jjb} \ppoti{\jjb}   \conj{\zetaeqni{\jjb}} \right\} ,
\end{equation}
which corresponds to the usual expression of the work per unit time done by the tidal force on the oceanic tidal motions \citep[e.g.][]{Webb1980,Farhat2022b}. 

Similarly, the tidally dissipated power\mynom{$\powerdissoc$}{Time-averaged tidally dissipated power in the ocean} within the oceanic layer is defined as 
\begin{equation}
\powerdissoc \define \timeav{  \integ{\rhowater \Hoc \fdrag \Vvect  \dotp \Vvect }{\surface}{\oceancap}{} },
\end{equation}
By making use of the orthogonality properties of the oceanic eigenmodes described by \eqs{gradphi}{gradpsi}, this power is expressed as \citep[e.g.][]{Webb1980}
\begin{equation}
\label{powerdissoc}
\powerdissoc = \frac{1}{2} \fdrag \rhowater \Rpla^4 \Hoc \sum_{\ftide} \sum_{\jjb = 1}^{+ \infty} \ftide^2 \left( \eigenvpoti{\jjb} \abs{\ppoti{\jjb}}^2  + \eigenvstri{\jjb} \abs{\pstreami{\jjb}}^2 \right).
\end{equation}
The dissipated power $\powerdissoc$ is equal to the input tidal power given by \eq{powertideoc} if the solid regions are rigid, as assumed in earlier studies \citep[e.g.][]{Webb1980,ADLML2018}. 

When the solid part is deformable, the total input tidal power given by \eq{powertide_klove}, $\powertide$, and the total tidally dissipated power, $\powerdiss$\mynom{$\powerdiss$}{Total time-averaged tidally dissipated power}, are the sum of the solid and oceanic contributions,
\begin{align}
\powertide & =  \powertidesol + \powertideoc , \\
\powerdiss  & = \powerdisssol + \powerdissoc,
\end{align} 
where $\powertidesol  \neq \powerdisssol $\mynom{$\powertidesol$}{Time-averaged tidal input power in the solid regions}\mynom{$\powerdisssol$}{Time-averaged tidally dissipated power in the solid regions} and $\powertideoc  \neq \powerdissoc$ in the general case. Particularly, we note that the two components of the input tidal power can take negative values due to the energy exchanges between the ocean and solid regions resulting from the loading and gravitational coupling, while $\powerdisssol  \geq 0$ and $\powerdissoc \geq 0$. In the following, the input and tidally dissipated powers of the solid regions, $\powertidesol$ and $\powerdisssol$, are evaluated from the oceanic contributions given by \eqs{powertideoc}{powerdissoc}, and the total input tidal power, expressed in \eq{powertide_klove}, by using the fact that $\powerdiss = \powertide$.

At this stage, we have established the expressions of all the tidal quantities that characterise the dissipative tidal response of the planet in the general case. In the following, we apply the theory to specific cases. The above expressions are thus used to evaluate the tidally dissipated power as a function of the forcing frequency. 

\begin{table*}[h]
\centering
\caption{\label{tab:param_refcase} Values of parameters used in the reference case. }
\begin{tabular}{lccc} 
 \hline 
 \hline 
\textsc{Parameter} & \textsc{Symbol} & \textsc{Value} & \textsc{Reference}  \\ 
 \hline 
  \multicolumn{4}{c}{\textit{Planet's Solid part}} \\
  Planet mass & $\Mpla$ & $1.0 \ {\rm \Mearth}$ & 1 \\  
  Planet radius & $\Rpla$ & $1.0 \ {\rm \Rearth}$ & 1\\  
  Effective shear modulus & $\mupla$ & 25.1189~GPa & 2 \\ 
  Maxwell time & $\tauM$ & 685~yr & 2\\ 
  Andrade time & $\tauA$ & 12897.1~yr & 2 \\ 
  Rheological parameter  & $\alphaA$ & 0.25 & 2 \\ 
  \multicolumn{4}{c}{\textit{Planet's ocean basin}} \\
  Oceanic depth & $\Hoc$ & 4.0~km & \\
  Continental angular radius & $\colcont$ & $90^\degree$ & \\ 
  Colatitude of the continental centre & $\colscont$ & $90^\degree$ & \\ 
  Longitude of the continental centre & $\lonscont$ & $0^\degree$ \\ 
  Seawater density & $\rhowater$ & $1022 \ {\rm kg \ m^{-3}}$ & \\ 
  Rayleigh drag frequency & $\fdrag$ & $1.0 \times 10^{-5} \ {\rm s^{-1}}$ &  \\ 
   \multicolumn{4}{c}{\textit{Perturber}} \\
   Mass of the perturber & $\Mpert$ & $1.0 \ \Mmoon$  & 1 \\
   Orbital frequency of the perturber & $\norb$ & $1.0 \ \nmoon$ & 1\\ 
   \multicolumn{4}{c}{\textit{Spectral method}} \\
  Truncation degree of the SPHs & $\nsph$ & 100 & \\
  Truncation degree of the oceanic SCHs & $\npot$ & 30 & \\ 
\hline
 \end{tabular}
 \tablebib{(1)~\citet{Mamajek2015}; (2) \citet{Bolmont2020}.}
 \end{table*}

\section{Study of a reference case}
\label{sec:reference_case}

Before making any attempt to explore the parameter space, it seems appropriate to elucidate the tidal response of the planet for a well-chosen reference configuration. In this section, we detail the physical setup of this configuration and we compute the frequency spectra of the quantities that characterise the semidiurnal tidal response of the planet. The obtained results are used as a starting point in the parametric study achieved in \sect{sec:parametric_study}.

\subsection{Physical setup}

Since the Earth-Moon system is the planet-satellite system that we know best, it appears as a privileged option for the aforementioned reference configuration. Therefore, we consider a simplified Earth-Moon system where the Moon orbits an Earth-sized planet with a circular and coplanar motion. In this framework, the Moon's trajectory is a circle in the planet's equatorial plane. The values used for the model parameters in this reference case are summarised in Table~\ref{tab:param_refcase}.

\def\wpanel{0.35\textwidth}
\def\wkey{0.11\textwidth}
\begin{figure*}[htb]
   \centering
  \hspace{2.5cm}  \textsc{Tidal input power}  \hspace{3.5cm} \textsc{Tidally dissipated power}   \hspace{2.5cm}~   \\[-0.7cm]
   \includegraphics[width=\wpanel,trim =0.3cm 0.7cm 3.cm 0.cm,clip]{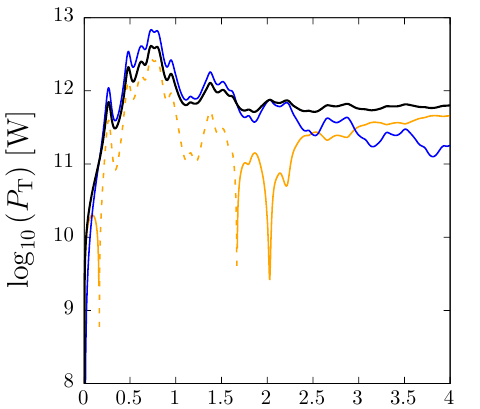} 
   \includegraphics[width=\wpanel,trim = 0.3cm 0.7cm 3.cm 0.cm,clip]{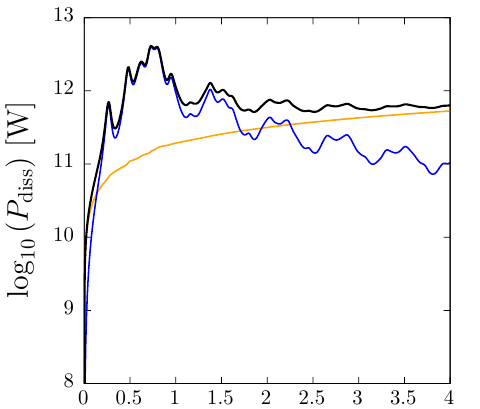}
   \includegraphics[width=\wkey,trim = 8.8cm 0.7cm 0.cm 0.cm,clip]{auclair-desrotour_fig3b.pdf}  \\[0.3cm]
   \hspace{2.5cm} \textsc{Love number}  \hspace{4.5cm} \textsc{Tidal torque}   \hspace{3cm}~  \\[-0.8cm]
    \includegraphics[width=\wpanel,trim =0.3cm 0.cm 3.cm 0.cm,clip]{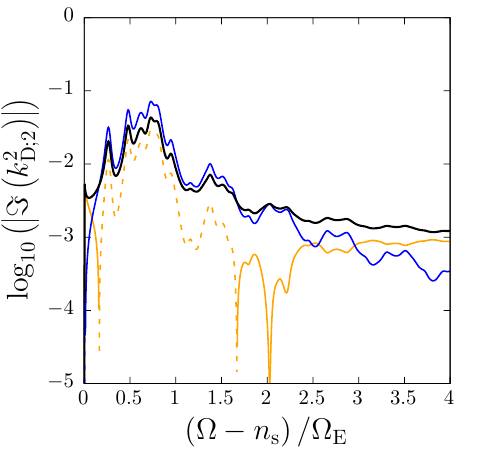} 
   \includegraphics[width=\wpanel,trim = 0.31cm 0.1cm 3.1cm 0.1cm,clip]{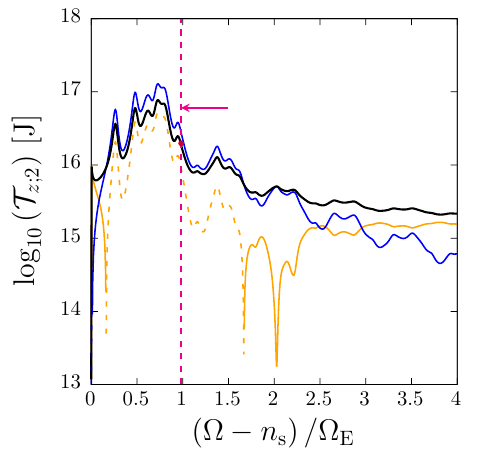} \hspace{\wkey}
      \caption{Frequency spectra of tidal quantities and their components for the reference case (Table~\ref{tab:param_refcase}) and the semidiurnal tide. {\it Top:} Tidal input power (left panel) and tidally dissipated power (right panel). {\it Bottom:} Imaginary part of the quadrupolar Love number (left panel) and tidal torque exerted about the planet's spin axis (right panel). All quantities are plotted in logarithmic scale (vertical axis) as functions of the normalised semidiurnal tidal frequency $\ftidequadn = \left( \spinrate - \norb \right) / \srearth$ (horizontal axis). In each plot, the orange, blue and black lines designate the response of the solid part, ocean, and whole planet, respectively. Solid lines indicate decelerating tidal torque ($\Im \left( \kpquad \right) \leq 0$ and $\torquezquad \leq 0$) and positive tidal input power ($\powertide\geq0$), while dashed lines indicate accelerating tidal torque ($\Im \left( \kpquad \right) >0$ and $\torquezquad >0$) and negative tidal input power ($\powertide<0$). In the plot of the tidal torque, the magenta vertical dashed line and arrow indicate the position of the present day Earth in the spectrum and the direction of evolution of the semidiurnal tidal frequency, respectively. The red dot designates the corresponding tidal torque.}
       \label{fig:spectra_refcase}%
\end{figure*}

Following \cite{Webb1980}, the planet's surface is assumed to be divided into continental and oceanic hemispheres centred at the equator. Namely, the continental angular radius is set to $\colcont = 90^\degree$ and the colatitude of the continental centre to $\colscont=90^\degree$. In this configuration, the coastlines correspond exactly to meridians. The ocean basin may be regarded as a geometrically simplified version of the Pacific ocean. Its depth is set to $\Hoc = 4$~km and its Rayleigh drag frequency to $\fdrag = 10^{-5} \ {\rm s^{-1}}$, consistently with the values found in earlier studies from constraints on the actual tidal dissipation rate of the Earth-Moon system and its age \citep[][]{Webb1980,Webb1982,Farhat2022b}. The mean density of seawater is set to $\rhowater = 1022 \ {\rm kg \ m^{-3}}$, which is a typical value of seawater density on Earth \citep[][]{GZ2008}. Finally, the solid part of the planet is assumed to behave as a visco-elastic body described by the Andrade model (see \append{app:solid_love_numbers}). In this model, the solid tidal response is parametrised by an effective shear modulus, $\mupla$\mynom{$\mupla$}{Effective shear modulus of the solid regions}, the Maxwell time associated with viscous friction $\tauM$\mynom{$\tauM$}{Maxwell time}, the Andrade time associated with the inelastic component of the solid elongation, $\tauA$\mynom{$\tauA$}{Andrade time}, and the dimensionless parameter $\alphaA$\mynom{$\alphaA$}{Parameter determining the duration of the transient response in the primary creep (Andrade model)}. The values used for these rheological parameters are adopted from \cite{Bolmont2020}, Table~2. 

In the considered system, the tidal perturber is a dimensionless body of mass $\Mpert = \Mmoon$ and mean motion $\npert = \nmoon$\mynom{$\npert$}{Mean motion of the perturber}, with $\Mmoon$\mynom{$\Mmoon$}{Lunar mass} and $\nmoon$\mynom{$\nmoon$}{Lunar mean motion} being the Lunar mass and mean motion, respectively. Since its orbit is circular and coplanar, the induced tidal perturbation reduces to the semidiurnal tide, which is associated with the quadrupolar tidal potential given by \eq{U22_norm} and the forcing frequency $\freq = 2 \left( \spinrate - \npert \right)$ \citep[e.g.][]{Ogilvie2014}. This quadrupole is coupled with the SCHs describing the oceanic tidal response, which are determined by the geometry of the ocean basin. To account for this coupling properly, the truncation degrees of the two sets of basis functions must be such that the spatial resolution reached with SPHs is higher than that reached with SCHs. Besides, the truncation degree of the SCHs shall be chosen sufficiently high to describe all the excited oceanic eigenmodes. 

Preliminary tests with various truncation degrees show that convergence is reached, in all cases, for truncation degrees $\npot$ less than ${\sim}30$. Therefore, we set the truncation degree of the SCHs to $\npot=30$. Similarly, a truncation degree of $\nsph = 100$ is sufficient for the SPHs. We note that an estimate of $\nsph$ can be obtained by considering the fact that the degrees of the SCHs approximately scale as the inverse of the ocean's angular radius (see \append{app:sch}). Typically, the SCHs exactly correspond to the hemispherical harmonics in the reference case, meaning that their degrees are roughly twice the degrees of the SPHs. Since the number of SPHs needed to represent accurately the sets of oceanic eigenfunctions increases as the size of the basin decreases, configurations dominated by the continental part would require truncation degrees greater than $100$ for the SPHs.

\subsection{Dissipation frequency spectra}

By making use of the \texttt{TRIP} computer algebra system \citep[][]{GL2011}, we numerically solved the algebraic LTEs established in \sect{ssec:tidal_response} for 1001 uniformly sampled values of the normalised semidiurnal frequency $\ftidequadn = \left( \spinrate - \npert \right) / \srearth$\mynom{$\ftidequadn$}{Normalised semidiurnal tidal frequency}, with $\srearth$\mynom{$\srearth$}{Spin rate of the actual Earth} being the spin rate of the present day Earth. For simplicity, we let the values of the spin period $\Prot = 2 \pi / \spinrate$\mynom{$\Prot$}{Spin period} vary with the tidal frequency while the mean motion of the satellite is fixed. In reality, both $\spinrate$ and $\npert$ would vary in the meantime due to exchanges of angular momentum between the planet's spin and the satellite's orbit. However, this approximation is relevant as long as $\npert \ll \spinrate$. The adopted frequency interval of the tidal forcing ranges between~0 (spin-orbit synchronisation; $\Prot \approx 27.2$~days) and~4 (fast spin rotation; $\Prot \approx 5.9$~hr). We note that the present Earth-Moon system corresponds to the normalised frequency $\ftidequadn  \approx 0.963$. 

Figure~\ref{fig:spectra_refcase} shows the obtained frequency spectra in the reference case for four tidal quantities and their components for each layer (solid part and ocean basin): (i) the input tidal power, (ii) the tidally dissipated power, (iii) the imaginary part of the quadrupolar Love number, and (iv) the tidal torque exerted on the planet about its spin axis. These quantities are evaluated by making use of the expressions given by \eqsfour{powertide_klove}{powerdissoc}{planet_love_number}{torquez}, respectively.

First of all, one observes that the plotted quantities simply related to each other, the input and dissipated powers differing from the other two quantities by a scaling factor of $\ftide$, as established in the tidal theory \citep[e.g.][]{Ogilvie2014}. Moreover, the tidally dissipated power predicted by the model for the Lunar semidiurnal tide ($M_2$) in the actual Earth-Moon system is close to the ${\sim} 2.5$~TW estimated from altimetric measurements and lunar laser ranging \citep[e.g.][]{ER2001, ER2003}. Considering this point, we shall emphasise that we do not expect to recover exactly the estimates obtained from more sophisticated numerical models owing to the numerous simplifications made in the present approach. Nevertheless, as demonstrated in the next section, the value obtained from measurements could actually be retrieved by slightly tuning the free parameters of the model, though this is not the purpose of this work. 

According to the frequency spectra of \fig{fig:spectra_refcase}, the present parameters of the Earth-Moon system (dashed grey line at $\ftidequadn  \approx 0.963$) places the oceanic tidal response close to a resonance, which corresponds to a maximum of the tidal dissipation rate. This feature of the spectra is in agreement with the increasing dissipation rate inferred from geological data in the past tens of millions of years \citep[see e.g.][and references therein]{Farhat2022b} and the predictions of more sophisticated models including realistic bathymetries and land-ocean distributions \citep[][]{Green2017,Daher2021}. In the low-frequency range, the tidal response of the planet is dominated by the oceanic dynamical tide, which is characterised by resonant peaks. Conversely, in the high-frequency range, the resonances associated with oceanic modes are attenuated by the visco-elastic adjustment of the solid part, and the latter becomes the predominant contributor to the tidally dissipated energy. This change of regime can be simply understood by considering the fact that the solid tidal torque scales as $\scale \ftide^{-\alphaA}$ with $\alphaA = 0.25$ (see Table~\ref{tab:param_refcase}), while the non-resonant background of the oceanic tidal torque scales as $\scale \ftide^{-3}$ \citep[e.g.][Eqs.~(47) and~(55), respectively]{Auclair2019}. 

We note that the tidal input power and tidally dissipated power are not equivalent for a given layer -- solid part or oceanic basin -- whereas they are strictly equal for the whole planet. This discrepancy is due to the solid-ocean energy transfer allowed by the elasticity of the solid part through ocean loading and self-attraction variations. The energy transfer  strongly affects the tidal elongation of the solid part when the oceanic resonances predominate because this elongation is essentially controlled by the ocean loading in this regime. As a consequence, the tidal torque exerted on the solid part can be accelerating instead of being decelerating as observed in the signs of the tidal input power, imaginary part of the Love number, and tidal torque. 

Another change of sign can be noticed near the zero-frequency limit ($\ftide \rightarrow 0$). This behaviour actually results from the maximum reached by the solid tidal torque for tidal periods close to the Maxwell and Andrade times, which are both much larger than the typical timescales associated with the ocean dynamics \citep[e.g.][]{Bolmont2020}. At these timescales, the ocean basin is no longer resonant and the dissipative component of its tidal response thus tends to vanish, which allows the solid tide to be predominant in the very low-frequency range. However, the solid tidal torque also tends to zero for $\ftide \ll \min \left( \tauM^{-1}, \tauA^{-1} \right)$, and $\torquez=0$ at $\ftide=0$.

Finally, \fig{fig:spectra_refcase} highlights the symmetry breaking effect induced by the continent. Whereas the tidal response of a global ocean takes the form of a regular series of harmonics \citep[see e.g.][Fig.~5]{ADLML2018}, the continent interferes with the forced wave by deviating tidal flows. As a consequence, additional oceanic eigenmodes are excited by the tidal forcing and the energy concentrated into one mode of the global ocean is spread over several modes. As they are each associated with one resonant peak, these new modes alter the general aspect of the spectra by breaking their regularity and by flattening the resonances of the dominating tidal modes. In the next sections, we show quantitatively that the former effect can actually be identified as the typical signature of a supercontinent in the tidal response of an ocean planet.

\section{Parametric study}
\label{sec:parametric_study}
As the reference case is now well understood, we can apply the harmonic analysis performed in \sect{sec:reference_case} to various configurations. The fact that the tidal response is controlled by a relatively small number of physical parameters in our approach (see Table~\ref{tab:param_refcase}) makes it possible to explore the parameter space in an exhaustive way. This is the purpose of the present section. 

\subsection{Preliminary physical analysis}

Before all, we shall briefly proceed to a selection of key parameters representing well the parameter space. We recall that the oceanic tidal response is fully determined by the dimensionless control parameters summarised in Table~\ref{tab:control_param}. Notwithstanding the deformation of the solid part, which intervenes as a corrective effect through the deformation factors ($ \tiltdi{},  \tiltgi{} $), this set of parameters can essentially be reduced to five numbers: three numbers governing the dynamics of oceanic tidal flows ($\ftiden , \Rowave, \fdragn$) and two numbers defining the geometry of the ocean basin ($\colcont,\colscont$). The parameter space of the oceanic tidal response thus simplifies to five dimensions. 

Moreover, $\ftiden  \approx 1$ for the semidiurnal tide as long as $\npert \ll \spinrate$, indicating that this number can be roughly considered as invariant over time for planet-satellite systems similar to the Earth-Moon system. Therefore, the parameter space can be limited to the set $\left(\Rowave, \fdragn,\colcont,\colscont \right)$ for the semidiurnal oceanic tide of a rapidly rotating Earth-like planet. For convenience, instead of varying the dimensionless parameters governing the ocean dynamics, we choose to vary the physical parameters they depend on, namely the oceanic depth $\Hoc$ for $\Rowave$, and the Rayleigh drag frequency $\fdrag$ for $\fdragn$. The first parameter defines the water volume of the ocean, while the second one characterises the efficiency of the dissipative mechanisms.

Concerning the geometry of the ocean basin, we note that the size of the continent is linked to the oceanic depth through the volume of sea water. Since one expects the water volume to be conserved as the size of the continent evolves, it seems more appropriate to vary $\colcont$ at constant volume rather than at constant depth. Consequently, when we vary the continental radius, we shall vary the oceanic depth by using the expression of the seawater volume in the thin layer approximation,
\begin{equation}
\volume = 2 \pi \Hoc \Rpla^2 \left( 1 + \cos \colcont \right). 
\end{equation}
Contrary to its size, the position of the continent on the globe can be changed independently of the other parameters.

\def\wpanel{0.47\textwidth}
\begin{figure*}[htb]
  \RaggedRight \twolabelsat{5.15cm}{\textsc{Colatitude of the continental centre}}%
                           {13.9cm}{\textsc{Angular radius of the continent}} \\[-0.2cm]
  \centering
   \includegraphics[width=\wpanel,trim =0.0cm 0.7cm 0cm 0.0cm,clip]{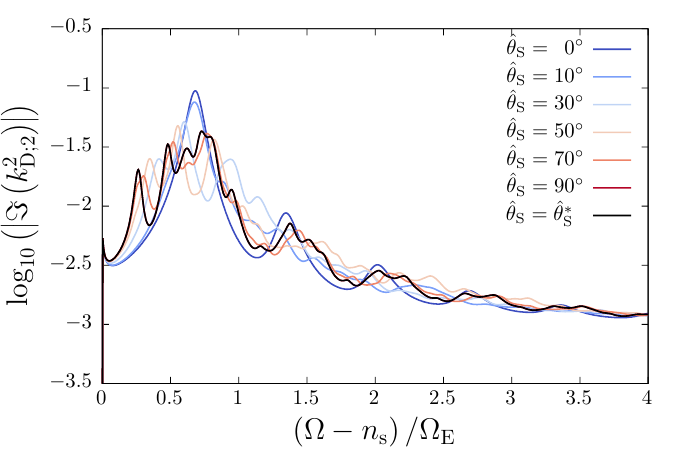} 
   \includegraphics[width=\wpanel,trim = 0.0cm 0.7cm 0cm 0.cm,clip]{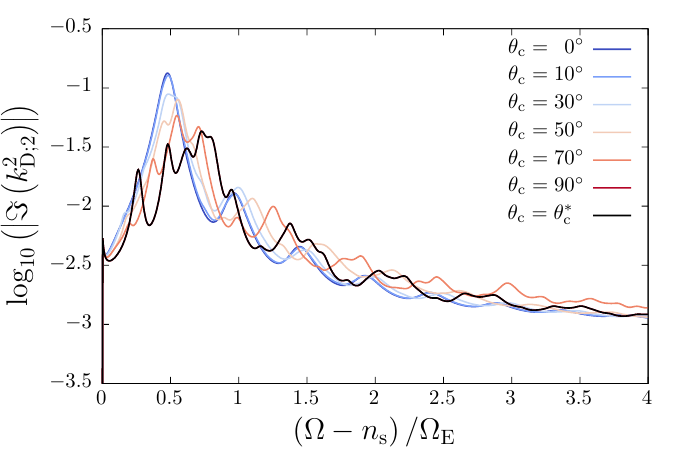}  \\[0.2cm]
   \RaggedRight \twolabelsat{5.15cm}{\textsc{Oceanic depth}}%
                           {13.9cm}{\textsc{Rayleigh drag frequency}} \\[-0.2cm]
  \centering
  \raisebox{0.cm}[0cm][0pt]{\includegraphics[width=\wpanel,trim =0.0cm 0.7cm 0cm 0.cm,clip]{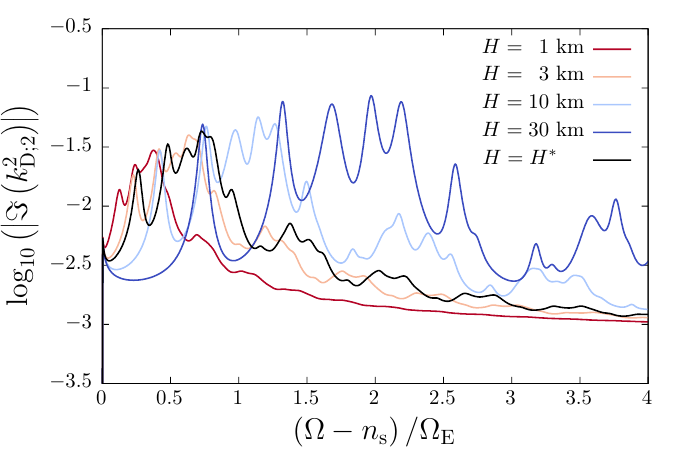}}
    \includegraphics[width=\wpanel,trim = 0.0cm 0.7cm 0.cm 0.cm,clip]{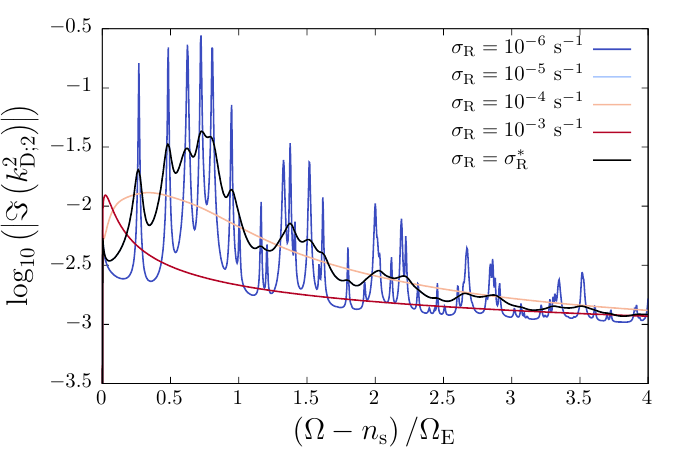} \\[0.2cm]
  \RaggedRight \twolabelsat{5.15cm}{\textsc{Shear modulus of the solid part}}%
                           {13.9cm}{\textsc{Andrade time of the solid part}} \\[-0.2cm]
  \centering
    \includegraphics[width=\wpanel,trim = 0.0cm 0cm 0cm 0.cm,clip]{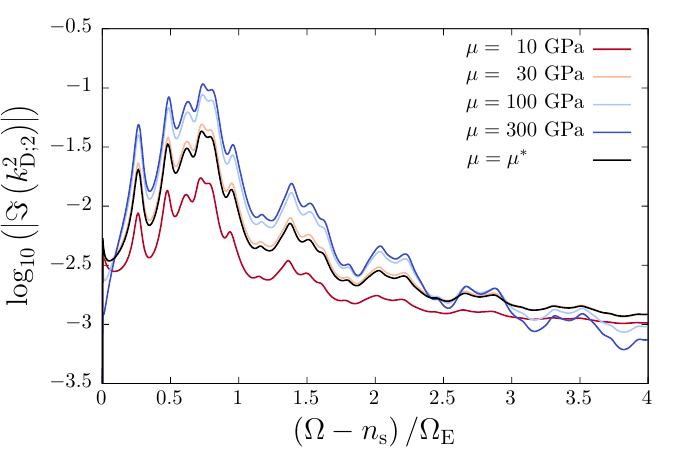} 
    \includegraphics[width=\wpanel,trim =0.0cm 0.0cm 0cm 0.cm,clip]{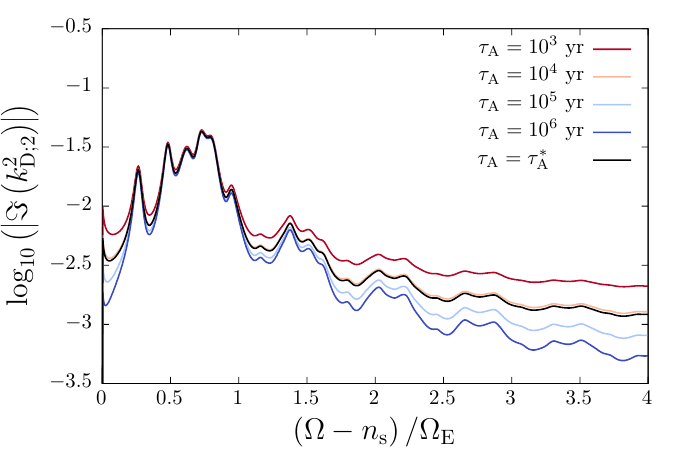}  
      \caption{Frequency spectra of the imaginary part of the quadrupolar Love number associated with the semidiurnal tide, $\kpquad$, for the considered reference case and various values of the model key parameters. {\it Top:} Variation of the colatitude of the continental centre (left panel) and continental angular radius at constant seawater volume (right panel). {\it Middle:} Variation of the oceanic depth (left panel) and Rayleigh drag frequency (right panel). {\it Bottom:} Variation of the effective shear modulus (left panel) and Andrade time (right panel) of the solid part. In each panel, the imaginary part of the Love number is plotted in logarithmic scale (vertical axis) as a function of the normalised semidiurnal frequency $\ftidequadn = \left( \spinrate - \norb \right) / \srearth $ (horizontal axis), with~$\srearth$ being the spin angular velocity of the actual Earth. The solid black line designates the reference case defined by Table~\ref{tab:param_refcase}: $\colscontrc = 90^\degree$, $\colcontrc = 90^\degree$, $\Hocrc = 4.0$~km, $\fdragrc = 10^{-5}~{\rm s^{-1}}$, $\muplarc = 25.1189 $~GPa, and $\tauArc = 12897.1$~yr.}
       \label{fig:spectra_parameters}%
\end{figure*}

A set of four additional parameters is brought by the tidal response of the solid part through the Andrade model: the effective shear modulus ($\mupla$), the Maxwell time ($\tauM$), the Andrade time ($\tauA$), and the dimensionless rheological parameter that accounts for the inelastic component of the solid deformation ($\alphaA$). A quick dimensional analysis allows the Maxwell and Andrade times to be eliminated from the list of parameters affecting the regime of the planet's tidal response. These times are indeed much larger than the typical times characterising the ocean dynamics and tidal forcing. Owing to this net separation of scales, varying them over several orders of magnitude would not induce any change of regime in the tidal dynamics. Typically, for $\abs{\sigma} \gg \tauM^{-1}, \tauA^{-1}$, the imaginary part of the quadrupolar Love number defined by the Andrade model for the solid part scales as \citep[e.g.][Eq.~(47)]{Auclair2019}
\begin{equation}
\label{imk2_solid_asymp}
\Im \left\{ \kquad \right\} \sim \frac{3}{2} \frac{A_2}{\left( 1 + A_2 \right)^2} \GammaF \left( 1 + \alphaA \right) \sin \left( \frac{\alphaA \pi}{2} \right) \left( \abs{\ftide} \tauA \right)^{-\alphaA}, 
\end{equation} 
with $A_2 = 38 \pi \Rpla^2 \mupla / \left(3 \Ggrav \Mpla^2 \right)$\mynom{$A_2$}{Dimensionless constant parametrising the quadrupolar solid tidal Love number} being a constant, and $\GammaF$\mynom{$\GammaF$}{Gamma function} the Gamma function \citep[][Chapter~6]{AS1972}. The above expression entails that, in the limit of $\abs{\sigma} \gg \tauM^{-1}, \tauA^{-1}$, the solid tidal response is dependent on $\tauA$, but independent of $\tauM$. We thus consider only the former in the parametric study.


Similarly as the Maxwell time, the rheological parameter $\alphaA$ can almost be considered as a constant parameter owing to the strong constraints provided by laboratory experiments on olivine minerals and ices, which suggest it takes a value between 0.20 and 0.40 \citep[][]{CR2011}. Due to these constraints, we choose to let it fixed to the value used in the reference case, $\alphaA = 0.25$ (see Table~\ref{tab:param_refcase}). Nevertheless, we note from \eq{imk2_solid_asymp} that varying $\alphaA$ would essentially modify the rheological behaviour of the solid part in the high-frequency range by acting on the scaling law $\Im \left\{ \kquad \right\} \scale \abs{\ftide}^{-\alphaA}$. Finally, the effective shear modulus of the solid part, $\mupla$, accounts for the elasticity of the solid part, which is responsible for the solid-ocean coupling. Thus, as discussed for $\tauA$, we shall consider various values of~$\mupla$ in the parametric study. 

In the above analysis we have examined the role played by the main parameters of the model, and we have selected the six of them that seem to be the most relevant to map the parameter space, ($\colscont$, $\colcont$, $\Hoc$, $\fdrag$, $\mupla$, $\tauA$). Other parameters occur in the model, such as the planet mass ($\Mpla$) or radius ($\Rpla$), the seawater density ($\rhowater$), or the mass of the perturber ($\Mpert$). However, they either affect the same dimensionless control numbers as the selected parameters, or they act as obvious scaling factors. For instance, the tidally dissipated power and torque scale as $\scale \Mpert^2$ since the tidal response is proportional to the forcing gravitational potential -- given by \eq{U22_norm} -- in the used linear tidal theory. Similarly, we note that the planet mass and radius act on the speed of gravity waves through the planet's surface gravity, which shifts the frequencies of resonant oceanic modes. Such dependences can be characterised analytically from the derivations detailed in \sect{sec:oceanic_response}.

\begin{table*}[h]
\centering
\caption{\label{tab:param_effects} Evolution of the frequency spectrum of the quadrupolar tidal Love number as one parameter increases while the values of other parameters are fixed.  }
\begin{tabular}{lcccccc} 
 \hline 
 \hline 
\textsc{Altered features of spectra} & $\colscont$ & $\colcont$ & $\Hoc$ & $\fdrag$ & $\mupla$ & $\tauA$ \\ 
 \hline
 Frequencies of resonant peaks & -- & $\nearrow$ & $\nearrow$ & -- & -- & --  \\ 
 Heights of resonant peaks & -- & -- & $\nearrow$  & $\searrow$ & $\nearrow$ & -- \\ 
 Widths of resonant peaks & -- & -- & -- & $\nearrow$ & -- & -- \\
 Non-resonant background & -- & -- & -- & $\nearrow$ & $\nearrow$ & $\searrow$ \\
 Irregularity of spectra & $\nearrow$ & $\nearrow$ & -- & $\searrow$ & -- & -- \\
\hline
 \end{tabular}
 \tablefoot{The table is limited to the quasi-adiabatic regime, where oceanic resonances are not completely damped by dissipative mechanisms. In the frictional regime, there is no oceanic resonances, and the quadrupolar Love number varies smoothly with the forcing tidal frequency.}
 \end{table*}
 
\subsection{Calculations}

Using the reference case defined in Table~\ref{tab:param_refcase} (hemispherical equatorial ocean) as a template configuration, we performed calculations for various values of the colatitude of the continental centre ($\colscont$), angular radius of the continent at constant volume ($\colcont$), oceanic depth ($\Hoc$), Rayleigh drag frequency ($\fdrag$), effective shear modulus of the solid part ($\mupla$), and Andrade time ($\tauA$). In these computations, the parameter values are defined so as to describe the transitions between the asymptotic regimes identified in Table~\ref{tab:control_param}. The obtained results are shown by \fig{fig:spectra_parameters}, where the imaginary part of the quadrupolar Love number is plotted as a function of the normalised semidiurnal frequency for variations of each parameter. The values of the reference case (Table~\ref{tab:param_refcase}) are superscripted by $^{\irefcase}$\mynom{$^{\irefcase}$}{Superscript referring to the values of the reference case defined in Table~\ref{tab:param_refcase}}:  $\colscontrc = 90^\degree$\mynom{$\colscontrc$}{Colatitude of the continental centre in the reference case}, $\colcontrc = 90^\degree$\mynom{$\colcontrc$}{Angular radius of the continent in the reference case}, $\Hocrc = 4.0$~km\mynom{$\Hocrc$}{Oceanic depth in the reference case}, $\fdragrc = 10^{-5}~{\rm s^{-1}}$\mynom{$\fdragrc$}{Rayleigh drag frequency in the reference case}, $\muplarc = 25.1189 $~GPa\mynom{$\muplarc$}{Effective shear modulus in the reference case}, and $\tauArc = 12897.1$~yr\mynom{$\tauArc$}{Andrade time in the reference case}.

\paragraph{Effect of the continental position.} First, we consider the role played by the geometry of the ocean basin. As the colatitude of the continental centre evolves from $0^\degree$ (polar continent) to $90^\degree$ (equatorial continent), the frequency spectrum of the Love number rapidly becomes irregular (see \fig{fig:spectra_parameters}, top left panel). For $\colscont = 0^\degree$, the coastline of the hemispherical continent exactly corresponds to the planet's equator. As a consequence, it does not interfere with the latitudinal component of tidal flows, which is already zero at the equator in the absence of a continent. This explains why we recover approximately the spectrum obtained for the global ocean \citep[][]{ADLML2018} up to a factor of 2, the noticed slight discrepancy being due to the contribution of the solid part. This spectrum is composed of a regular series of harmonics that results from the coupling -- caused by Coriolis forces -- between the oceanic normal modes and the quadrupolar tidal potential. For a non-rotating planet, only one resonance would be observed in that configuration. A slight inclination of the continental centre with respect to the pole (\fig{fig:spectra_parameters}, top left panel, $\colscont = 10^\degree$) appears to be sufficient to break the axis symmetry and to alter significantly the regularity of the spectrum. The harmonics of the polar case are split into more modes, which induces intermediate relative maxima and decreases the dissipation rate reached while crossing the main resonances. 

\paragraph{Effect of the continental size.} Such an effect can also be observed as the angular radius of the continent varies from $0^\degree$ (global ocean) to $90^\degree$ (hemispherical ocean) at constant water volume (\fig{fig:spectra_parameters}, top right panel). As the size of the continent increases, the peaks of the global ocean get progressively altered by the interferences of coastlines with tidal flows. However, we remark that this effect really becomes significant only for $\colcont \geq 30^\degree$, which suggests that continents smaller than South America hardly alter the tidal response of the global ocean even if they are located at the equator. Moreover, the spectra highlight the fact that oceanic resonances are shifted to the right as the size of the continent increases, which also means resonances become of smaller amplitudes, so less dissipative. This effect is directly related to the ocean's geometry. Acoustically, the ocean basin is analogous to a vibrating drumhead, with its size and depth standing for the drumhead's diameter and tension, respectively. On one hand, the eigenfrequencies of the oceanic normal modes increase as the size of the basin decreases since the latter corresponds to the wavelength of the lowest tidal modes. On the other hand, the speed of gravity waves scales as $\scale \sqrt{\Hoc}$, meaning that it increases as the size of the ocean basin decreases at constant volume. 

\paragraph{Effect of the oceanic depth.} Consistently with the above drumhead analogy, increasing the oceanic depth essentially induces a dilation of the frequency spectrum without changing its global aspect (\fig{fig:spectra_parameters}, middle left panel). This is related to the way the resonances are controlled by the oceanic depth. Both the heights and frequencies of resonant peaks scale as $\propto \sqrt{\Hoc}$ \citep[e.g.][Eq~(55)]{Auclair2019}, which explains why their relative positions and amplitudes are conserved as $\Hoc$ varies. 

\paragraph{Effect of dissipative mechanisms.} Similarly, we analyse the evolution of the frequency spectrum with the Rayleigh drag frequency (\fig{fig:spectra_parameters}, middle right panel) using the closed-form solutions provided by the analytical theory \citep[e.g.][]{ADLML2018}. The plot highlights the transition from the quasi-adiabatic regime ($\fdrag = 10^{-6} \ {\rm s^{-1}}$) to the frictional regime ($\fdrag = 10^{-3} \ {\rm s^{-1}}$). In the former, the resonances of oceanic modes are weakly damped, while they are completely annihilated in the latter. 

The two asymptotic regimes can be considered separately. In the quasi-adiabatic regime, the frequencies of resonances are almost independent of the Rayleigh drag frequency, whereas their heights scale as $\scale \fdrag^{-1}$, their widths as $\scale \fdrag$, and the non-resonant background of the oceanic dissipation rate as $\scale \fdrag$ \citep[][]{ADMLP2015,ADLML2018,Auclair2019}. As a consequence, the number of visible peaks increases as $\fdrag$ decreases. The new peaks appear in the high-frequency range since they are associated with oceanic harmonics of high degrees. In the zero-drag limit ($\fdrag \rightarrow 0$), the number of resonant peaks exactly corresponds to the number of harmonics with eigenfrequencies less than the upper bound of the considered frequency interval. In the frictional regime, the spectrum is regular and only one maximum is visible, for a frequency scaling as $\scale \fdrag^{-1}$. This behaviour is described by a closed-form solution that will be detailed in a forthcoming article. Considering the estimate of ${\sim} 10^{-5} \ {\rm s^{-1}} $ found for the Earth's Rayleigh drag frequency from model adjustments to the current dissipation rate \citep[][]{Webb1980,Farhat2022b}, we note that this value is just slightly less than the critical value marking the transition between the quasi-adiabatic and frictional regimes in \fig{fig:spectra_parameters} (middle right panel), namely $\fdrag \approx 3.2 \times 10^{-5} \ {\rm s^{-1}}$ (see also \sect{sec:metric_continent}).

\paragraph{Effect of the solid elasticity and anelasticity.} We finally examine the role played by the parameters of the solid part (\fig{fig:spectra_parameters}, bottom panels). As discussed above, the elasticity of the solid part acts to attenuate the resonances of oceanic tidal modes by elastic adjustment. The load caused by a local water mass surplus lowers the oceanic floor, which decreases the gravitational potential of the planet's tidal response. If the solid part is rigid, this tidal response corresponds to the oceanic tide, as observed for $\mupla = 300$~GPa. The Andrade time only affects the non-resonant background of the spectrum through the scaling factor $\tauA^{-\alphaA}$ of the solid dissipation rate, as shown by \eq{imk2_solid_asymp}. As $\tauA$ increases, the non-resonant background decreases. This effect is visible in the high-frequency range, in particular. Table~\ref{tab:param_effects} summarises the effects of all the studied parameters on specific features of the frequency spectra. 

\paragraph{Ignored effects and model limitations.} \rec{As the spectral features detailed above for tidal dissipation have been obtained from the simplified theoretical framework of \sect{sec:oceanic_response}, we shall end this section by discussing the main ignored effects that may alter them. First, although the single continent configuration appears as a convenient setup to introduce the anisotropic effect induced by continentality, realistic land-ocean distributions are more complex. Consequently, the oceanic tidal response cannot be simply related to the geometry of the ocean basin in the general case, as highlighted by numerical studies performed for Earth and exoplanetary land-ocean distributions \citep{ER2001,ER2003,Green2017,Blackledge2020}. However, since the oceanic tidal waves are planetary scale, one may reasonably expect that they shall mainly interfere with large scale continental features. The predominant modes that shape the tidal dissipation spectrum of the planet are thus related to the largest length scales of the continental distribution.} 

\rec{That said, as shown by \cite{Green2017}, the smoothness of the continental geometry tends to yield overestimates of the tidally dissipated energy. The coastline fractality thus both acts to attenuate and broaden the resonances. Similarly, we remark that the eigenfrequencies of the oceanic tidal modes vary with the oceanic depth, given that the phase speed of gravity waves scale as $\scale \Hoc^{1/2}$. As a consequence, planetary scale variations of the oceanic depth presumably act to attenuate and broaden the resonant peaks. Complex bathymetries are associated with spatially dependent dissipation rates, the dissipation coefficients being strongly determined by small-scale topographic features \citep[e.g.][]{ER2001,Carter2008,Arbic2010,Green2013,Merdith2021}. It is also noteworthy that predominant dissipative mechanisms such as turbulent bottom drag scale non-linearly with the speed of tidal flows, which precludes resonances to reach very high amplitudes. Analogously, the stabilising effect of an hypothetic overlying ice shell would damp the resonant peaks, thus reducing tidal dissipation, as observed in the case of icy satellites \citep[e.g.][]{Beuthe2016,Matsuyama2018}.}

\def\wpanel{0.47\textwidth}
\begin{figure*}[htb]
  \RaggedRight \twolabelsat{5.15cm}{\textsc{Colatitude of the continental centre}}%
                           {13.9cm}{\textsc{Angular radius of the continent}} \\[-0.2cm]
  \centering
   \includegraphics[width=\wpanel,trim =0.0cm 0.7cm 0cm 0.0cm,clip]{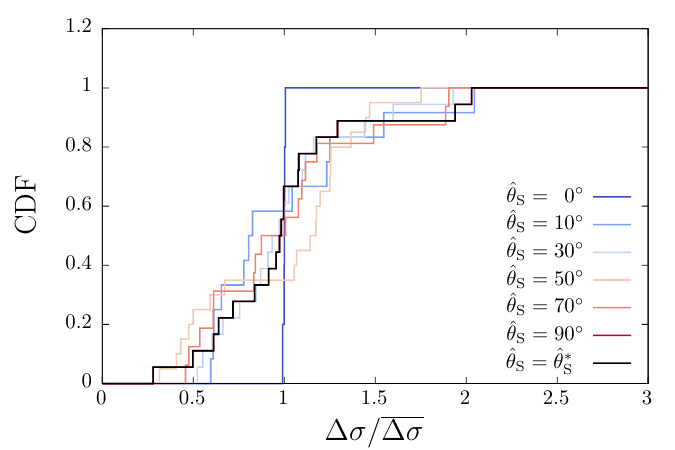} 
   \includegraphics[width=\wpanel,trim = 0.0cm 0.7cm 0cm 0.cm,clip]{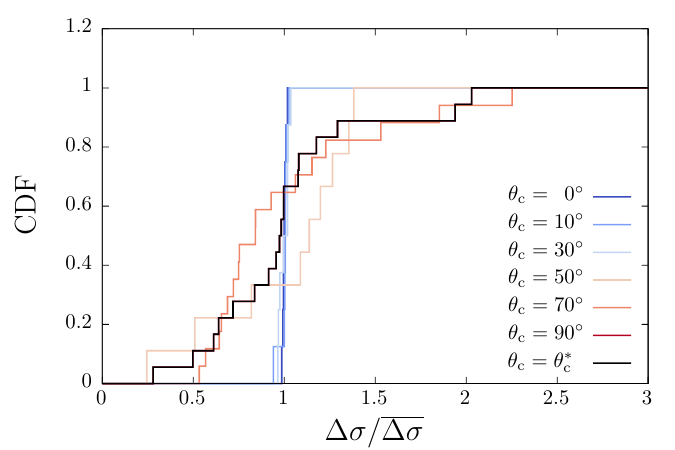} \\[0.2cm]
   \RaggedRight \twolabelsat{5.15cm}{\textsc{Oceanic depth}}%
                           {13.9cm}{\textsc{Rayleigh drag frequency}} \\[-0.2cm]
  \centering
  \raisebox{0.cm}[0cm][0pt]{\includegraphics[width=\wpanel,trim =0.0cm 0.7cm 0cm 0.cm,clip]{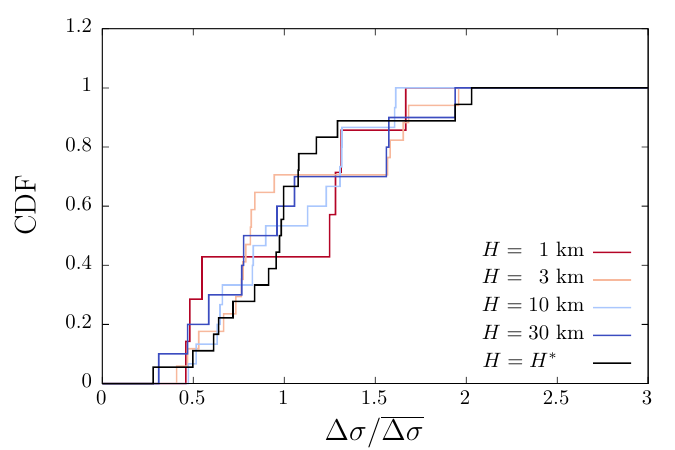}}
    \includegraphics[width=\wpanel,trim = 0.0cm 0.7cm 0.cm 0.cm,clip]{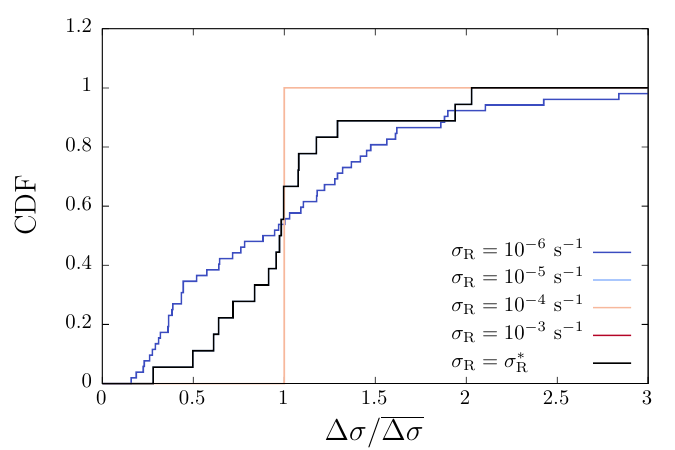} \\[0.2cm]
  \RaggedRight \twolabelsat{5.15cm}{\textsc{Shear modulus of the solid part}}%
                           {13.9cm}{\textsc{Andrade time of the solid part}} \\[-0.2cm]
  \centering
    \includegraphics[width=\wpanel,trim = 0.0cm 0cm 0cm 0.cm,clip]{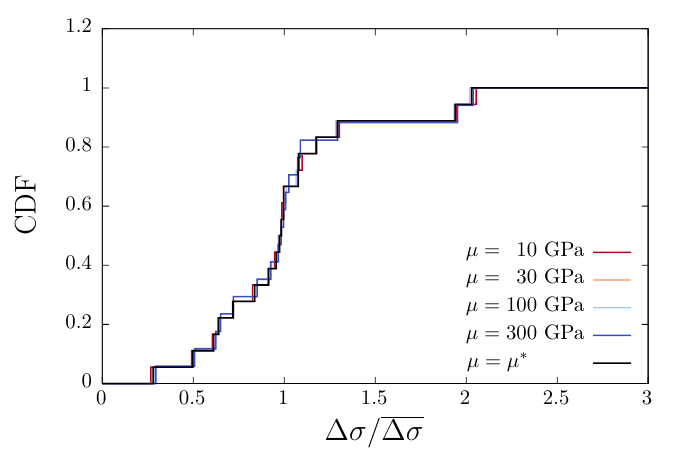} 
    \includegraphics[width=\wpanel,trim =0.0cm 0.0cm 0cm 0.cm,clip]{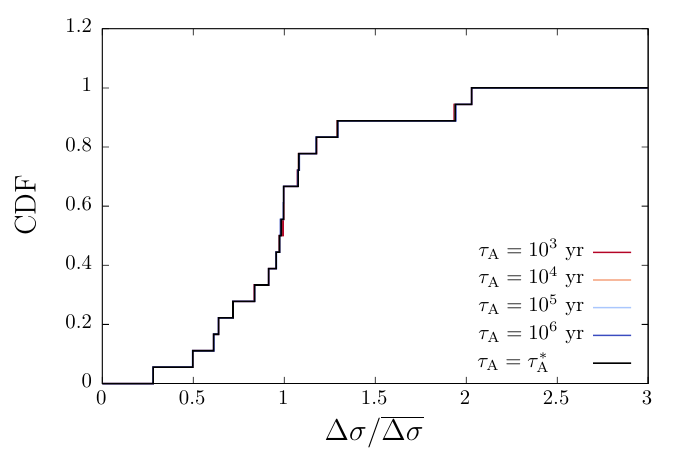}  
      \caption{\rec{CDF of the frequency intervals separating two consecutive relative maxima in the spectra of \fig{fig:spectra_parameters}. {\it Top:} Variation of the colatitude of the continental centre (left panel) and continental angular radius at constant seawater volume (right panel). {\it Middle:} Variation of the oceanic depth (left panel) and Rayleigh drag frequency (right panel). {\it Bottom:} Variation of the effective shear modulus (left panel) and Andrade time (right panel) of the solid part. For each spectrum, the frequency intervals are normalised by their average, and the CDF is normalised. The solid black line designates the reference case defined by Table~\ref{tab:param_refcase}: $\colscontrc = 90^\degree$, $\colcontrc = 90^\degree$, $\Hocrc = 4.0$~km, $\fdragrc = 10^{-5}~{\rm s^{-1}}$, $\muplarc = 25.1189 $~GPa, and $\tauArc = 12897.1$~yr.}}
       \label{fig:cdf_parameters}%
\end{figure*}

\def\wpanel{0.35\textwidth}
\def\wkey{0.12\textwidth}
\begin{figure*}[htb]
   \centering
   \includegraphics[width=\wpanel,trim =0.0cm 0.0cm 3.0cm 0.0cm,clip]{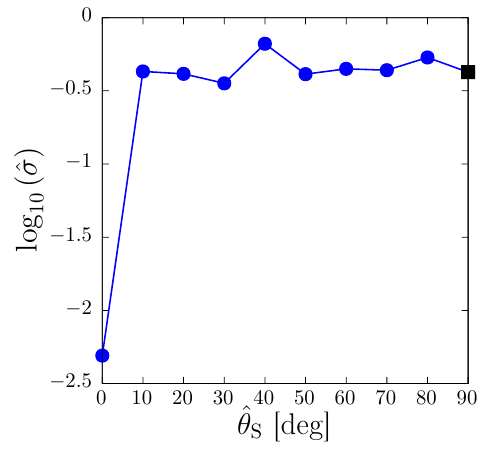} 
   \includegraphics[width=\wpanel,trim =0.0cm 0.0cm 3.0cm 0.0cm,clip]{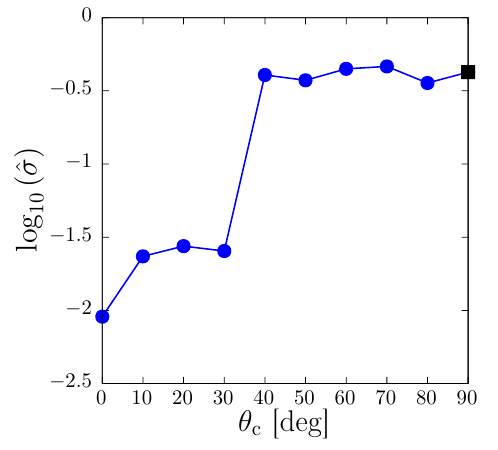}  \\
   \includegraphics[width=\wpanel,trim =0.0cm 0.0cm 3.0cm 0.0cm,clip]{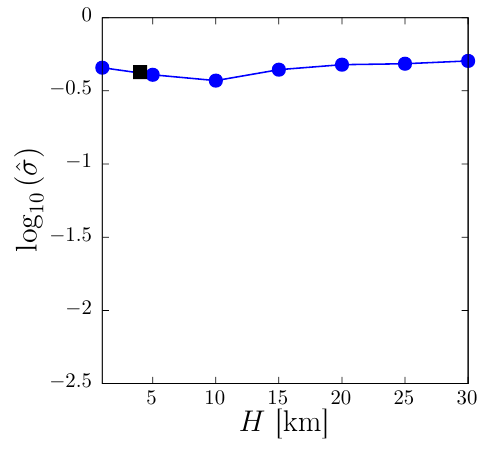} 
   \includegraphics[width=\wpanel,trim =0.0cm 0.0cm 3.0cm 0.0cm,clip]{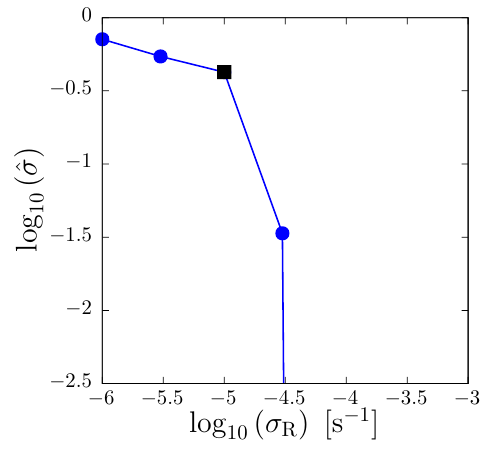}  \\
    \includegraphics[width=\wpanel,trim =0.0cm 0.0cm 3.0cm 0.0cm,clip]{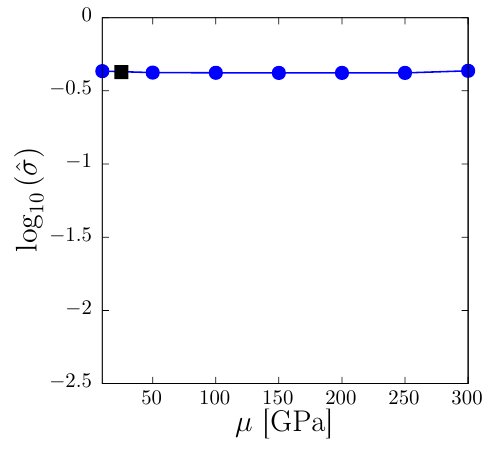} 
   \includegraphics[width=\wpanel,trim =0.0cm 0.0cm 3.0cm 0.0cm,clip]{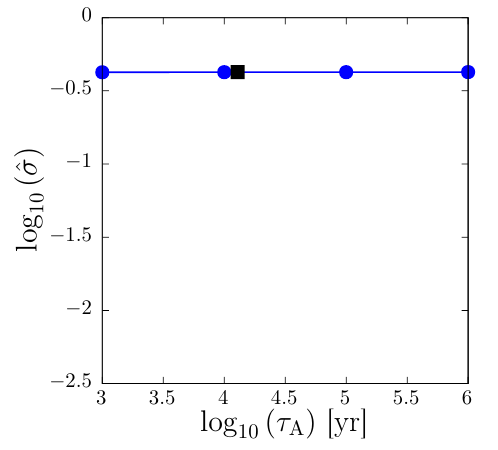} 
      \caption{Evolution of the normalised standard deviation of the frequency interval separating two relative maxima with the parameters of \fig{fig:spectra_parameters}. The metric, given by \eq{sdnorm}, is plotted in logarithmic scale (vertical axis) as a function of the parameters (horizontal axis). {\it Top:} Dependence on the colatitude of the continental centre (left panel) and angular radius of the continent (right panel). {\it Middle:} Dependence on the oceanic depth (left panel) and Rayleigh drag frequency (right panel). {\it Bottom:} Dependence on the effective shear modulus (left panel) and Andrade time (right panel) of the solid part. Blue dots indicate the values obtained for various values of the parameters, while black squares designate the reference case defined in Table~\ref{tab:param_refcase}\rec{: $\colscontrc = 90^\degree$, $\colcontrc = 90^\degree$, $\Hocrc = 4.0$~km, $\fdragrc = 10^{-5}~{\rm s^{-1}}$, $\muplarc = 25.1189 $~GPa, and $\tauArc = 12897.1$~yr.}}
       \label{fig:metric_parameters}%
\end{figure*}

\section{A metric for the signature of supercontinents}
\label{sec:metric_continent}

Until now, we have been investigating the direct problem that consists in generating frequency spectra of the tidal quantities from sets of values assigned to the model parameters. Particularly, in \sect{sec:parametric_study}, we made an attempt to infer the tidal response of planets hosting ocean basins from knowledge of their geometry. The relationships between spectral and geometric features actually correspond to a very broad field of mathematical problems known as spectral geometry. Through popular case studies of this field, such as Kac's drum \citep[][]{Kac1966}, it was shown that the geometry of a vibrating system can be predicted to a certain extent if its spectral features are known. This is the so-called inverse problem that we aim to address in the present section by investigating the extent to which one can infer the geometry of the ocean basin from the spectral features of the planet's tidal response. Particularly, we define further a metric that quantifies the specific signature of a supercontinent in the evolution of the tidal torque with the tidal frequency. This frequency dependence of the tidal torque is straightforwardly connected to the orbital evolution of the satellite or the rotational evolution of the planet, as any crossed resonance induces a rapid variation of the orbital parameters over a short time period \citep[see e.g.][]{ADLPM2014,Farhat2022b}.  

The parametric study of \sect{sec:parametric_study} shows evidence of the numerous degeneracies characterising the dependences of the frequency spectra on the system's parameters (see Table~\ref{tab:param_effects}). Several features -- such as the height of resonant peaks or the non-resonant background -- appear to be sensitive to three key parameters at the same time. Particularly, these results highlight the predominant role played by the Rayleigh drag frequency, which affects almost all of the spectral features listed in Table~\ref{tab:param_effects}. Nevertheless, we also remark that the only parameters having an impact on the regularity of the spectra -- notwithstanding the effect of the Rayleigh drag frequency -- are those defining the geometry of the ocean basin, namely the colatitude of the continental centre and the angular radius of the continent. This argues for the existence of spectral quantities sensitive to these parameters only. Such quantities are metrics of the geometry. Ideally, they shall vary with $\colscont$, $\colcont$, and $\fdrag$, while being insensitive to the other parameters. 

In \fig{fig:spectra_parameters}, the irregularity of spectra seems to be mainly characterised by a significant dispersion of the frequency intervals separating consecutive relative maxima of the tidal torque, denoted by $\intftidei{\kk} \define \ftidei{\kk+1} - \ftidei{\kk}$ in the following, with $\ftidei{\kk}$\mynom{$\ftidei{\kk}$}{Tidal frequency of the $\kk$-th maximum}\mynom{$\ftidei{\kk}$}{$\kk$-th frequency interval between two consecutive maxima} being the frequency of the $\kk$-th maximum. \rec{In order to make these frequency intervals insensitive to the dilation-contraction effect induced by oceanic depth variations (see \fig{fig:spectra_parameters}, middle left panel), they are normalised by their average\mynom{$\intftidemean$}{Mean frequency interval separating two consecutive relative maxima of the tidal torque}, given by
\begin{equation}
\intftidemean \define \frac{1}{\Nint} \sum_{\kk=1}^{\Nint} \intftidei{\kk},
\end{equation}
where $\Nint$ designates the total number of intervals. We thus denote by $\intftideni{\kk} \define \intftidei{\kk}/\intftidemean $ the normalised intervals\mynom{$\intftideni{\kk}$}{$\kk$-th frequency interval normalised by the mean interval}. Finally, sorting the intervals in ascending order and using the subscript $\jj$ such that $\intftidei{\jj} < \intftidei{\jj+1}$ for $\jj = 1 , \ldots , \Nint$, we introduce the cumulative distribution function (CDF) of the normalised frequency intervals\mynom{CDF}{Cumulative distribution function},}
\begin{equation}
\CDF \left( \intftiden \right) \define  \frac{1}{\Nint} \underset{\intftideni{\jj} \leq \intftiden}{\argmax} \left\{ \intftideni{\jj} \right\}.
\end{equation}

\rec{Figure~\ref{fig:cdf_parameters} shows the CDFs of the frequency intervals separating two consecutive maxima for the spectra plotted in \fig{fig:spectra_parameters}. In these plots, the quasi-uniformly spaced peaks of the global and polar hemispherical oceans are characterised by a sharp step in the CDF, which abruptly switches from 0 to 1. Conversely, the symmetry breaking effect of continental shelves induces a smooth transition. As noted qualitatively with the spectra of \fig{fig:spectra_parameters}, the slope of the CDF essentially varies with the position of the continent on the globe, its size, and the dissipative time scale. This leads us to consider, as an appropriate metric for the continentality effect, the normalised standard deviation of the frequency intervals separating consecutive maxima of the tidal torque, \mynom{$\standarddevn$}{Normalised standard deviation of the intervals separating consecutive relative maxima of the tidal torque}}
\begin{equation}
\standarddevn \define \sqrt{ \frac{1}{\Nint} \sum_{\kk=1}^{\Nint} \left( \frac{\intftidei{\kk}}{\intftidemean}  - 1\right)^2  }.
\label{sdnorm}
\end{equation}
\rec{The thus defined normalised standard deviation is such that $0 \leq \standarddevn \leq 1$, where $\standarddevn=0$ corresponds to uniformly spaced relative maxima or to $\Nint = 1$ (two maxima), and $\standarddevn=1$ to an extreme dispersion of the resonant peaks. By convention, we set $\standarddevn$ to zero when one unique maximum is observed.}

By reproducing the methodology detailed in \sect{sec:parametric_study}, we computed the tidal response of the reference planet defined in Table~\ref{tab:param_refcase} for various values of the colatitude of the continental centre, angular radius of the continent at constant seawater volume, oceanic depth, Rayleigh drag frequency, effective shear modulus of the solid part, and Andrade time, each parameter being modified independently of the others. In all cases, we calculated the values of $\standarddevn$ from the frequency spectra obtained for the imaginary part of the quadrupolar Love number. These values are plotted in \fig{fig:metric_parameters}. 

\rec{The quantity $\standarddevn$ appears to be} almost insensitive to variations of $\Hoc$, $\mupla$, and $\tauA$, whereas it varies over several orders of magnitude when $\colscont$, $\colcont$, or $\fdrag$ are modified. Moreover, we recover quantitatively the tendencies identified in the parametric study (Table~\ref{tab:param_effects}), namely $\standarddevn$ really accounts for the role played by the geometry of the ocean basin and the dissipative timescale in the planet's tidal response. Thus, the normalised standard deviation defined by \eq{sdnorm} can be considered as a spectral metric of the planet's geometric features, both on the global scale pertaining to the location or spread of land, if any, along with the local topographical scale dictating the dissipative timescale in the ocean. 

Interestingly, the dependence of $\standarddevn$ on $\colscont$, $\colcont$, or $\fdrag$, is not even. Instead, threshold effects may be observed\rec{, similar to those of \fig{fig:cdf_parameters}}. First, the evolution of the metric with $\colscont$ (\fig{fig:metric_parameters}, top left panel) illustrates the fact that a small inclination of the continent on the globe with respect to the pole is sufficient to break the axi-symmetry. By switching from $\colscont = 0^\degree$ (polar continent) to $\colscont= 10^\degree$, the metric skyrockets to reach a plateau around $\standarddevn \sim 0.4$. This plateau suggests that changing the position of the continental centre from $10^\degree$ to $90^\degree$ does not affect the regularity of the frequency spectrum. Basically, one observes here a saturation regime where the signature of the supercontinent through the chosen metric is the same regardless of its position on the globe. 

A similar behaviour is observed as the angular radius of the continent grows from $0^\degree$ (global ocean) to $90^\degree$ (hemispherical ocean) at constant water volume (\fig{fig:metric_parameters}, top right panel). While the metric does not evolve much with $\colcont$ in general, it increases drastically around $\colcont \approx 35^\degree$, which is the critical size identified from the plotted spectra in the parametric study. This value is consistent with the highest value of $\colcont$ for which the frequency spectrum still exhibits a regular series of resonant peaks that resembles the tidal response of the global ocean (\fig{fig:spectra_parameters}, top right panel). Beyond $35^\degree$, the spectrum becomes complex and irregular due to the strong interference of the continent with tidal flows. We note that the metric remains stable for $\colcont \gtrsim 40^\degree$, which suggests that the regularity of the frequency spectrum is insensitive to the size of the continent beyond this value. This highlights another saturation regime, similar to that observed for the position of the continent, in which the signature of the supercontinent is the same regardless of its size. \rec{It is noteworthy that the way the continent interferes with the oceanic tidal flows is related to its size with respect to the typical horizontal wavelengths of the predominating tidal modes, which are planetary scale. As a consequence, the critical angular radius marking the regime transition for the studied Earth-sized planet would be similar for a super-Earth or a smaller planet.}

Finally, the metric decays monotonically as the Rayleigh drag frequency increases, with a sharp variation occurring around $\fdrag \sim 3.2 \times 10^{-5} \ {\rm s^{-1}}$ (\fig{fig:metric_parameters}, middle right panel). This critical value marks the transition between the quasi-adiabatic and frictional regimes (see Table~\ref{tab:control_param}). In the latter, the resonances associated with oceanic modes are damped by the strong friction of tidal flows with the oceanic floor, which makes the tidal torque evolve smoothly with the tidal frequency, as shown by \fig{fig:spectra_parameters} (middle right panel). Therefore, the regularity of the spectrum conveys no information about the geometry of the ocean basin in this regime. Conversely, in the quasi-adiabatic regime, $\standarddevn$ increases as $\fdrag$ decays due to the growing number of visible resonant peaks. 

\section{Conclusions}
\label{sec:conclusions}
In this paper we have examined the linear response of an ocean planet hosting a supercontinent to gravitational tidal forcing. This problem may be regarded as a simplified model of the tides raised on Earth by the Moon and their interactions with the land-ocean distribution, which predominantly drives the evolution of the Earth-Moon system over long time scales \citep[e.g.][]{BR1999,Daher2021,Tyler2021,Farhat2022b}. The theory developed in the present study expands on earlier works that investigated the tidal response of hemispherical or global oceans \citep[][]{LH1968,LHP1970,Webb1980,Webb1982,ADLML2018,Farhat2022b}. In this approach, the continent is represented by a spherical cap of arbitrary size and position, and the LTEs are written in the shallow water approximation for an oceanic layer of uniform depth. These simplifications allow the problem to be formulated analytically in terms of explicitly defined oceanic eigenmodes, which provides a deep insight into the physics governing the planet's tidal response. Additionally, such a formalism appears to be well suited to the harmonic analysis of the tidal dynamics since it relies on a small set of key parameters.

As a first step, we established the equations describing the dynamics of the forced tidal flows. These equations include the gravitational and surface couplings that result from the deformation of the solid part generated by the ocean loading and the tidal forces exerted by the perturber. The visco-elastic response of the solid part to tidal forcings is described in a generic way by gravitational and load Love numbers. We computed it in practice by making use of the closed-form solutions provided by the Andrade model for a homogeneous body, following the prescriptions obtained from 1D models taking the planet's internal structure into account. The LTEs were expressed as an algebraic system describing the evolution of coupled eigenmodes in the frequency domain. This system is controlled by a small set of clearly identified dimensionless numbers that characterise the regime of oceanic tidal waves and the geometry of the ocean basin. We finally used the harmonic expansion of tidal flows to express the corresponding input tidal power, tidally dissipated power, tidal Love numbers, and tidal torque exerted about the spin axis. 

As a second step, we applied the theory to a reference case, which is essentially defined as a simplified Earth-Moon system. In this configuration, the continent is hemispherical and located at the equator. By computing the evolution of the aforementioned tidal quantities with the forcing tidal frequency, we recovered the solutions obtained in \cite{Farhat2022b}. We showed that the planet's tidal dissipation rate is dominated by the contribution of the ocean, while the solid part determines it mainly in the high-frequency range, that is for a fast rotating planet. Moreover, the obtained results highlight the symmetry breaking effect of the supercontinent, which splits the modes of the global ocean into more modes by interfering with tidal flows. As a result, the spectra of tidal quantities are flattened and they exhibit additional relative maxima corresponding to the resonances of the basin modes. 

As a third step, we carried out a parametric study for a set of six key parameters, by using the previously defined reference case as a template configuration. The obtained results reveal the degeneracies affecting the evolution of the main spectral features with the selected parameters. However, they also indicate that the specific signature of the geometry can be unravelled to a certain extent by characterising the erraticity of the spectra. As the position or angular radius of the continent increase, the frequency spectrum of the tidal torque becomes irregular, while the other parameters only alter the frequencies of the resonant peaks, their widths, their heights, and the non-resonant background. The only exception to this statement is the Rayleigh drag frequency, which affects the regularity of frequency spectra as well. 

The above qualitative results were refined in a quantitative way as a last step. By following an inverse problem approach inspired from spectral geometry, we showed that some geometric features of the ocean basin may be inferred from the spectral irregularity, which is quantified by the normalised standard deviation of the frequency intervals separating consecutive relative maxima in the frequency spectrum of the tidal Love number. The sensitivity of this quantity to the geometrical parameters is characterised by threshold effects. As regards the position of the continent on the globe, the metric only allows the poles to be distinguished from other locations. Similarly, it seems to indicate that the signatures of continents with angular radii larger than $40^\degree$ are indistinguishable from each other. 

Nevertheless, the harmonic analysis also suggests that a continent similar to South America (${\sim} 30^\degree$-angular radius) or smaller would not alter, qualitatively, the oceanic tidal response even if it is located at the equator, where the interference of the coastlines with tidal flows tends to be maximal. In such configurations, the tidally dissipated energy predicted by the harmonic analysis is approximately the same as in the absence of continent. Finally the metric appears to be sensitive to the transition between the quasi-adiabatic and frictional regimes, which occurs for $\fdrag \approx 3.2 \times 10^{-5} \ {\rm s^{-1}}$. This critical value is slightly greater than that estimated for Earth \citep[${\sim} 10^{-6}-10^{-5} \ {\rm s^{-1}}$, e.g.][]{GM1971,Webb1973,Tyler2021,Farhat2022b}, which is consistent with the frequency-resonant behaviour of the Earth's tidal dissipation rate inferred from Lunar laser ranging and geological data \citep[e.g.][]{BR1999,Tyler2021}.

Both the spectral method and the metric for continentality effects elaborated in the present work are highly relevant to studies dealing with the long-term tidal evolution of bodies with surface oceans, such as extrasolar rocky planets orbiting in the habitable zone of their host star. Particularly, the studied spectral irregularity of the tidal torque has direct implications on the history of planet-satellite systems. Tidal resonances cause irregular evolutions of the satellite's orbit and planet's rotation, thus leading to abrupt LOD or obliquity variations that might result in significant climatic effects. \rec{We will examine specifically the consequences of continentality on the long-term evolution of the Earth-Moon system in a separate dedicated study by using the theory established in the present work.}

\rec{As this theory forms a basis for more sophisticated descriptions of continentality effects, we will consider refining the geometry in the future in order to account for the role played by the fractality of non-smooth coastlines, the complex bathymetry of the ocean basin, or the spatial dependence of energy dissipation.} Also, while we have characterised the signature of one unique supercontinent in the planet's tidal response, the combined effects of several continents still needs to be elucidated. All these questions require further work improving meanwhile the theory and the quality of observational constraints on the past history of the Earth-Moon system. \rec{We will address them through several forthcoming studies.}

\begin{acknowledgements}
\rec{The authors thank the anonymous referee for his/her insightful comments that significantly enhanced the manuscript.} This project has been supported by the French Agence Nationale de la Recherche (AstroMeso ANR-19-CE31-0002-01) and by the European Research Council (ERC) under the European Union's Horizon 2020 research and innovation program (Advanced Grant AstroGeo-885250). This research has made use of NASA's Astrophysics Data System. 
\end{acknowledgements}

\bibliographystyle{aa}  
\bibliography{references} 

\begin{appendix}


\section{Nomenclature}
\label{app:nomenclature}

The notations introduced in the main text are listed below in order of appearance.

\vspace{-1.2cm}

\printnomenclature

\section{Associated Legendre Functions (ALFs)}
\label{app:legendre_functions}

The ALFs, usually denoted by $\LegF{\llat}{\mm}$, are the solutions of the equation
\begin{equation}
\label{helmholtz_col}
\frac{1}{\sin \col} \DD{}{\col} \left( \sin \col \DD{\hfunc}{\col} \right) + \left[  \llat \left( \llat + 1 \right) - \frac{\mm^2}{\sin^2 \col} \right] \hfunc = 0,
\end{equation}
which is found by separating the longitude ($\lon$) and the colatitude ($\col$) in the eigenvalues problems associated with the Laplace equation in spherical coordinates (see \eq{helmholtz}). The function $\hfunc \left( \col \right) $ in \eq{helmholtz_col} is assumed to be a regular function of the colatitude. The parameters $\llat$ and $\mm$ are commonly referred to as the degree and the order of the Legendre function, respectively. They may be integral, non-integral, or complex depending of the eigenvalues of the problem, $\mm^2$ and $\llat \left( \llat + 1 \right)$, which are determined both by the boundary conditions and the geometry of the domain of definition of the solutions on the unit sphere. 

In the general case, the ALFs are defined, for $\zz \in \Cset$ such that $\left| \zz \right| < 1$ and arbitrary complex constants, $\llat$ and $\mm$, as \citep[][Eq.~(8.1.2)]{AS1972}
\begin{equation}
\Plm \left( \zz \right) \define \frac{1}{\GammaF \left( 1 - \mm \right)} \left( \frac{1 + \zz }{1 - \zz} \right)^{\frac{\mm}{2}}  \hypergtwoFone \left( - \llat , \llat +1 ; 1 - \mm ; \frac{1 - \zz}{2} \right).
\label{alfgen}
\end{equation}
In the above equation, $\GammaF$ is the Gamma function introduced in \eq{imk2_solid_asymp}, which is defined for $\Re \left(\zz \right) >0$ as \citep[][Eq.~(6.1.1)]{AS1972}
\begin{equation}
\GammaF \left( \zz \right) \define \integ{t^{\zz-1} \expo{-t} }{t}{0}{\infty}.
\label{GammaF}
\end{equation}
The hypergeometric function $ \hypergtwoFone$ is defined as \citep[][Eq.~(15.1.1)]{AS1972}
\begin{equation}
\label{hyperg2F1}
\hypergtwoFone \left( a , b ; c ; d \right) = \sum_{\jj = 0}^{\infty} \frac{\pochham{a}{\jj} \pochham{b}{\jj} }{ \pochham{c}{\jj}} \frac{\xx^{\jj}}{\jj !} ,  
\end{equation}
where $\pochham{a}{\jj}$ designates the Pochhammer symbol \citep[][Eq.~(6.1.22)]{AS1972}, 
\begin{align}
& \pochham{a}{0} \define 1, \\
& \pochham{a}{\jj} \define a \left( a + 1 \right) \ldots \left( a + \jj  - 1 \right) = \frac{\GammaF \left(a + \jj \right)}{\GammaF \left( a \right)}.
\end{align}
The derivatives of the ALFs with respect to $\zz$ are given by 
\begin{equation}
\dd{\LegF{\llat}{\mm}}{\zz} = \frac{1}{\zz^2 - 1} \left[ \zz \llat \LegF{\llat}{\mm} \left( \zz \right) - \left( \mm + \llat \right) \LegF{\llat-1}{\mm} \left( \zz \right)  \right]. 
\end{equation}
We note that the functions $\GammaF$ and $\hypergtwoFone$ in \eq{alfgen} are defined only if their arguments satisfy certain conditions. In practice, the ALFs are computed for any values of these arguments by using the symmetry properties of the regularised hypergeometric function $\hypergtwoFonereg$, which is defined as \citep[e.g.][]{Johansson2016}
\begin{equation}
\hypergtwoFonereg \left( a , b ; c ; \zz \right)  \define \frac{\hypergtwoFone \left( a , b ; c ; \zz \right) }{\GammaF \left( c \right)}. 
\end{equation}

For $\mm \in \Zset$, $\llat \in \Rset$ and $\zz=\xx \in \Rset$ such that $-1 \leq \xx \leq1$, which corresponds to the case of the SCHs, the general definition given by \eq{alfgen} simplifies to \citep[e.g.][]{Thebault2006}
\begin{align}
\label{alfscha}
\Plm \left( \xx \right) = & \frac{\left( -1 \right)^{\frac{\mm + \abs{\mm}}{2}} }{2^{\abs{\mm}} \abs{\mm}!} \frac{\GammaF \left( \llat + \abs{\mm} +1 \right)}{\GammaF \left( \llat - \mm +1 \right) } \left( 1 - \xx^2 \right)^{\abs{\mm}/2}  \\
 & \times \hypergtwoFonereg \left( \abs{\mm} - \llat , \llat + \abs{\mm} + 1; \abs{\mm} +1 ; \frac{1 - \xx}{2} \right). \nonumber
\end{align}
Here, the hypergeometric function is expressed as 
\begin{equation}
\label{hyperg2F1reg_Plm}
\hypergtwoFonereg \left( \abs{\mm} - \llat , \llat + \abs{\mm} + 1;  \abs{\mm} +1 ; \frac{1 - \xx}{2} \right) = \sum_{\jj=0}^{\infty} \hyperci{\jj}  \left( \frac{1 - \xx}{2} \right)^\jj,
\end{equation}
with the coefficients $\hyperci{\jj}$ recursively defined as 
\begin{align}
& \hyperci{0} = 1, \\
& \hyperci{\jj+1} = \frac{\left( \jj + \abs{\mm} - \llat \right) \left( \jj + \abs{\mm} + \llat + 1 \right)}{\left( \jj + 1 \right) \left( \jj + 1 + \abs{\mm} \right)} \hyperci{\jj}. 
\end{align}
In the general case, the sum of \eq{alfscha} converges only for $-1<\xx \leq 1$, which corresponds to the inequality $0 \leq \col < \pi$ in the change of coordinates $\xx = \cos \col$. However the sum of \eq{alfscha} is not infinite any more when $\llat$ is an integer since $\hyperci{\jj+1} = 0$ as long as $\jj >  \llat - \abs{\mm} $. As a consequence it converges also for $\col = \pi$ in this case. 

Since the degrees $\llat$ tend to integral values ($\llat = 0 , 1 , 2, \ldots$) as $\col \rightarrow \pi$, the generalised ALFs given by \eqs{alfgen}{alfscha} continuously converge towards the well-known ALFs of the SPHs, which are defined for $ -1 \leq \xx \leq 1 $, and $\llat$ and $\mm$ integers such that $\llat \geq \abs{\mm}$, as \citep[][]{AS1972,Arfken2005} 
\begin{equation}
\Plm \left( x \right) \define \left(-1 \right)^\mm \left( 1 - x^2 \right)^{\mm/2}  \DDn{}{x}{\mm} \Pl \left( x \right),
\label{Plmstd}
\end{equation}
the $\Pl$ designating the Legendre polynomials, given by 
\begin{equation}
\Pl \left( x \right) \define \frac{1}{2^\llat \llat !} \DDn{}{x}{\llat} \left[ \left( x^2 - 1 \right)^\llat \right]. 
\end{equation}
Owing to the large number of terms that is sometimes required to compute the sum of \eq{hyperg2F1reg_Plm} at machine precision, the numerical evaluation of the generalised ALFs given by \eq{alfscha} may be tricky in some cases. Particularly, the series converges very slowly when $\xx \approx -1$, which occurs if the ocean is almost global. In practice, the evaluation is achieved by making use of the build-in hypergeometric function $\hypergtwoFone$ of the \texttt{TRIP} algebraic manipulator \citep[][]{GL2011}.

\section{Spherical Cap Harmonics (SCHs)}
\label{app:sch}
\begin{figure*}[htb]
   \centering
   \includegraphics[width=0.8\textwidth,trim = 0.cm 0.cm 0.5cm 5.4cm,clip]{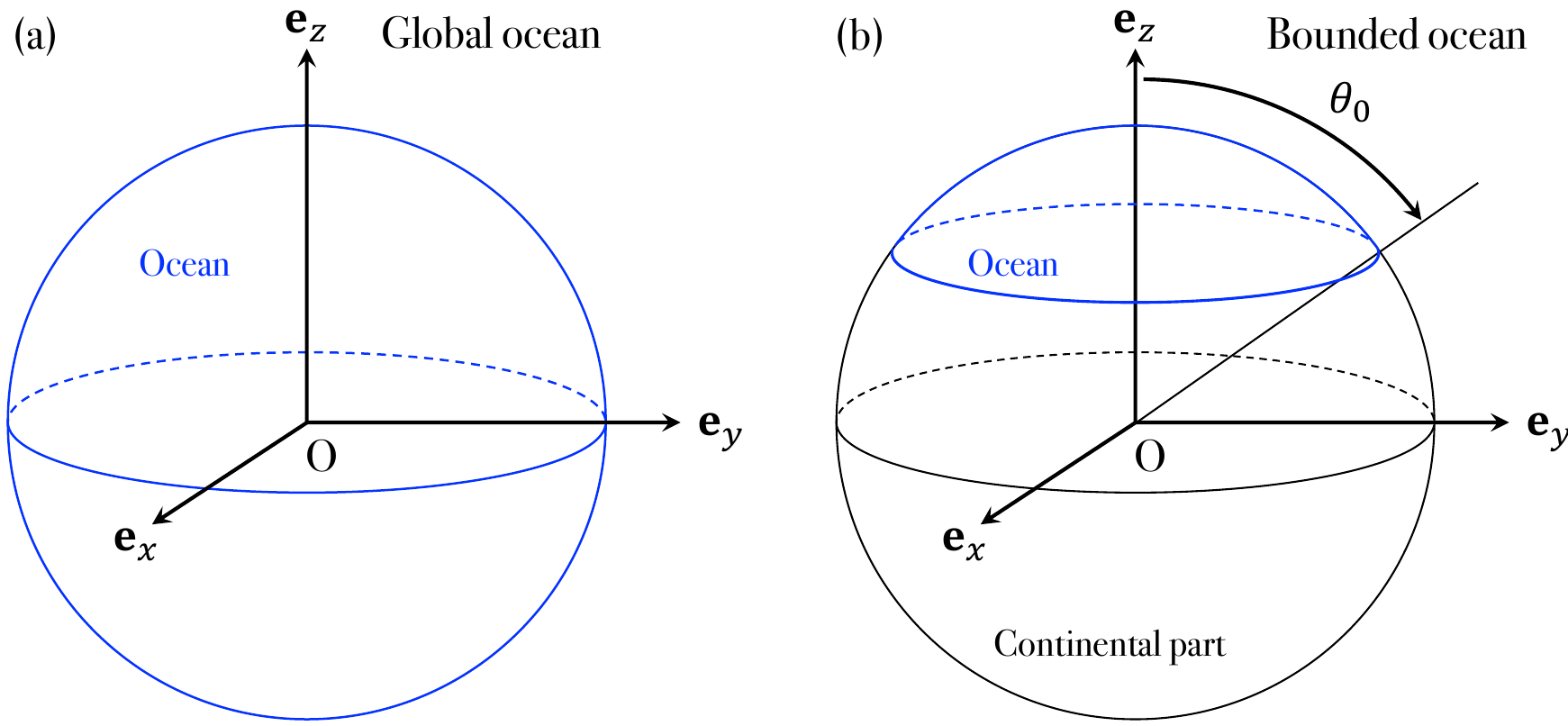}
      \caption{Geometry of the land-ocean distribution in the model. {\it Left:} global ocean distribution where the eigenfunctions are the SPHs. {\it Right:} bounded circular ocean distribution where the eigenfunctions are the SCHs. In the bounded case, the frame of reference $\rframe{\iocean}{O}{\ex}{\ey}{\ez}$ introduced in \sect{ssec:geometry_ocean} is such that $\ez$ corresponds to the symmetry axis going through the centre of the ocean basin, while it is not specified in the global case due to the spherical symmetry. }
       \label{fig:truncated_sphere}%
\end{figure*}

In the same way that the SPHs are the harmonics of the entire sphere, the SCHs designate the harmonics of the sphere truncated at a colatitude~$\colbd$ \citep[e.g.][and references therein]{Haines1985,Thebault2004,Thebault2006,Torta2019}. Figure~\ref{fig:truncated_sphere} shows the two geometrical configurations. We note that the natural coordinate system of the SCHs is such that the z-axis is the axis going through the centre of the bounded ocean, which is the symmetry axis of the system, while the coordinate system of the SPHs does not need to be specified due to the spherical symmetry of the global ocean. 

\begin{table*}[h]
\centering
\caption{\label{tab:sch_degrees} Values of the real degrees $\llati{\nk}$ of the SCHs with Neumann (SCHNs) or Dirichlet (SCHDs) boundary conditions for various angular radii of the supercontinent ($\colcont$). }
\begin{tabular}{crrrrrrrrrrrr} 
 \hline 
 \hline 
$\colcont$ (deg) & $\left( 0,0 \right)$ & $\left( 1,0 \right)$ & $\left( 1,1 \right)$ & $\left( 2,0 \right)$ & $\left( 2,1 \right)$ & $\left( 2,2 \right)$  \\ 
 \hline 
  \multicolumn{7}{c}{\textit{Neumann boundary conditions (SCHNs)}} \\
0 &     0.000000 &     1.000000 &     1.000000 &     2.000000 &     2.000000 &     2.000000 \\ 
10 &     0.000000 &     1.014533 &     0.984857 &     2.040535 &     1.957224 &     1.999318 \\ 
20 &     0.000000 &     1.054431 &     0.945331 &     2.139648 &     1.889718 &     1.989826 \\ 
30 &     0.000000 &     1.115646 &     0.900510 &     2.277970 &     1.885701 &     1.956707 \\ 
40 &     0.000000 &     1.196823 &     0.867186 &     2.450068 &     1.948680 &     1.901867 \\ 
50 &     0.000000 &     1.298769 &     0.851964 &     2.657126 &     2.062763 &     1.851211 \\ 
60 &     0.000000 &     1.424123 &     0.856313 &     2.904340 &     2.220869 &     1.827443 \\ 
70 &     0.000000 &     1.577470 &     0.880927 &     3.200567 &     2.423975 &     1.840734 \\ 
80 &     0.000000 &     1.765876 &     0.927569 &     3.559164 &     2.679212 &     1.896224 \\ 
90 &     0.000000 &     2.000000 &     1.000000 &     4.000000 &     3.000000 &     2.000000 \\ 
100 &     0.000000 &     2.296162 &     1.105011 &     4.553241 &     3.408276 &     2.162574 \\ 
110 &     0.000000 &     2.680266 &     1.254256 &     5.266546 &     3.939640 &     2.402474 \\ 
120 &     0.000000 &     3.195691 &     1.467987 &     6.219529 &     4.654189 &     2.752588 \\ 
130 &     0.000000 &     3.920704 &     1.783375 &     7.555648 &     5.660659 &     3.273969 \\ 
140 &     0.000000 &     5.012005 &     2.275396 &     9.561941 &     7.176982 &     4.090469 \\ 
150 &     0.000000 &     6.835398 &     3.119597 &    12.908284 &     9.712069 &     5.492825 \\ 
160 &     0.000000 &    10.488504 &     4.843239 &    19.604445 &    14.793105 &     8.355251 \\ 
170 &     0.000000 &    21.459763 &    10.083479 &    39.699467 &    30.056683 &    17.051968 \\ 
\hline 
\multicolumn{7}{c}{\textit{Dirichlet boundary conditions (SCHDs)}} \\
0 &     0.000000 &     1.000000 &     1.000000 &     2.000000 &     2.000000 &     2.000000 \\ 
10 &     0.201220 &     1.286874 &     1.014533 &     2.359807 &     2.040535 &     2.000661 \\ 
20 &     0.274502 &     1.424749 &     1.054431 &     2.561533 &     2.139648 &     2.009153 \\ 
30 &     0.346184 &     1.568297 &     1.115646 &     2.777734 &     2.277970 &     2.037781 \\ 
40 &     0.422281 &     1.726592 &     1.196823 &     3.019705 &     2.450068 &     2.095353 \\ 
50 &     0.506293 &     1.905895 &     1.298769 &     3.296147 &     2.657126 &     2.186686 \\ 
60 &     0.601509 &     2.112863 &     1.424123 &     3.616943 &     2.904340 &     2.315699 \\ 
70 &     0.711801 &     2.355851 &     1.577470 &     3.994876 &     3.200567 &     2.487779 \\ 
80 &     0.842253 &     2.646183 &     1.765876 &     4.447503 &     3.559164 &     2.711416 \\ 
90 &     1.000000 &     3.000000 &     2.000000 &     5.000000 &     4.000000 &     3.000000 \\ 
100 &     1.195606 &     3.441373 &     2.296162 &     5.690014 &     4.553241 &     3.374653 \\ 
110 &     1.445566 &     4.008030 &     2.680266 &     6.576628 &     5.266546 &     3.869485 \\ 
120 &     1.777288 &     4.762779 &     3.195691 &     7.758259 &     6.219529 &     4.542151 \\ 
130 &     2.240037 &     5.818619 &     3.920704 &     9.412016 &     7.555648 &     5.497155 \\ 
140 &     2.932276 &     7.401491 &     5.012005 &    11.892079 &     9.561941 &     6.944195 \\ 
150 &     4.083687 &    10.038551 &     6.835398 &    16.024836 &    12.908284 &     9.373283 \\ 
160 &     6.383235 &    15.311204 &    10.488504 &    24.289410 &    19.604445 &    14.255320 \\ 
170 &    13.275607 &    31.126398 &    21.459763 &    49.081367 &    39.699467 &    28.946234 \\ 
\hline
 \end{tabular}
 \tablefoot{The values are given for the doublets $\left(\nk , \abs{\mm} \right) $ such that $ \nk=0,1,2$ and $\abs{\mm} \leq \nk $.}
 \end{table*}
 
 \begin{figure}[htb]
   \centering
   \includegraphics[width=0.48\textwidth,trim = 0.cm 0.cm 0.0cm 0cm,clip]{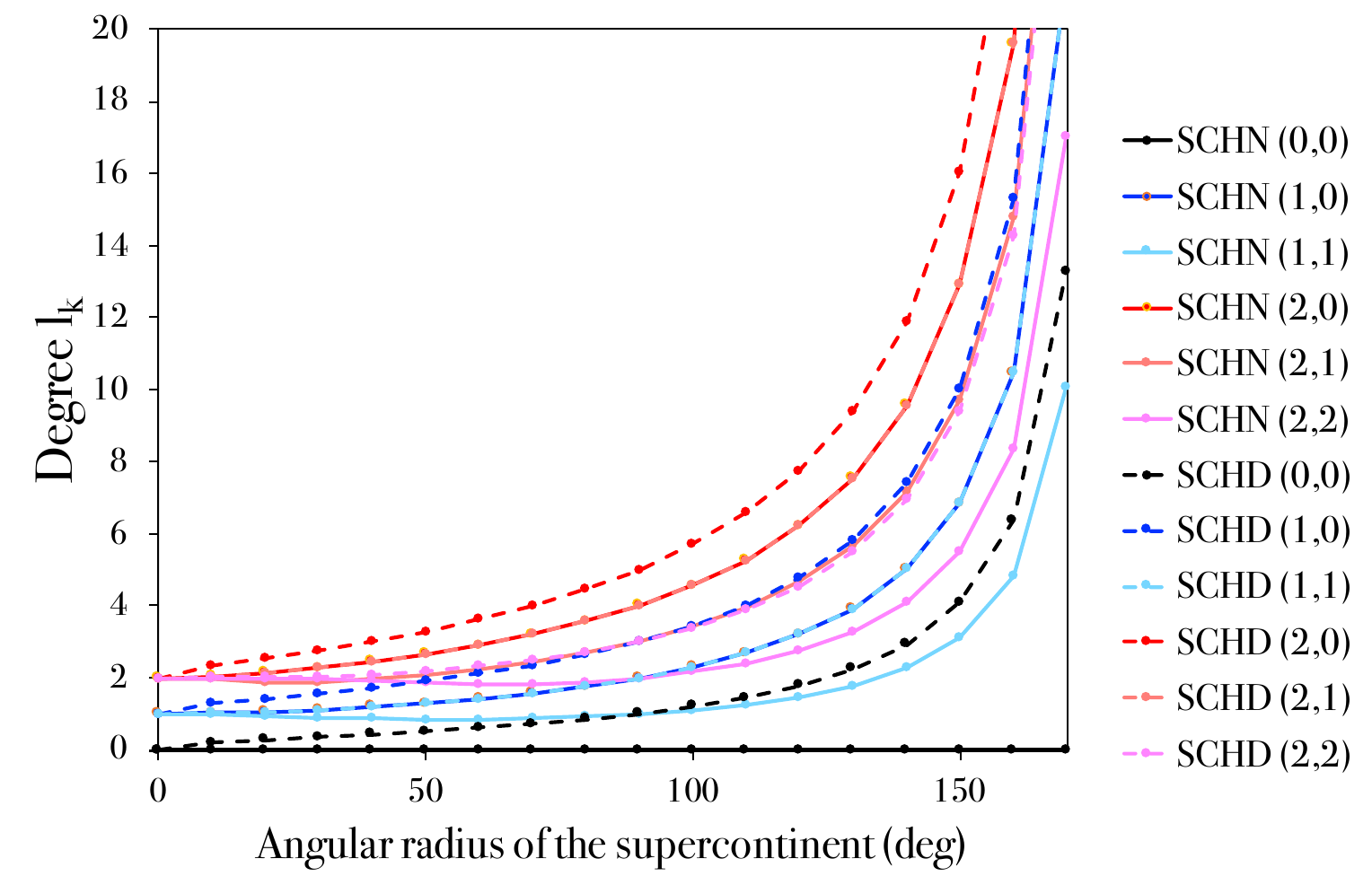}
      \caption{Real degrees $\llati{\kk}$ of the SCHs for doublets $\left( \kk , \abs{\mm} \right)$ with $0 \leq \abs{\mm} \leq \kk \leq 2$ as functions of the angular radius of the supercontinent. The set of basis functions obtained for the Neumann condition given by \eq{neucond} (SCHNs) are designated by solid lines, and those obtained for the Dirichlet condition given by \eq{dircond} (SCHDs) by dashed lines. }
       \label{fig:degrees_size}%
\end{figure}
 
 \begin{table*}[h]
\centering
\caption{\label{tab:sch_eigenv} Eigenvalues of the SCHs with Neumann (SCHNs) or Dirichlet (SCHDs) boundary conditions for various angular radii of the supercontinent. }
\begin{tabular}{crrrrrrrrrrrr} 
 \hline 
 \hline 
$\colcont$ (deg) & $\left( 0,0 \right)$ & $\left( 1,0 \right)$ & $\left( 1,1 \right)$ & $\left( 2,0 \right)$ & $\left( 2,1 \right)$ & $\left( 2,2 \right)$  \\ 
 \hline 
  \multicolumn{7}{c}{\textit{Neumann boundary conditions (SCHNs)}} \\
0 &     0.000000 &     2.000000 &     2.000000 &     6.000000 &     6.000000 &     6.000000 \\ 
10 &     0.000000 &     2.043810 &     1.954799 &     6.204319 &     5.787950 &     5.996589 \\ 
20 &     0.000000 &     2.166257 &     1.838982 &     6.717744 &     5.460750 &     5.949235 \\ 
30 &     0.000000 &     2.360311 &     1.711429 &     7.467118 &     5.441569 &     5.785407 \\ 
40 &     0.000000 &     2.629208 &     1.619197 &     8.452903 &     5.746033 &     5.518966 \\ 
50 &     0.000000 &     2.985569 &     1.577808 &     9.717446 &     6.317753 &     5.278194 \\ 
60 &     0.000000 &     3.452250 &     1.589586 &    11.339534 &     7.153126 &     5.166990 \\ 
70 &     0.000000 &     4.065883 &     1.656959 &    13.444194 &     8.299630 &     5.229035 \\ 
80 &     0.000000 &     4.884195 &     1.787952 &    16.226814 &     9.857388 &     5.491887 \\ 
90 &     0.000000 &     6.000000 &     2.000000 &    20.000000 &    12.000000 &     6.000000 \\ 
100 &     0.000000 &     7.568523 &     2.326061 &    25.285246 &    15.024623 &     6.839299 \\ 
110 &     0.000000 &     9.864090 &     2.827415 &    33.003048 &    19.460404 &     8.174357 \\ 
120 &     0.000000 &    13.408133 &     3.622974 &    44.902072 &    26.315661 &    10.329330 \\ 
130 &     0.000000 &    19.292626 &     4.963802 &    64.643464 &    37.703724 &    13.992842 \\ 
140 &     0.000000 &    30.132199 &     7.452826 &   100.992648 &    58.686045 &    20.822406 \\ 
150 &     0.000000 &    53.558065 &    12.851483 &   179.532083 &   104.036347 &    35.663952 \\ 
160 &     0.000000 &   120.497214 &    28.300199 &   403.938705 &   233.629059 &    78.165464 \\ 
170 &     0.000000 &   481.981173 &   111.760030 &  1615.747119 &   933.460872 &   307.821577 \\ 
\hline 
\multicolumn{7}{c}{\textit{Dirichlet boundary conditions (SCHDs)}} \\
0 &     0.000000 &     2.000000 &     2.000000 &     6.000000 &     6.000000 &     6.000000 \\ 
10 &     0.241710 &     2.942918 &     2.043810 &     7.928496 &     6.204319 &     6.003307 \\ 
20 &     0.349854 &     3.454660 &     2.166257 &     9.122983 &     6.717744 &     6.045851 \\ 
30 &     0.466027 &     4.027851 &     2.360311 &    10.493540 &     7.467118 &     6.190333 \\ 
40 &     0.600603 &     4.707711 &     2.629208 &    12.138321 &     8.452903 &     6.485855 \\ 
50 &     0.762626 &     5.538331 &     2.985569 &    14.160734 &     9.717446 &     6.968283 \\ 
60 &     0.963323 &     6.577054 &     3.452250 &    16.699223 &    11.339534 &     7.678163 \\ 
70 &     1.218462 &     7.905884 &     4.065883 &    19.953910 &    13.444194 &     8.676820 \\ 
80 &     1.551642 &     9.648467 &     4.884195 &    24.227788 &    16.226814 &    10.063195 \\ 
90 &     2.000000 &    12.000000 &     6.000000 &    30.000000 &    20.000000 &    12.000000 \\ 
100 &     2.625081 &    15.284420 &     7.568523 &    38.066269 &    25.285246 &    14.762935 \\ 
110 &     3.535228 &    20.072333 &     9.864090 &    49.828657 &    33.003048 &    18.842401 \\ 
120 &     4.936042 &    27.446847 &    13.408133 &    67.948839 &    44.902072 &    25.173291 \\ 
130 &     7.257803 &    39.674941 &    19.292626 &    97.998058 &    64.643464 &    35.715869 \\ 
140 &    11.530519 &    62.183561 &    30.132199 &   153.313622 &   100.992648 &    55.166033 \\ 
150 &    20.760187 &   110.811047 &    53.558065 &   272.820192 &   179.532083 &    97.231717 \\ 
160 &    47.128920 &   249.744161 &   120.497214 &   614.264854 &   403.938705 &   217.469478 \\ 
170 &   189.517352 &   999.979043 &   481.981173 &  2458.061916 &  1615.747119 &   866.830706 \\  
\hline
 \end{tabular}
 \tablefoot{The values of $\eigenvali{\llat}$ are were computed from \eq{sch_eigenv} for the doublets $\left(  \nk , \abs{\mm} \right) $ such that $\nk=0,1,2$ and $\abs{\mm} \leq \nk$ using the values of real degrees given by Table~\ref{tab:sch_degrees}. }
 \end{table*}

The SPHs and the SCHs are found by solving -- for specific boundary conditions discussed further -- the eigenvalue-eigenfunction problem described by the Helmholtz equation \citep[e.g.][]{Haines1985},
\begin{equation}
\label{helmholtz}
\lap \ffunc = - \eigenval \ffunc, 
\end{equation}
where $\lap$ designates the Laplace operator, defined for any function $\ffunc \left( \col, \lon \right)$ as
\begin{equation}
\lap \ffunc \define \frac{1}{\sin \col} \dd{}{\col} \left( \sin \col \dd{\ffunc}{\col} \right) + \frac{1}{\sin^2 \col} \ddd{\ffunc}{\lon}{\lon},
\end{equation}
the notation $\dd{}{\xx}$ or $\ddd{}{\xx}{\xx}$ referring to the first or second partial derivatives with respect to the $\xx$ coordinate. In both case, the variables $\col$ and $\lon$ can be separated, which allows the problem to be solved as two decoupled eigenvalue problems. 

Since the functions $\ffunc \left( \col, \lon  \right)$ in \eq{helmholtz} are defined for all $\lon$, the boundary conditions on $\lon$ are those of continuity,
\begin{align}
\ffunc \left( \col, \lon \right) & = \ffunc \left( \col , \lon + 2 \pi \right) , \\
\dd{\ffunc}{\lon} \left( \col , \lon \right) & = \dd{\ffunc}{\lon} \left( \col , \lon + 2 \pi \right),
\end{align}
which restricts the values of $\mm$ to be integral ($\mm \in \Zset$). These conditions hold both for the ordinary SPHs and for the SCHs. Analogously, the two sets of basis functions share the same regularity condition on $\col$ at the pole ($\col =0$), 
\begin{align}
 \dd{\ffunc}{\col} \left( 0 , \lon \right) = 0 && \mm=0, \\
 \ffunc \left( 0 , \lon \right) = 0 && \mm \neq 0. 
\end{align}
However they differ as regards the second boundary condition. The SPHs are obtained by solving the Helmholtz equation given by \eq{helmholtz} on the entire sphere, $\sphere{:}\left( \col , \lon \right) \in \left[ 0, \pi \right] \times \left[ 0 , 2 \pi \right[$, with regularity boundary conditions at $\col=\pi$ similar to those applied at $\col=0$,
\begin{align}
\dd{\ffunc}{\col} \left( \pi, \lon \right) = 0 && \mm=0 , \\
\ffunc \left( \pi , \lon \right) = 0 && \mm \neq 0. 
\end{align}
They are associated with integral degrees $\llat \in \Nset$ such that $\llat \geq \abs{\mm}$.

Following \cite{Varshalovich1988}, we use, for the complex SPHs, the convention given by \citep[][Sect.~5.2, Eq.~(1)]{Varshalovich1988}
\begin{equation}
\label{sphc}
\Ylm{\llat}{\mm} \left( \col , \lon \right) = \sqrt{\frac{\left( 2 \llat + 1 \right) \left( \llat - \mm \right)!}{4 \pi \left( \llat + \mm \right)!}} \LegF{\llat}{\mm} \left( \cos \col \right) \expo{\inumber \mm \lon},
\end{equation}
or, equivalently,
\begin{equation}
\Ylm{\llat}{\mm} \left( \col , \lon \right) \define \left( -1 \right)^{\frac{\abs{\mm} - \mm}{2}}  \sqrt{\frac{\left( 2 \llat + 1 \right) \left( \llat - \abs{\mm} \right) !  }{4 \pi \left( \llat + \abs{\mm} \right)! }} \LegF{\llat}{\abs{\mm}} \left( \cos \col \right) \expo{\inumber \mm \lon},
\end{equation}
where the the $\Plm$ are the standard ALFs defined in \eq{Plmstd}. The eigenvalues associated with the SPHs only depend on the degree $\llat$, and are expressed as
\begin{equation}
\eigenvali{\llat} = \llat \left( \llat + 1 \right). 
\end{equation}
In the used convention (\eq{sphc}), the SPHs form a set of orthonormal basis functions through the scalar product defined, for any complex functions $\ffunc$ and $\gfunc$, as
\begin{align}
\label{sprodYlm}
\scal{\ffunc}{\gfunc} &  \define \integ{\conj{\ffunc} \gfunc }{\surface}{\sphere}{}, \\
 & = \integ{\integ{ \conj{\ffunc \left( \col , \lon \right)} \gfunc \left( \col , \lon \right) \sin \col \,}{\col}{\col=0}{\pi} }{\lon}{\lon=0}{2 \pi}. \nonumber
\end{align} 
the notation $\infvar{\surface}$ referring to an infinitesimal surface element of the unit sphere, and $\conj{\ffunc}$ to the conjugate of $\ffunc$. This scalar product allows the $\ltwonorm$-norm of the space of functions of $\sphere$ to be defined,
\begin{equation}
\label{ltwonorm}
\normtwo{f} \define \sqrt{\scal{\ffunc}{\ffunc}} .
\end{equation}
Hence, for any pair of SPHs $\Ylm{\llat}{\mm}$ and $\Ylm{\pp}{\qq}$, 
\begin{equation}
\label{sphscal}
\scal{\Ylm{\llat}{\mm}}{\Ylm{\pp}{\qq}} = \kron{\llat}{p} \kron{\mm}{\qq}.
\end{equation}

The SCHs are obtained by solving the Helmholtz equation given by \eq{helmholtz} on the domain filled by the spherical cap, $\sphcap:\left( \col , \lon \right) \in \left[0 , \colbd \right] \times \left[0 , 2 \pi \right[$, together with the boundary conditions specified at $\sphcapbd$, namely $\col = \colbd$. In the present model, the applied boundary conditions are either Neumann (fixed value for $\dd{\ffunc}{\col} $ at $\col= \colbd$) or Dirichlet (fixed value for the function $\ffunc$ itself) conditions, which are respectively formulated as
\begin{align}
 \label{neucond}
 \dd{\ffunc}{\col} \left( \colbd , \lon \right) = 0  && \mbox{(Neumann condition)},  \\
 \label{dircond}
 \ffunc \left( \colbd, \lon \right) = 0  && \mbox{(Dirichlet condition)}. 
\end{align}
Each condition leads to a set of orthogonal basis functions. These sets are referred to as the `SCHNs' for \eq{neucond} and the `SCHDs' for \eq{dircond}.

As the Helmholtz equation is the same for both the SCHs and the SPHs, solving the eigenfunction-eigenvalue problem defined by \eqs{helmholtz}{neucond} (or \eq{dircond}) amounts to finding the set of ALFs (see \eq{alfscha}) that satisfy the specified boundary conditions for a given $\mm \in \Zset$. As a consequence, the eigenfunction-eigenvalue problem becomes a root-finding problem where one computes the series of real degrees $\llat $ such that, alternately,
\begin{align}
\label{neucond_plm}
\DD{\LegF{\llat}{\mm} }{\col} \left( \cos \colbd \right) = 0, && \mbox{(Neumann condition)} , \\
\label{dircond_plm}
\LegF{\llat}{\mm} \left( \cos \colbd \right) = 0 && \mbox{(Dirichlet condition)}.
\end{align} 
This can be achieved numerically using a standard Newton-Raphson algorithm \citep[e.g.][Sect.~9.4]{press2007numerical} since the ALFs defined by \eq{alfscha} and their derivatives are regular functions of $\llat$. 

For each condition (\eq{neucond} or \eq{dircond}), we end up with the degrees of the distorted ALFs, $\llati{\nk} \left( \mm \right)$, which are subscripted with indices $\nk \in \Nset$ such that $\nk \geq \abs{\mm}$. These indices\footnote{Other conventions exist for the indices $\nk$. For instance, \cite{Haines1985} introduces a convention where the parity of $\kk - \abs{\mm}$ allows the two sets of basis functions (with Neumann or Dirichlet conditions) to be distinguished. In this convention, the roots of \eq{neucond_plm} (Neumann condition) are designated by even $\nk - \abs{\mm}$, while those of \eq{dircond_plm} (Dirichlet condition) are designated by odd $\nk - \abs{\mm}$.} correspond to the integral degrees of the SPHs towards which the SCHs converge for the $\ltwonorm$-norm defined by \eq{ltwonorm} in the global ocean limit: $\llati{\nk} \rightarrow \nk$ as $\colbd \rightarrow 180^\degree$ independently of the applied boundary condition. Similarly as integral degrees, the real degrees $\llati{\nk} $ are ranked in ascending order. For any $\nk$ and $\jj$ such that $\nk < \jj$, the inequality $\llati{\nk} < \llati{\jj}$ is verified. The values of the real degrees are given by Table~\ref{tab:sch_degrees} for $0 \leq \abs{\mm} \leq 2$ and various values of angular radius of the continent, $\colcont = 180 - \colbd$ (in degrees). They are plotted on \fig{fig:degrees_size}. 

We note that $\llati{\nk} \geq \abs{\mm}$ when Dirichlet conditions are applied, similarly as in the case of SPHs \citep[e.g.][]{Thebault2006}. However, when the Neumann conditions are applied, this is only valid for $\colbd \leq 90^\degree$ since there exists degrees $\llati{\abs{\mm}} < \abs{\mm}$ as long as $\colbd > 90^\degree$. Asymptotic scaling laws may be used to estimate the values of the degrees $\llati{\nk} \gg \abs{\mm}$ as functions of the indices $\nk$ and the angle $\colbd$ \citep[see e.g.][Eq.~(40)]{Thebault2006}. Particularly, these laws show that $\llati{\nk} \scale \colbd^{-1}$ for $\colbd \ll 1$ consistently with what may be observed in \fig{fig:degrees_size}. This results from the fact that the degree $\llat$ of an harmonic scales as the inverse of the corresponding wavelength, which is determined by the size of the spherical cap. 

The resulting SCHs are defined on the unit sphere, for $\abs{\mm} = 0, \ldots , + \infty$ and $\llat = \llati{\abs{\mm}}, \llati{\abs{\mm}+1}, \ldots$, as 
\begin{equation}
\label{schc}
\Hlm{\llat}{\mm} \left( \col , \lon \right) \define 
\left\{
\begin{array}{ll}
\alm{\llat}{\mm} \LegF{\llat}{\abs{\mm}} \left( \cos \col \right) \expo{\inumber \mm \lon} & \mbox{if} \ \left( \col , \phi \right) \in \sphcap, \\
0 & \mbox{if} \ \left( \col , \phi \right) \in \sphere \setminus \sphcap
\end{array}
\right.
\end{equation}
where the normalisation coefficients $\alm{\llat}{\mm} \in \Rset$ are to be specified. The corresponding eigenvalues $\eigenvali{\llat}$ are expressed as 
\begin{equation}
\label{sch_eigenv}
\eigenvali{\llat} \left( \mm \right) = \llat \left( \mm \right) \left[ \llat \left( \mm \right) + 1 \right],
\end{equation}
similarly as in the case of SPHs, though they now also depend on the order $\mm$ through $\llat$. The eigenvalues associated with the SCHNs and SCHDs are given in Table~\ref{tab:sch_eigenv} for $0 \leq \abs{\mm} \leq \nk \leq 2$ and various values of the continental angular radius ($\colcont$). They are computed straightforwardly from the values of the real degrees given in Table~\ref{tab:sch_degrees} using the expression given by \eq{sch_eigenv}.

The SCHs obtained for a given boundary condition are orthogonal basis functions of the space of regular functions on $\sphcap$ through the scalar product introduced in \eq{sprodYlm}. Since the harmonics are set to zero for $\col>\colbd$, this scalar product reads
\begin{equation}
\label{scal_sch}
\scal{\ffunc}{\gfunc}  = \integ{\integ{ \conj{\ffunc \left( \col , \lon \right)} \gfunc \left( \col , \lon \right) \sin \col \,}{\col}{\col=0}{\colbd} \, }{\lon}{\lon=0}{2 \pi}.
\end{equation}
For any pair of SCHs belonging to the same set, $\Hlm{\llat}{\mm}$ and $\Hlm{\pp}{\qq}$, the property given by \eq{sphscal} for the SPHs,
\begin{equation}
\label{schortho}
\scal{\Hlm{\llat}{\mm}}{\Hlm{\pp}{\qq}} = \kron{\llat}{\pp} \kron{\mm}{\qq},
\end{equation} 
is enforced by defining the normalisation coefficients $\alm{\llat}{\mm} \in \Rset$ as
\begin{equation}
\label{normcoeff}
\alm{\llat}{\mm} = \left( -1 \right)^{\frac{\abs{\mm} - \mm}{2}}  \left[ 2 \pi \integ{\left[ \LegF{\llat}{\abs{\mm}} \left( \cos \col \right) \right]^2 \sin \col \,}{\col}{0}{\colbd} \right]^{-1/2}. 
\end{equation}
However, the functions in one set are not orthogonal to those in the other since \citep[e.g.][]{Haines1985} 
\begin{equation}
\scal{\Hlm{\llat}{\mm}}{\Hlm{\pp}{\qq}} = \kron{\mm}{\qq} \frac{2 \pi \sin \colbd \LegF{\llat}{\mm} \left( \cos \colbd \right) }{\left( \pp - \llat \right) \left( \pp + \llat + 1 \right)} \DD{\LegF{\pp}{\qq}}{\col} \left( \cos \colbd \right) ,
\end{equation}
for $\Hlm{\llat}{\mm}$ and $\Hlm{\pp}{\qq}$ taken in the sets of SCHNs and SCHDs, respectively. In the general case, $\LegF{\llat}{\mm} \left( \cos \colbd \right) \neq 0 $ and $\DD{\LegF{\pp}{\qq}}{\col} \left( \cos \colbd \right) \neq 0$, which implies $\scal{\Hlm{\llat}{\mm}}{\Hlm{\pp}{\mm}} \neq 0$. This highlights the fact that the SCHDs and the SCHNs are two distinct complete sets of basis functions for the space of functions defined on $\sphcap$.

In the case of real eigenfunctions defined as 
\begin{equation}
\Hlm{\llat}{\mm} \left( \col , \lon \right) \define \left\{
\begin{array}{ll}
\alm{\llat}{\mm} \LegF{\llat}{\abs{\mm}} \left( \cos \col \right) \cos \left( \mm \lon + \lonm \right) & \mbox{if} \ \left( \col , \phi \right) \in \sphcap, \\[0.1cm]
0 & \mbox{if} \ \left( \col , \phi \right) \in \sphere \setminus \sphcap,
\end{array} 
\right.
\end{equation}
with $\lonm =0$ if $\mm \geq 0$ and $\lonm = \pi/2$ if $\mm<0$, the orthogonality property given by \eq{schortho} still holds and the normalisation coefficients defined in \eq{normcoeff} become
\begin{equation}
\alm{\llat}{\mm} = \left( -1 \right)^{\frac{\abs{\mm} + \mm}{2}} \left[ \left( 1 + \kron{\mm}{0} \right) \pi \integ{\left[ \LegF{\llat}{\abs{\mm}} \left( \cos \col \right) \right]^2 \sin \col \,}{\col}{0}{\colbd} \right]^{-1/2}.
\end{equation}
We note that the factors $\left( - 1 \right)^k$ with $k = \left( \abs{\mm} - \mm \right)/2 $ or $k = \left( \abs{\mm} + \mm \right)/2$ in the expressions of the normalisation coefficients are introduced to make the SCHs consistent with the SPHs in the used convention (\eq{sphc}).

As $\colbd$ increases, the ALFs smoothly evolve from the Bessel functions ($\colbd \rightarrow 0$) to the standard ALFs of integer parameters ($\colbd \rightarrow \pi$). Similarly, the SCHs evolve from the cylindrical harmonics, where there is no curvature effect (small-sea configuration), to the SPHs given by \eq{sphc} (global-ocean configuration). Nevertheless, we emphasise that both the eigenvalues and the eigenfunctions depend on the specified boundary condition through the degrees $\llat$, different conditions leading to different sets of eigenfunctions. Besides, the limit $\colbd \rightarrow \pi$ is not strictly equivalent to the global-ocean configuration since boundary conditions are still applied at $\sphcapbd$, instead of the periodic conditions of the SPHs. In other words, the presence of a small island still alters the eigenmodes of the oceanic tide locally although this effect is negligible at a global scale. The impact of the island on the horizontal structure of tidal flows is quantified by the differences of degrees, $\abs{\llati{\nk} - \nk}$, and overlap coefficients, $\abs{1-\scal{\Hlm{\llati{\nk}}{\mm}}{\Ylm{\nk}{\mm}}} $, both of them tending to zero as the size of the island decreases.

\section{Euler rotation matrix}
\label{app:euler_rotation}

\begin{figure}[t]
   \centering
   \includegraphics[width=0.48\textwidth,trim = 0.cm 0.cm 11.7cm 6.2cm,clip]{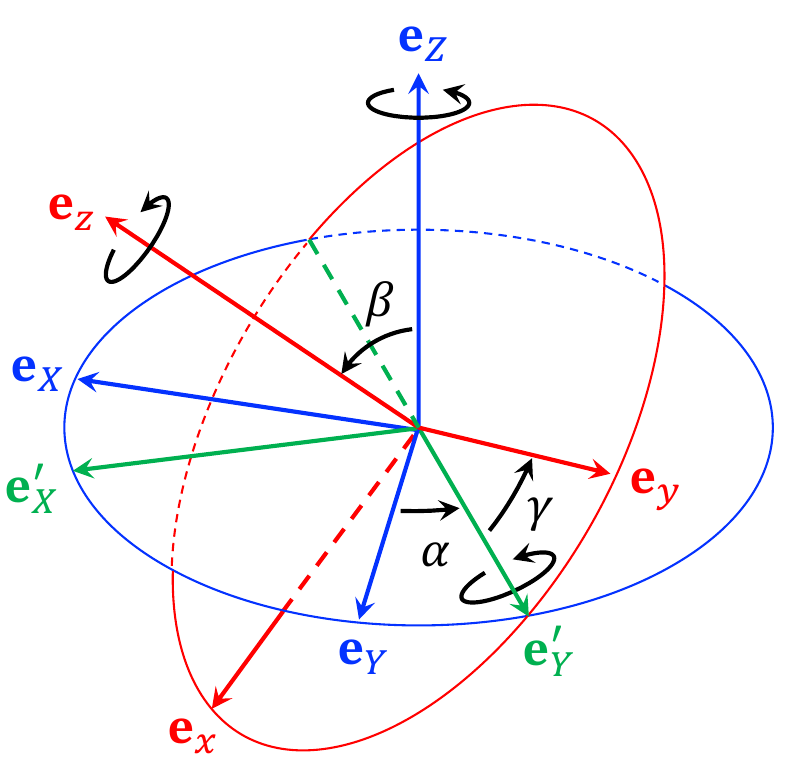}
      \caption{Euler angles corresponding to the standard Euler rotation matrix defined by \eq{eulermat}.}
       \label{fig:euler_angles}%
\end{figure}

We consider the frame of reference $\rframe{}{O}{\eX}{\eY}{\eZ} $ having the planet's centre of gravity as origin and associated with the Cartesian coordinates $\left( \XX, \YY, \ZZ \right)$. As shown by \fig{fig:euler_angles}, any rotation $\rframe{}{O}{\eX}{\eY}{\eZ} \rightarrow \rframe{\irotation}{O}{\ex}{\ey}{\ez}$ of this frame of reference may be performed by three rotations about the coordinate axes \citep[e.g.][Sect.~1.4]{Varshalovich1988}: (i) a rotation about the $\ZZ$-axis through an angle $\angalp$ such that $0 \leq \angalp < 2 \pi$, (ii) a rotation about the new $\YY^\prime$-axis through an angle $\angbet$ such that $0 \leq \angbet \leq \pi$, and (iii) a rotation about the new axis $\zz $ through an angle $\anggam$ such that $0 \leq \anggam \leq 2 \pi$. The frame of reference $\rframe{\irotation}{O}{\ex}{\ey}{\ez}$ is associated with the Cartesian coordinates $\left( \xx, \yy , \yy \right)$. The three rotation angles, $\angalp$, $\angbet$, and $\anggam$, are the so-called Euler angles. 

Introducing the notations $\cosi{\angrot} = \cos \angrot$ and $\sini{\angrot} = \sin \angrot$ for $\angrot = \angalp, \angbet, $ or $\anggam$, the standard Euler rotation matrix is defined in terms of the Euler angles as \citep[e.g.][Eq.~(54) p.~30]{Varshalovich1988}
\begin{equation}
\eulermat_{ \angalp , \angbet , \anggam} \define 
\begin{bmatrix}
\cosi{\angalp} \cosi{\angbet} \cosi{\anggam} - \sini{\angalp} \sini{\anggam} & 
- \cosi{\angalp} \cosi{\angbet} \sini{\anggam} - \sini{\angalp} \cosi{\anggam} & 
\cosi{\angalp} \sini{\angbet} \\
\sini{\angalp} \cosi{\angbet} \cosi{\anggam} + \cosi{\angalp} \sini{\anggam} & 
- \sini{\angalp} \cosi{\angbet} \sini{\anggam} + \cosi{\angalp} \cosi{\anggam} & 
\sini{\angalp} \sini{\angbet} \\ 
- \sini{\angbet} \cosi{\anggam} & 
\sini{\angbet} \sini{\anggam} & \cosi{\angbet} 
\end{bmatrix},
\label{eulermat}
\end{equation}
which is formed by the product of three rotation matrices,
\begin{align}
\eulermat_{ \angalp , \angbet , \anggam} = &  \rotmati{3}{\angalp} \rotmati{2}{\angbet} \rotmati{3}{\anggam} \\
 = &
\begin{bmatrix}
\cosi{\angalp} &  - \sini{\angalp}  & 0 \\
\sini{\angalp} & \cosi{\angalp} & 0 \\
0 & 0 & 1
 \end{bmatrix}
\begin{bmatrix}
\cosi{\angbet} & 0 &  \sini{\angbet}  \\
0 & 1 & 0 \\
- \sini{\angbet} & 0 & \cosi{\angbet} 
\end{bmatrix}
\begin{bmatrix}
\cosi{\anggam} & - \sini{\anggam}  & 0 \\
\sini{\anggam} & \cosi{\anggam} & 0 \\
0 & 0 & 1
 \end{bmatrix}.
\nonumber
\end{align}
The inverse rotation matrix $\eulermat^{-1}$ is given by 
\begin{equation}
\eulermat^{-1}_{ \angalp, \angbet, \anggam} = \transp{\eulermat}_{\angalp, \angbet, \anggam}= \eulermat_{- \anggam, - \angbet , - \angalp}. 
\end{equation}
The Euler rotation matrix and its inverse correspond to the change of basis matrices between the initial and rotated vectorial bases, $\left( \eX, \eY , \eZ \right) \rightarrow \left( \ex, \ey , \ez \right)$, which is expressed as 
\begin{align}
\label{change_basis_rotation}
&
\begin{bmatrix}
\ex \\
\ey \\
\ez 
\end{bmatrix}
= \eulermat_{\angalp,\angbet,\anggam}^{-1} 
\begin{bmatrix}
\eX \\
\eY \\
\eZ
\end{bmatrix},
& 
\begin{bmatrix}
\eX \\
\eY \\
\eZ
\end{bmatrix}
= \eulermat_{\angalp,\angbet,\anggam}
\begin{bmatrix}
\ex \\
\ey \\
\ez 
\end{bmatrix}. 
\end{align}
Besides the rotation $\Uvect \rightarrow \Uvectrot $ of any vector $\Uvect $ through the Euler angles $\angalp, \angbet,$ and $\anggam$ in the Cartesian coordinate system $\coordsini \left\{ \XX , \YY , \ZZ \right\}$ is expressed as $\Uvectrot = \eulermat_{\angalp,\angbet,\anggam} \Uvect $.

In spherical coordinates, the transformation relations are deduced from \eq{change_basis_rotation} by making use of trigonometric identities. We denote by $\left( \rrpla, \colpla , \lonpla \right)$ the spherical coordinates associated with the initial frame of reference, $\framestd$, where $\rrpla$ designates the radial coordinate, $\colpla$ the colatitude, and $\lonpla$ the longitude. The analogous coordinates in the rotated frame of reference, $\framerotation$, are denoted by $\left( \rr, \col , \lon \right)$, with $\rr = \rrpla$. In the forward transformation, $\left( \rrpla, \colpla , \lonpla \right) \rightarrow \left( \rr, \col , \lon \right) $, the relations between angles $\left( \colpla,\lonpla \right)$ and $\left( \col,\lon \right)$ are given by \citep[][Sect.~1.4.1, Eqs~(2-3)]{Varshalovich1988}
\begin{align}
\cos \col & = \cos \colpla \cos \angbet + \sin \colpla \sin \angbet \cos \left( \lonpla - \angalp \right) , \\
\cot \left( \lon + \anggam \right) & = \cot \left( \lonpla - \angalp \right) \cos \angbet - \frac{\cot \colpla \sin \angbet}{ \sin \left( \lonpla - \angalp \right)},
\end{align}
while the inverse relations are 
\begin{align}
\cos \colpla & = \cos \col \cos \angbet - \sin \col \sin \angbet \cos \left( \lon + \anggam \right) , \\ 
\cot \left( \lonpla - \angalp \right) & = \cot \left( \lon + \anggam \right) \cos \angbet + \frac{\cot \col \sin \angbet}{\sin \left( \lon + \anggam \right)}. 
\end{align}

As an example, we consider the rotation generating the coordinate system of the ocean basin from the geocentric coordinate system rotating with the planet. From the definition of the ocean centre in the usual geocentric frame of reference, we obtain 
\begin{equation}
\ez = \cos \lonoc \sin \coloc \eX + \sin \lonoc \sin \coloc \eY + \cos \coloc \eZ.
\label{ez1}
\end{equation}
Besides, by considering the same vector as the image of $\eZ$ through a rotation of Euler angles $\angalp$, $\angbet$, and $\anggam$, we can write $\ez = \eulermat_{\angalp, \angbet, \anggam} \transp{\left( 0 , 0 , 1 \right)}$, where we made use of the standard Euler rotation matrix defined in \eq{eulermat}. It follows
\begin{equation}
\ez = \cos \angalp \sin \angbet \eX + \sin \angalp \sin \angbet \eY + \cos \angbet \eZ. 
\label{ez2}
\end{equation}
By comparing \eqs{ez1}{ez2}, we identify $\angbet = \coloc$ and $\angalp = \lonoc$. Considering the inverse transformation, $\eZ = \transp{\eulermat_{\angalp, \angbet, \anggam}} \transp{\left( 0 , 0 , 1 \right)}$, we get
\begin{equation}
\eZ = - \sin \angbet \cos \anggam \ex + \sin \angbet \sin \anggam \ey + \cos \angbet \ez,
\end{equation}
which shows that $\eZ$ does not depend on the rotation through the angle $\angalp$ in the oceanic frame of reference. 

We note that the angle $\anggam$ describes a rotation of coordinates around the axis defined by $\ez$. Since the ocean is circular, the geometry remains unchanged through this rotation. The angle $\anggam$ can therefore be set to $\anggam = 0$, which allows the position of the North pole in the ocean's coordinate system to be simplified to $\eZ = - \sin \angbet \ex + \cos \angbet \ez$. The coordinates of the North pole in the coordinate system defined from the centre of the ocean basin are thus given by $\left( \colnp , \lonnp \right) = \left( \angbet , \pi \right)$. At the end, the ocean's coordinate system is simply obtained from the usual geocentric coordinate system through a rotation of Euler angles $\left( \angalp , \angbet , \anggam \right)= \left( \lonoc , \coloc , 0 \right)$. We shall emphasise that the gyroscopic coefficients of the tidal theory only depend on the colatitude of the supercontinent's centre since $\angbet$ is the only angle necessary to define the position of the spin axis in the coordinate system of the ocean basin. Moreover, the rotation of angle $\angalp$ can be ignored owing to the periodicity of the tidal response across the longitudinal coordinate, $\lonpla$, which implies that the longitude of the supercontinent's centre can be set to $\lonscont=0$ without any loss of generality.

\section{Rotation of spherical harmonics}
\label{app:rotation_sph}

We consider a rotation $\rframe{}{O}{\eX}{\eY}{\eZ} \rightarrow \rframe{\irotation}{O}{\ex}{\ey}{\ez}$ described by the Euler angles introduced in \append{app:euler_rotation}, which are denoted by $\angalp, \angbet$, and $\anggam$ (see \fig{fig:euler_angles}). The initial frame of reference, $\framestd$, is associated with the spherical coordinate system $\left( \rrpla, \colpla, \lonpla  \right) $, while the rotated frame of reference, $\framerotation$, is associated with the coordinate system $\left( \rr , \col , \lon \right)$. In the convention defined by \eq{sphc}, the complex SPHs of the rotated coordinate system are expressed as functions of those defined in the initial coordinate system \citep[][Sect.~4.1]{Varshalovich1988},
\begin{equation}
\Ylm{\llat}{\mm} \left( \col , \lon \right) = \sum_{\qq=-\llat}^\llat \wignerD{\qq}{\mm}{\llat} \left( \angalp, \angbet, \anggam \right) \Ylm{\llat}{\qq} \left(  \colpla, \lonpla \right),
\label{Ylm_Ylq}
\end{equation} 
and, conversely, the SPHs of the initial coordinate system are expressed as
\begin{equation}
\Ylm{\llat}{\mm} \left( \colpla , \lonpla \right) = \sum_{\qq=-\llat}^\llat \conj{\wignerD{\mm}{\qq}{\llat}} \left( \angalp, \angbet, \anggam \right) \Ylm{\llat}{\qq} \left( \col, \lon \right).
\label{Ylq_Ylm}
\end{equation}

\begin{figure*}[htb]
   \centering
   \includegraphics[width=0.3\textwidth,trim = 3.cm 1.5cm 19.5cm 3cm,clip]{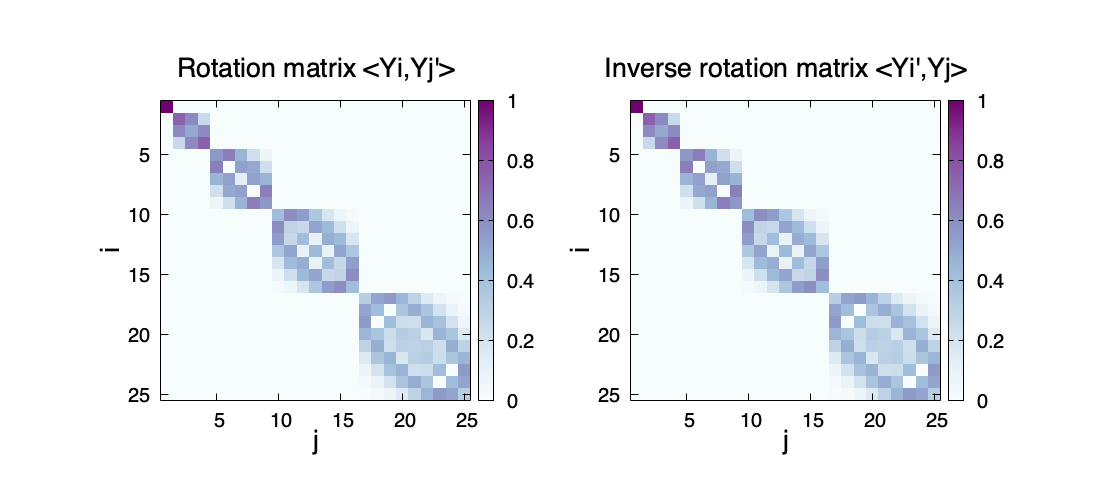} 
   \includegraphics[width=0.3\textwidth,trim = 3.cm 1.5cm 19.5cm 3cm,clip]{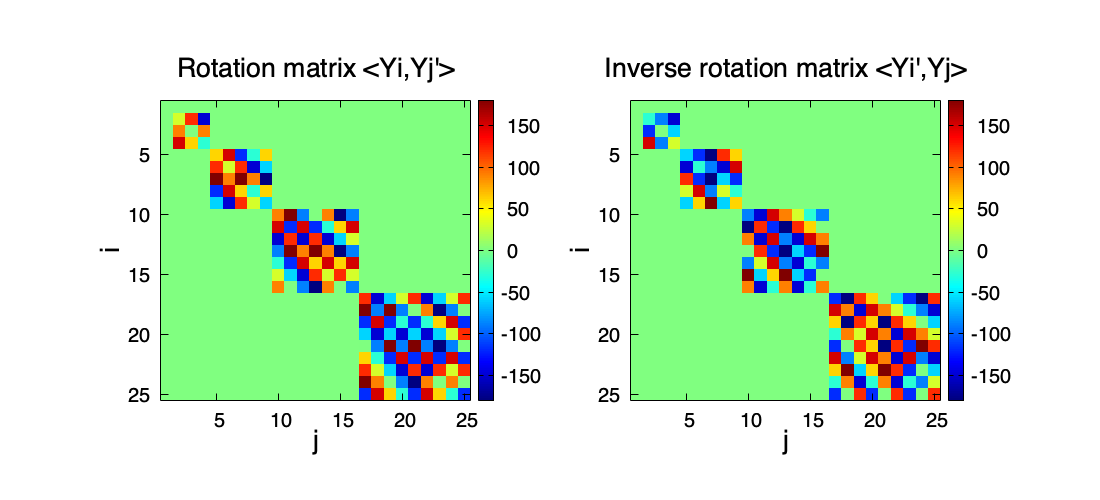} 
      \caption{Rotation matrix $\wignermat$ of \eq{wignermat} for Euler angles $\left( \angalp, \angbet, \anggam \right) = \left(120^\degree, 60^\degree, 270^\degree \right) $. {\it Left:} Modulus of the matrix coefficients. {\it Right:} Argument of the matrix coefficients (degrees). The truncation degree of the set of SPHs is set to $\lmax = 4$. The SPHs are sorted in ascending order of degrees ($\llat=0,1,2,3,4$), with orders ($-\llat \leq \mm \leq \llat$) grouped together.}
       \label{fig:rmat}%
\end{figure*}

The notation $\wignerD{\qq}{\mm}{\llat}$ refers to the Wigner D-functions \citep[][Sect.~4.3, Eq.~(1)]{Varshalovich1988}, 
\begin{equation}
\wignerD{\qq}{\mm}{\llat} \left( \angalp, \angbet, \anggam \right)  \define \expo{- \inumber \qq \angalp} \wignerd{\qq}{\mm}{\llat} \left( \angbet \right) \expo{- \inumber \mm \anggam},
\end{equation}
where $\wignerd{\qq}{\mm}{\llat}$ is a real function expressed as \citep[][Sect.~4.3.1, Eq.~(2)]{Varshalovich1988}
\begin{align}
\wignerd{\qq}{\mm}{\llat} \left( \angbet \right) = & \left( -1 \right)^{\llat - \mm} \sqrt{\left( \llat + \qq \right)! \left( \llat - \qq \right)! \left( \llat + \mm \right) ! \left( \llat - \mm \right)! } \\
& \times \sum_{\kk} \left( -1 \right)^\kk \frac{ \left( \cos \frac{\angbet}{2} \right)^{\qq + \mm + 2 \kk} \left( \sin \frac{\angbet}{2} \right)^{2 \llat - \qq - \mm - 2 \kk} }{\kk ! \left( \llat - \qq - \kk \right) ! \left( \llat - \mm - \kk \right) ! \left( \qq + \mm + \kk \right) ! }. \nonumber
\end{align}
In the above equation, $\kk$ runs over all integer values for which the factorial arguments are non-negative, namely $\max \left( 0 , - \qq - \mm \right) \leq \kk \leq \min \left( \llat - \qq , \llat - \mm \right)$. 

The Wigner D-functions introduced in \eqs{Ylm_Ylq}{Ylq_Ylm} are the matrix elements of the rotation operator, 
\begin{equation}
\label{wignermat}
\wignermat \define
\begin{bmatrix}
\wignerDmati{0}{0}{0} & 0 & \ldots & \ldots & \ldots & 0  \\
0 & \wignerDmati{\qq}{\mm}{1} &  \ddots & & & \vdots   \\
 \vdots  &  \ddots & \ddots & \ddots & &\vdots  \\
 \vdots &  & \ddots & \wignerDmati{\qq}{\mm}{\llat} & \ddots & \vdots \\
 \vdots &  & & \ddots & \ddots & 0  \\
 0 & \ldots & \ldots & \ldots & 0 & \wignerDmati{\qq}{\mm}{\lmax} 
\end{bmatrix},
\end{equation}
where $\lmax$ is the chosen truncation degree, and $\wignerDmati{\qq}{\mm}{\llat} $ is the matrix defined as 
\begin{equation}
\wignerDmati{\qq}{\mm}{\llat} \define 
\begin{bmatrix}
\wignerD{-\llat}{-\llat}{\llat} & \ldots & \wignerD{-\llat}{\mm}{\llat} & \ldots &  \wignerD{-\llat}{\llat}{\llat} \\
\vdots & & \vdots & & \vdots \\
\wignerD{\qq}{-\llat}{\llat} & \ldots & \wignerD{\qq}{\mm}{\llat} & \ldots & \wignerD{\qq}{\llat}{\llat} \\
\vdots & & \vdots & & \vdots \\
\wignerD{\llat}{-\llat}{\llat} & \ldots & \wignerD{\llat}{\mm}{\llat} & \ldots & \wignerD{\llat}{\llat}{\llat} 
\end{bmatrix} .
\end{equation}
Thus, the set of SPHs associated with the rotated coordinate system, denoted by $\vectSPHrot \define \transp{\left[ \Ylmi{1}, \ldots ,\Ylmi{\nsph}  \right]}$, is expressed as a function of the initial set, $\vectSPHini \define \transp{\left[ \Ylmrefi{1}, \ldots ,\Ylmrefi{\nsph}  \right]}$, through the algebraic relation
\begin{equation}
\vectSPHrot = \transp{\wignermat } \vectSPHini,
\end{equation}
which is analogous to the change of basis equation given by \eq{change_basis_rotation}. The inverse transformation is given by 
\begin{equation}
\vectSPHini = \conj{\wignermat } \vectSPHrot.
\end{equation}
We remark that $\transp{\wignermat }\conj{\wignermat } = \idmat$, the symbol $\idmat$ referring to the identity matrix. As a consequence, the rotation transformation is conservative: going back and forth through the transformation does not induce any loss of information. Figure~\ref{fig:rmat} shows the rotation matrix obtained for a rotation of Euler angles $\left( \angalp, \angbet, \anggam \right) = \left(120^\degree, 60^\degree, 270^\degree \right) $ with a truncation degree set to $\lmax = 4$. As previously observed in \eq{Ylm_Ylq}, the rotated SPHs are expressed in terms of the non-rotated SPHs having the same degree ($\llat$). The orders ($\mm$) only are mixed up by the rotation of the coordinate system.

\section{Transition matrices}
\label{app:transition_matrices}

In this appendix, we establish the general transition relations between sets of basis functions of various types  (SPHs, SCHNs, SCHDs; see \sect{ssec:helmholtz_decomposition} and \append{app:sch}), and associated with different coordinate systems. These relations are used to compute the overlap coefficients between the forcing gravitational tidal potential and the tidal response in the calculation of the tidal torque.

\def\wpanel{0.22\textwidth}
\def\hraisebox{0.15\textwidth}
\def\hraisek{0.10\textwidth}
\def\hraiseboxps{0.12\textwidth}
\def\jspace{0.19\textwidth}
\begin{figure*}[htb]
   \centering
  \hspace{1.8cm}  \textsc{SCHNs}  \hspace{2.5cm} \textsc{SCHDs}   \hspace{2.5cm} \textsc{SPHs}  \hspace{0.7cm}~   \\[0.3cm]
 \raisebox{\hraisebox}[1cm][0pt]{%
   \begin{minipage}{1.4cm}%
   \textsc{SCHNs}
\end{minipage}}
  \raisebox{\hraisek}[1cm][0pt]{$\jjb$}
   \includegraphics[height=\wpanel,trim = 3.5cm 23.65cm 27.08cm 2.5cm,clip]{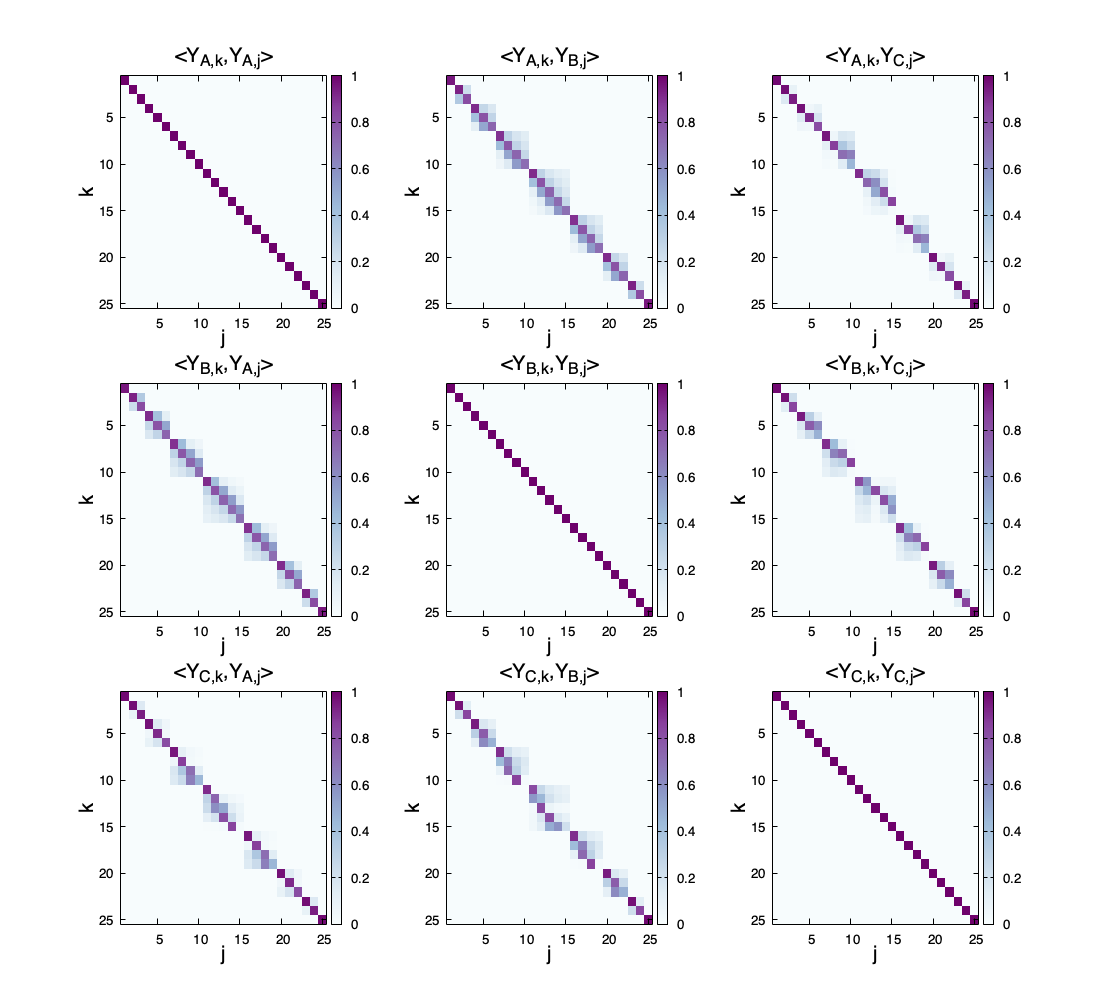}
  \includegraphics[height=\wpanel,trim = 15cm 23.65cm 15.58cm 2.5cm,clip]{auclair-desrotour_figF1.png}
   \includegraphics[height=\wpanel,trim = 26.5cm 23.65cm 4.08cm 2.5cm,clip]{auclair-desrotour_figF1.png}
    \includegraphics[height=\wpanel,trim = 34.7cm 23.65cm 2.6cm 2.5cm,clip]{auclair-desrotour_figF1.png} \\
    \raisebox{\hraisebox}[1cm][0pt]{%
   \begin{minipage}{1.4cm}%
   \textsc{SCHDs}
\end{minipage}}
  \raisebox{\hraisek}[1cm][0pt]{$\jjb$}
   \includegraphics[height=\wpanel,trim = 3.5cm 12.8cm 27.08cm 13.4cm,clip]{auclair-desrotour_figF1.png}
   \includegraphics[height=\wpanel,trim = 15cm 12.8cm 15.58cm 13.4cm,clip]{auclair-desrotour_figF1.png}
   \includegraphics[height=\wpanel,trim = 26.5cm 12.8cm 4.08cm 13.4cm,clip]{auclair-desrotour_figF1.png} \hspace{0.7cm}~ \\
   \raisebox{\hraisebox}[1cm][0pt]{%
   \begin{minipage}{1.4cm}%
   \textsc{SPHs}
\end{minipage}}
  \raisebox{\hraisek}[1cm][0pt]{$\jjb$}
   \includegraphics[height=\wpanel,trim = 3.5cm 1.9cm 27.08cm 24.3cm,clip]{auclair-desrotour_figF1.png} 
   \includegraphics[height=\wpanel,trim = 15cm 1.9cm 15.58cm 24.3cm,clip]{auclair-desrotour_figF1.png}
   \includegraphics[height=\wpanel,trim = 26.5cm 1.9cm 4.08cm 24.3cm,clip]{auclair-desrotour_figF1.png} \hspace{0.7cm}~ \\
    \hspace{2.cm}  $\kkb$ \hspace{\jspace} $\kkb$ \hspace{\jspace} $\kkb$ 
      \caption{Transition matrices between the SCHNs, SCHDs and SPHs for an ocean basin of angular radius $\colbd = 130^\degree$. The transition matrices are defined in \eq{transmat}. The SCHs (SCHNs and SCHDs) are calculated from the expressions given by \eq{schc}, and the SPHs from the expression given by \eq{sphc}, using the same coordinate system for all the sets of basis functions and the truncation degree $\lmax = 4$. In every set, the basis functions are sorted in ascending order of orders ($\mm=-4,-3,\ldots , 3,4$), with degrees ($\llat \geq \abs{\mm}$) grouped together.}
       \label{fig:transition_mat}%
\end{figure*}

\def\wpanel{0.28\textwidth}
\def\hraisebox{0.20\textwidth}
\def\hraiseboxps{0.12\textwidth}
\begin{figure*}[htb]
   \centering
  \hspace{1cm}  \textsc{Original function ($\Hlm{\llati{3}}{1}$)} \hspace{2cm} \textsc{Expanded in SPHs} \hspace{2cm} \textsc{Expanded in rotated SPHs} \\[0.3cm]
 \raisebox{\hraisebox}[1cm][0pt]{%
   \begin{minipage}{1.2cm}%
   $\lmax = 6$
\end{minipage}}
   \includegraphics[height=\wpanel,trim = 0.cm 1.5cm 3.0cm 1.5cm,clip]{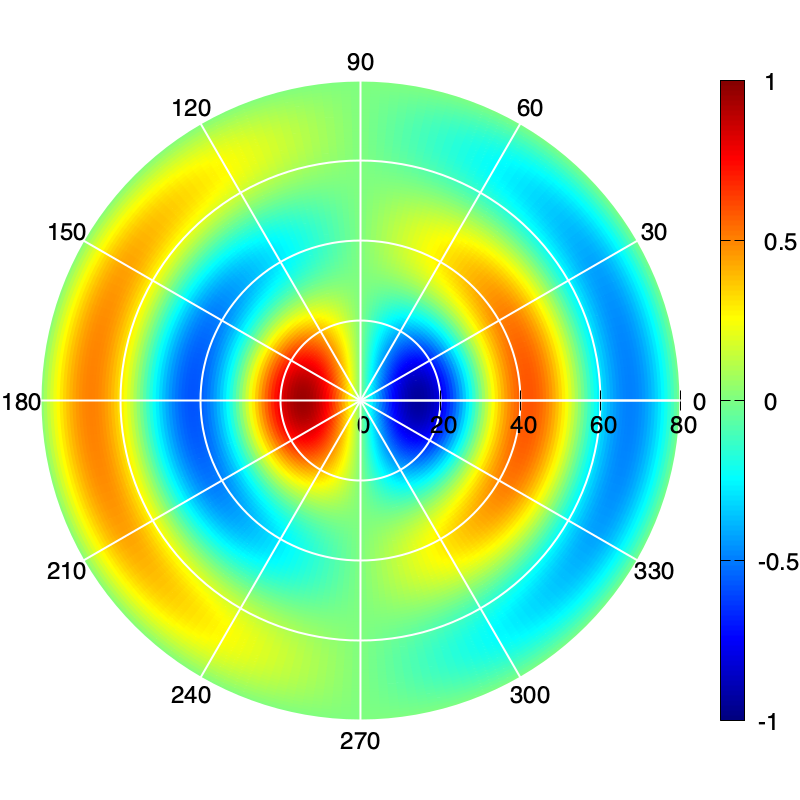}
   \includegraphics[height=\wpanel,trim = 0.cm 1.5cm 3.0cm 1.5cm,clip]{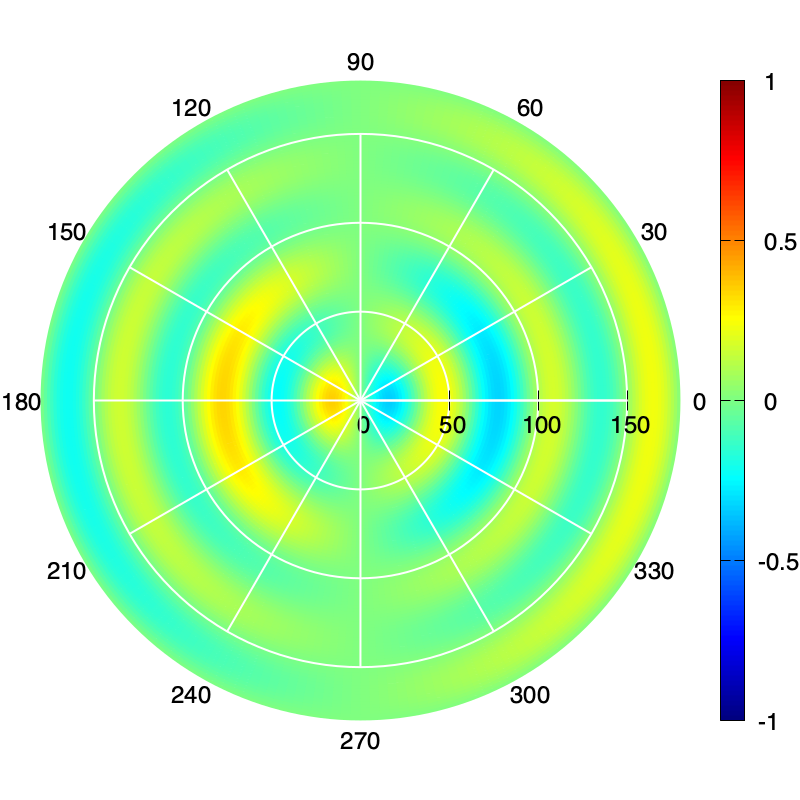}
   \includegraphics[height=\wpanel,trim = 0.cm 1.5cm .0cm 1.5cm,clip]{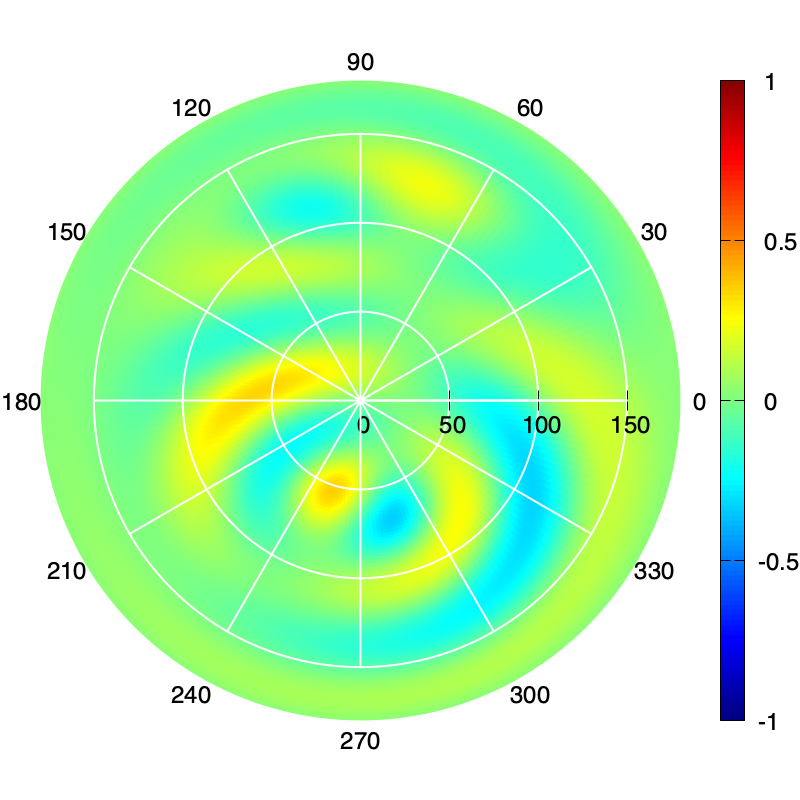} \\
 \raisebox{\hraisebox}[1cm][0pt]{%
   \begin{minipage}{1.2cm}%
   $\lmax = 8$
\end{minipage}}
   \includegraphics[height=\wpanel,trim = 0.cm 1.5cm 3.0cm 1.5cm,clip]{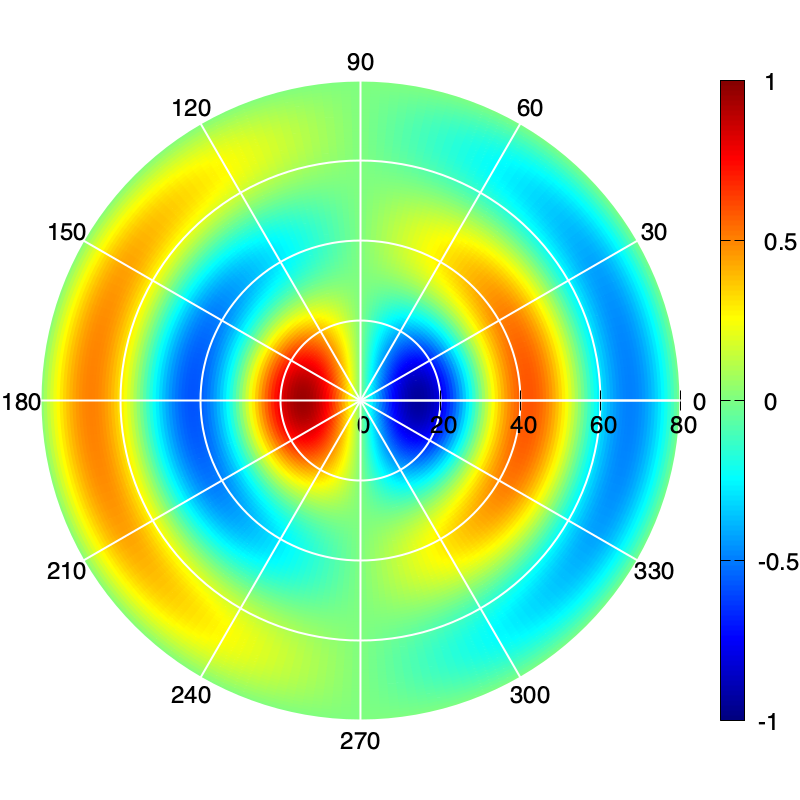}
   \includegraphics[height=\wpanel,trim = 0.cm 1.5cm 3.0cm 1.5cm,clip]{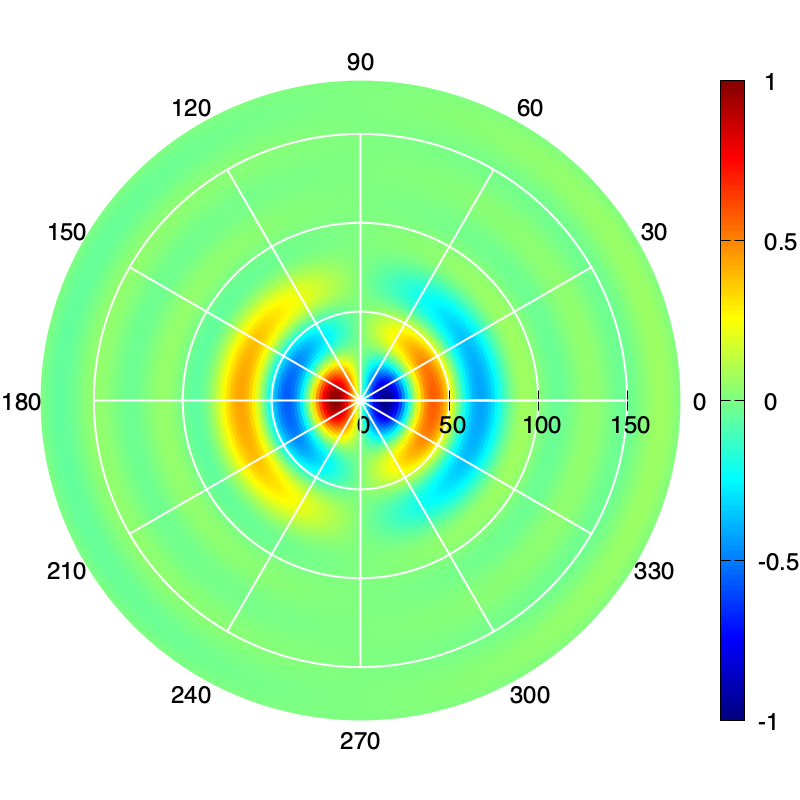}
   \includegraphics[height=\wpanel,trim = 0.cm 1.5cm .0cm 1.5cm,clip]{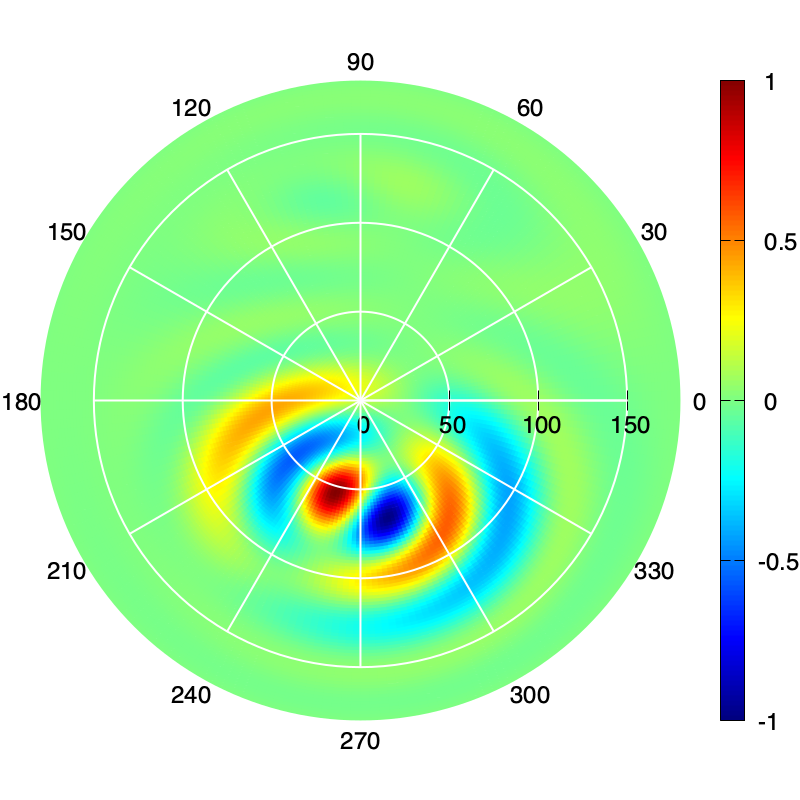} \\
 \raisebox{\hraisebox}[1cm][0pt]{%
   \begin{minipage}{1.2cm}%
   $\lmax = 16$
\end{minipage}}
   \includegraphics[height=\wpanel,trim = 0.cm 1.5cm 3.0cm 1.5cm,clip]{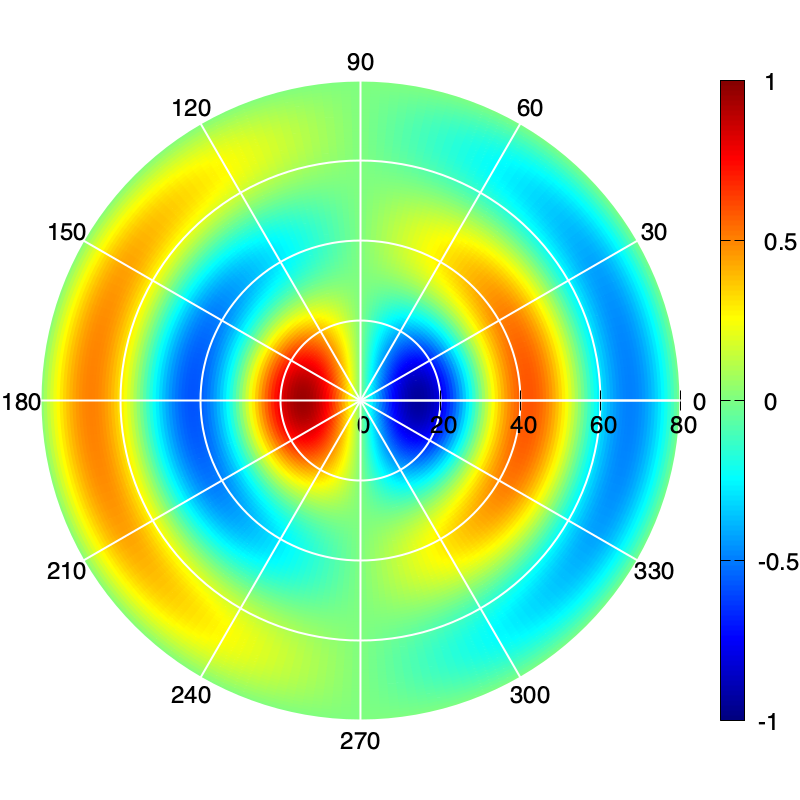}
   \includegraphics[height=\wpanel,trim = 0.cm 1.5cm 3.0cm 1.5cm,clip]{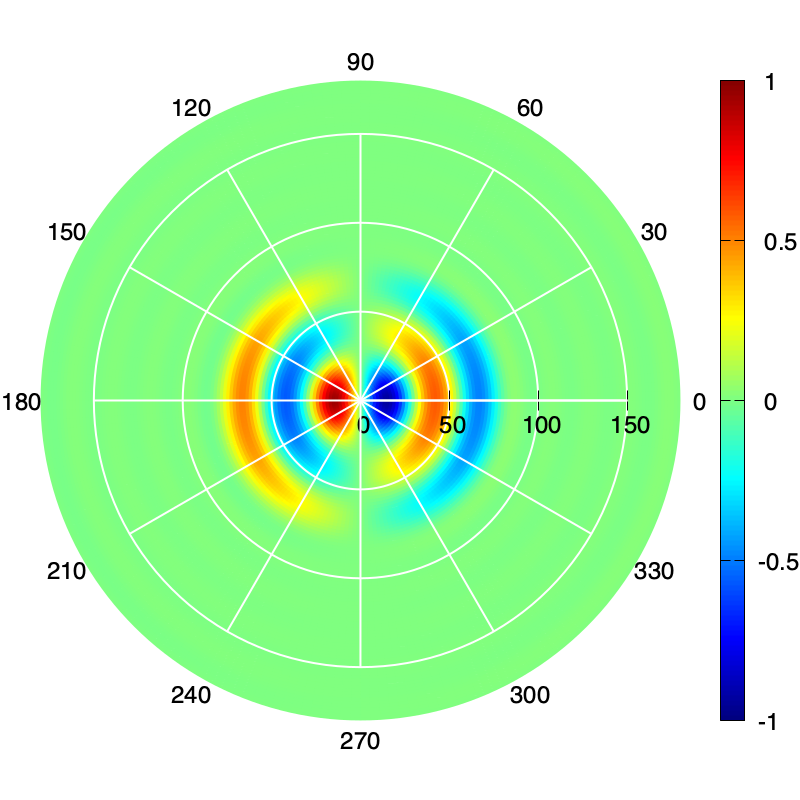}
   \includegraphics[height=\wpanel,trim = 0.cm 1.5cm .0cm 1.5cm,clip]{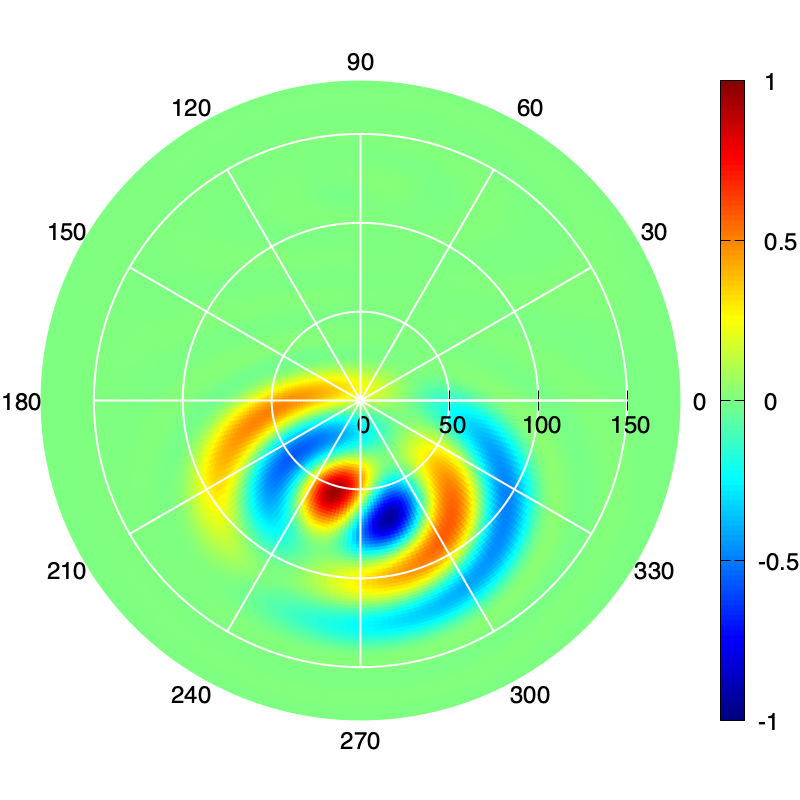} 
      \caption{Real part of complex SCH of order $\mm = 1$ and degree $\llati{3} = 6.806057$ corresponding to an ocean basin of angular radius $\colbd = 80^\degree$ (i.e. $\colcont = 100^\degree$) with Dirichlet boundary conditions (\eq{dircond}). {\it Left:} Function plotted in its coordinate system using the expression given by \eq{schc}. {\it Middle:} Function expanded in SPHs up to the truncation degree $\lmax = 6,8, 16$ (from top to bottom) in the same coordinate system using the transition relation of \eq{Yab_Ya}. {\it Right:} Function expanded in rotated SPHs ($\angalp = 120^\degree$, $\angbet = 60^\degree$, $\anggam=270^\degree$) following the relation given by \eq{vfuncAB}. All functions are plotted in polar coordinates, the radial coordinate corresponding to the colatitude of the considered coordinate system, and the angular coordinate to its longitude (in degrees).}
       \label{fig:transition}%
\end{figure*}

Let be two sets of basis functions, $\Abasisref$ and $\Bbasisref$, associated with the same coordinate system, $\framestd$, but not necessarily the same definition domains\footnote{By definition domain, we refer to the domain where the functions are non-zero, the SCHs being defined on the full unit sphere similarly as the SPHs in the used convention (see \eq{schc}).}, denoted by $\sphcapAref$ and $\sphcapBref$, respectively. Both $\Abasisref$ and $\Bbasisref$ can be either the SPHs, SCHNs, or SCHDs, whose expressions are given by \eqs{sphc}{schc}. The sets undergo the rotation transformations $\Abasisref \rightarrow \Abasis$ and $\Bbasisref \rightarrow \Bbasis$, where $\Abasis$ and $\Bbasis$ are assumed to have the same definition domains as $\Abasisref$ and $\Bbasisref$, respectively. The expressions of the standard Euler rotation matrices $\rmati{\Abasis}$ and $\rmati{\Bbasis}$ describing these transformations are given by \eq{eulermat}, and the corresponding Wigner rotation matrices for the functional bases, $\wignermatA$ and $\wignermatB$, are given by \eq{wignermat}. We note that the second Euler angle can only be set to $\angbet =0$ in the case of the SCHs since the definition domains of the initial and rotated function would not overlap otherwise in that case. However, the rotation matrix of the SPHs still holds for the rotations of angles $\angalp$ and $\anggam$ since these angles just introduce a phase lag in the $\expo{\inumber \mm \lon}$ component of the SCHs.  

The initial and rotated functional basis vectors are denoted by $\vfuncAref$, $\vfuncBref$, $\vfuncA$, and $\vfuncB$, the vector $\vfunci{\anybasis}$ of any set $\anybasis$ being expressed as 
\begin{equation}
\vfunci{\anybasis} = \transp{\left[ \funci{\anybasis}{1}, \funci{\anybasis}{2}, \ldots, \funci{\anybasis}{\kk} , \ldots , \funci{\anybasis}{\nbasisi{\anybasis}}  \right]},
\end{equation}
where the $\funci{\basis{H}}{\kk}$ designate the orthogonal functions of set $\anybasis$, and $\nbasisi{\anybasis}$ its total number of elements, which depends on the truncation degree $\lmaxi{\anybasis}$ introduced in \eq{wignermat}. To lighten expressions, we assume that the sets of  basis functions $\Abasisref$ and $\Abasis$ contain $\nbasisA$ elements, and that $\Bbasisref$ and $\Bbasis$ contain $\nbasisB$ elements, so that $\nbasisi{\Abasisref} = \nbasisi{\Abasis} = \nbasisA $ and $\nbasisi{\Bbasisref} = \nbasisi{\Bbasis} = \nbasisB$. Using the above notations, we can write down the relations
\begin{equation}
\label{Y_Yrot}
\begin{array}{ll}
\vfuncA = \transp{\wignermatA} \vfuncAref, & \vfuncB = \transp{\wignermatB} \vfuncBref,  \\[0.3cm]
\vfuncAref = \conj{\wignermatA} \vfuncA, & \vfuncBref = \conj{\wignermatB} \vfuncB. 
\end{array}
\end{equation}

As a next step, we establish the relationship between the bases $\Abasisref$ and $\Bbasisref$, which share the same coordinate system. The functions $\funcBrefi{\jj} $ and $\funcArefi{\kk} $ are expressed in terms of each other through the relations
\begin{align}
\label{Yj_series}
& \funcBrefi{\jj} = \sum_{\kk=1}^{\nbasisA} \frac{\scal{\funcArefi{\kk}}{\funcBrefi{\jj}} }{\scal{\funcArefi{\kk}}{\funcArefi{\kk}}} \funcArefi{\kk} + \resfi{\Bbasisref}{\jj} ,  \\
\label{Yk_series}
& \funcArefi{\kk} = \sum_{\jj=1}^{\nbasisB} \frac{\scal{\funcBrefi{\jj}}{\funcArefi{\kk}} }{\scal{\funcBrefi{\jj}}{\funcBrefi{\jj}}} \funcBrefi{\jj} + \resfi{\Abasisref}{\kk} .
\end{align}
In the above equations, the functions $\resfi{\Bbasisref}{\jj} $ and $\resfi{\Abasisref}{\kk}$ are the residuals of the functions $\funcBrefi{\jj}$ or $\funcArefi{\kk}$ when the latter are expanded in terms of the functions of set $\Abasisref$ or $\Bbasisref$, respectively. These residuals result both from the truncation of the sets $\Abasisref$ and $\Bbasisref$ -- which are mathematically defined as infinite-dimension sets of basis functions for the definition domains $\sphcapAref $ and $\sphcapBref$ --, and from the fact that $\sphcapAref $ and $\sphcapBref$ are different from each other in the general case. 

For instance, if $\sphcapAref \subsetneq \sphcapBref$, the functions of set $\Bbasisref$ can be expressed as linear combinations of the elements of set $\Abasisref$ on $\sphcapAref$ only. This is the case for SPHs, which cannot be approximated using SCHs while the opposite is true. Therefore, the function $\funcBrefi{\jj}$ tends to be fully described by the functions of set $\Abasisref$ as $\nbasisA \rightarrow \infty$ if $\sphcapBref \subset \sphcapAref$, and consequently $\normtwo{\resfi{\Bbasisref}{\jj} } \rightarrow 0$ in that case. Conversely, $\normtwo{\resfi{\Bbasisref}{\jj} }>0$ for $\nbasisA = \infty$ in the general case if $\sphcapAref \subsetneq \sphcapBref$. The residual function $\resfi{\Abasisref}{\kk}$ behaves in a similar way if the domains $\sphcapAref $ and $\sphcapBref$ are interchanged: $\normtwo{\resfi{\Abasisref}{\kk}} \rightarrow 0$ as $\nbasisB \rightarrow \infty$ if $\sphcapAref \subset \sphcapBref$, while $\normtwo{\resfi{\Abasisref}{\kk}} >0$ if $\sphcapBref \subsetneq \sphcapAref$. 


Since the functions of the two sets are assumed to be orthonormal by construction (see \append{app:sch}), $\scal{\funcArefi{\kk}}{\funcArefi{\kk}} = \scal{\funcBrefi{\jj}}{\funcBrefi{\jj}} = 1$ for all $\kk \in \left\{1 , \ldots , \nbasisA  \right\}$ and $\jj \in \left\{ 1 , \ldots , \nbasisB \right\}$. As a consequence, the expressions given by \eqs{Yj_series}{Yk_series} simplify to 
\begin{align}
\label{Yj_Ybak}
& \funcBrefi{\jj} = \sum_{\kk=1}^{\nbasisA} \scal{\funcArefi{\kk}}{\funcBrefi{\jj}} \funcArefi{\kk} + \resfi{\Bbasisref}{\jj} & \mbox{on} \ \sphcapBref,  \\
\label{Yk_Yabj}
& \funcArefi{\kk} = \sum_{\jj=1}^{\nbasisB} \scal{\funcBrefi{\jj}}{\funcArefi{\kk}} \funcBrefi{\jj} + \resfi{\Abasisref}{\kk} & \mbox{on} \ \sphcapAref,
\end{align}
which, introducing the vectors 
\begin{align}
 \resmBref & = \transp{\left[ \resfi{\Bbasisref}{1}, \ldots , \resfi{\Bbasisref}{\jj} , \ldots , \resfi{\Bbasisref}{\nbasisB}  \right]},  \\
\resmAref& = \transp{\left[ \resfi{\Abasisref}{1}, \ldots , \resfi{\Abasisref}{\kk} , \ldots , \resfi{\Abasisref}{\nbasisA}  \right]}  ,
\end{align}
may be rewritten as 
\begin{align}
\label{Yab_Ya}
& \vfuncBref = \transp{\transmati{\Abasisref}{\Bbasisref}} \vfuncAref + \resmBref,&  \vfuncAref = \conj{\transmati{\Abasisref}{\Bbasisref}} \vfuncBref + \resmAref.
\end{align}
In the above equations, the transition matrix is expressed as 
\begin{equation}
\label{transmat}
\transmati{\Abasisref}{\Bbasisref} \! \define \!
\begin{bmatrix}
\scal{\funcArefi{1}}{\funcBrefi{1}} & \ldots & \scal{\funcArefi{1}}{\funcBrefi{\jj}} & \ldots & \scal{\funcArefi{1}}{\funcBrefi{\nbasisB}} \\
\vdots &  & \vdots & & \vdots \\
\scal{\funcArefi{\kk}}{\funcBrefi{1}} & \ldots & \scal{\funcArefi{\kk}}{\funcBrefi{\jj}} & \ldots & \scal{\funcArefi{\kk}}{\funcBrefi{\nbasisB}} \\
\vdots &  & \vdots & & \vdots \\
\scal{\funcArefi{\nbasisA}}{\funcBrefi{1}} & \ldots & \scal{\funcArefi{\nbasisA}}{\funcBrefi{\jj}} & \ldots & \scal{\funcArefi{\nbasisA}}{\funcBrefi{\nbasisB}} 
\end{bmatrix} \! \!,
\end{equation}
where $\scal{\funcArefi{\kk}}{\funcBrefi{\jj}}$ is the scalar product of the $\kk$-th element of $\Abasisref$ with the $\jj$-th element of $\Bbasisref$ following the formulation given by \eq{sprodYlm}. We note that the integral of the scalar product is actually performed on the intersection of the domains associated with $\Abasisref$ and $\Bbasisref$, namely $\sphcap = \sphcapAref \cap \sphcapBref$. 

With the used normalisations, the transition matrix satisfies the properties $\transmati{\Bbasisref}{\Abasisref} = \transp{\conj{\transmati{\Abasisref}{\Bbasisref}}} $, $\transmati{\Abasisref}{\Abasisref} = \idmati{\nbasisA} $, and $\transmati{\Bbasisref}{\Bbasisref} = \idmati{\nbasisB}$, the notation $\idmati{\nbasisA}$ referring to the identity matrix of size $\nbasisA$. However, one should pay attention to the fact that $\transmati{\Abasisref}{\Bbasisref} \transmati{\Bbasisref}{\Abasisref} \neq \transmati{\Abasisref}{\Abasisref} $ and  $\transmati{\Bbasisref}{\Abasisref} \transmati{\Abasisref}{\Bbasisref} \neq \transmati{\Bbasisref}{\Bbasisref} $ due to the inability of the sets of basis functions to fully describe the space of functions defined on the unit sphere, as discussed above. In other words, going back and forth between the two sets necessarily induces a loss of information contrary to the rotation operator of the SPHs detailed in \append{app:rotation_sph}, which is conservative. 

Figure~\ref{fig:transition_mat} shows the transition matrices $\transmati{\Abasisref}{\Bbasisref}$ between the SCHNs, SCHDs, and SPHs for an ocean of angular radius $\colbd = 130^\degree$ (i.e. a continent of angular radius $\colcont = 50^\degree$). The basis functions are sorted in ascending order of degrees and orders in every set, with degrees grouped together, which differs from the SPHs of \fig{fig:rmat}, where orders are grouped together. Only basis functions with the same order overlap, due to the fact that the SCHs and SPHs have the same longitudinal component. However, some functions belonging to one set are poorly described by the basis functions of the other sets. 

By combining together the relations given by \eqs{Y_Yrot}{Yab_Ya}, we finally obtain
\begin{align}
\label{vfuncAB}
& \vfuncB = \transp{\transmati{\Abasis}{\Bbasis}} \vfuncA + \transp{\wignermatB} \resmBref , & \vfuncA = \conj{\transmati{\Abasis}{\Bbasis}} \vfuncB + \transp{\wignermatA} \resmAref , 
\end{align}
where the transition matrix $\transmati{\Abasis}{\Bbasis}$ is given by 
\begin{equation}
\label{transmatAB}
\transmati{\Abasis}{\Bbasis} = \transp{\conj{\wignermatA}} \transmati{\Abasisref}{\Bbasisref} \wignermatB.
\end{equation}
A given function $\ffunc$ defined on the sphere can thus be written both in terms of the basis functions of $\Abasis$ and those of $\Bbasis$ as
\begin{align}
& \ffunc = \transp{\vcoorA}  \vfuncA + \resci{\Abasis}{\ffunc}, &  \ffunc = \transp{\vcoorB}  \vfuncB + \resci{\Bbasis}{\ffunc},
\end{align}
where we have introduced the residual functions $\resci{\Abasis}{\ffunc}$ and $\resci{\Bbasis}{\ffunc}$, and the coordinate vectors of $\ffunc$ in the two sets of basis functions, $\vcoorA$ and $\vcoorB$, defined as
\begin{align}
 \vcoorA \define & \transp{\left[  \coorAi{1}, \ldots , \coorAi{\kk} , \ldots , \coorAi{\nbasisA} \right]} , \\
 \vcoorB \define & \transp{\left[  \coorBi{1}, \ldots , \coorBi{\jj} , \ldots , \coorBi{\nbasisB} \right]} .
\end{align}
The vectors $ \vcoorA$ and $ \vcoorB$ contain the coefficients of the expansion of $\ffunc$ on the set of basis functions $\Abasis$ or $\Bbasis$, respectively. They are related to each other by the transition matrix $\transmati{\Abasis}{\Bbasis}$ introduced in \eq{transmatAB} through the relation equations
\begin{align}
& \vcoorA = \transmati{\Abasis}{\Bbasis} \vcoorB, & \vcoorB = \transp{\conj{\transmati{\Abasis}{\Bbasis}}} \vcoorA.
\end{align}

Figure~\ref{fig:transition} shows the three steps of the transition from SCHs to rotated SPHs through the example of the eigenfunction $\Hlm{\llati{3}}{1}$ obtained with Dirichlet conditions (SCHD). First the function is evaluated in its associated coordinate system using the expression given by \eq{schc} (left panels). We note that the region where the function is set to zero is not represented in the plot. Second, the function is expanded in series of SPHs in the same coordinate system (middle panels). Finally, we apply an Euler rotation to the SPHs, as described by \eq{Ylm_Ylq}, and $\Hlm{\llati{3}}{1}$ is evaluated in terms of the rotated SPHs in the rotated coordinate system (right panel). These three steps are shown for various truncation degrees of the SPHs, $\lmax = 6,8,16$ (from top to bottom). We observe that the discontinuity of the gradient of the function at $\col = \colbd$ gives rise to `ringing' when $\Hlm{\llati{3}}{1}$ is expanded in series of SPHs. This effect also alters the expansions of the SCHNs due to their discontinuous transition at $\col = \colbd$. However it vanishes as the truncation degrees $\lmax$ increases.

\section{Gyroscopic coefficients}
\label{app:gyro_coeff}
In this section, we detail the calculation of the gyroscopic coefficients that couple the LTEs. Following earlier studies \citep[][]{Webb1980,Webb1982,Farhat2022b}, these coefficients are denoted by $\gyrocoeff$. They are defined as
\begin{align}
\label{gyro1}
\gyroci{\jjb}{\kkb} \define & - \integ{\cos \colpla \, \er \dotp \left( \conj{\grad \fpoti{\jjb} } \crossp \grad \fpoti{\kkb} \right) }{\surface}{\sphcap}{}, \\ 
\label{gyro2}
\gyroci{\jjb}{- \kkb} \define & \ \ \ \ \integ{\cos \colpla  \left( \conj{\grad \fpoti{\jjb} } \dotp \grad \fstreami{\kkb} \right)}{\surface}{\sphcap}{}, \\
\label{gyro3}
\gyroci{-\jjb}{\kkb} \define &  - \integ{\cos \colpla \left( \conj{\grad \fstreami{\jjb}} \dotp \grad \fpoti{\kkb} \right)}{\surface}{\sphcap}{}, \\
\label{gyro4}
\gyroci{-\jjb}{-\kkb} \define & - \integ{\cos \colpla \, \er \dotp \left(  \conj{\grad \fstreami{\jjb}} \crossp \grad \fstreami{\kkb} \right)}{\surface}{\sphcap}{},
\end{align}
where the $\fpoti{\jjb}$ and $\fstreami{\jjb}$ are the basis functions generating the potential and stream functions, respectively (see \sect{ssec:helmholtz_decomposition} and \append{app:sch}).

\def\wpanel{0.22\textwidth}
\def\hraisebox{0.15\textwidth}
\def\hraisek{0.10\textwidth}
\def\hraiseboxps{0.12\textwidth}
\def\jspace{0.19\textwidth}
\begin{figure*}[htb]
   \centering
  \hspace{1.8cm}  $\Im \left( \gyroci{\jjb}{\kkb} \right)$  \hspace{2.5cm} $\Re \left( \gyroci{\jjb}{-\kkb} \right)$ \hspace{2.5cm} $\Re \left( \gyroci{-\jjb}{\kkb} \right)$ \hspace{2.5cm} $\Im \left( \gyroci{-\jjb}{-\kkb} \right)$    \\[0.3cm]
 \raisebox{\hraisebox}[1cm][0pt]{%
   \begin{minipage}{1.4cm}%
   $\coloc = 0^\degree$
\end{minipage}}
  \raisebox{\hraisek}[1cm][0pt]{$\jjb$}
   \includegraphics[height=\wpanel,trim = 4.5cm 18.4cm 22.cm 3.3cm,clip]{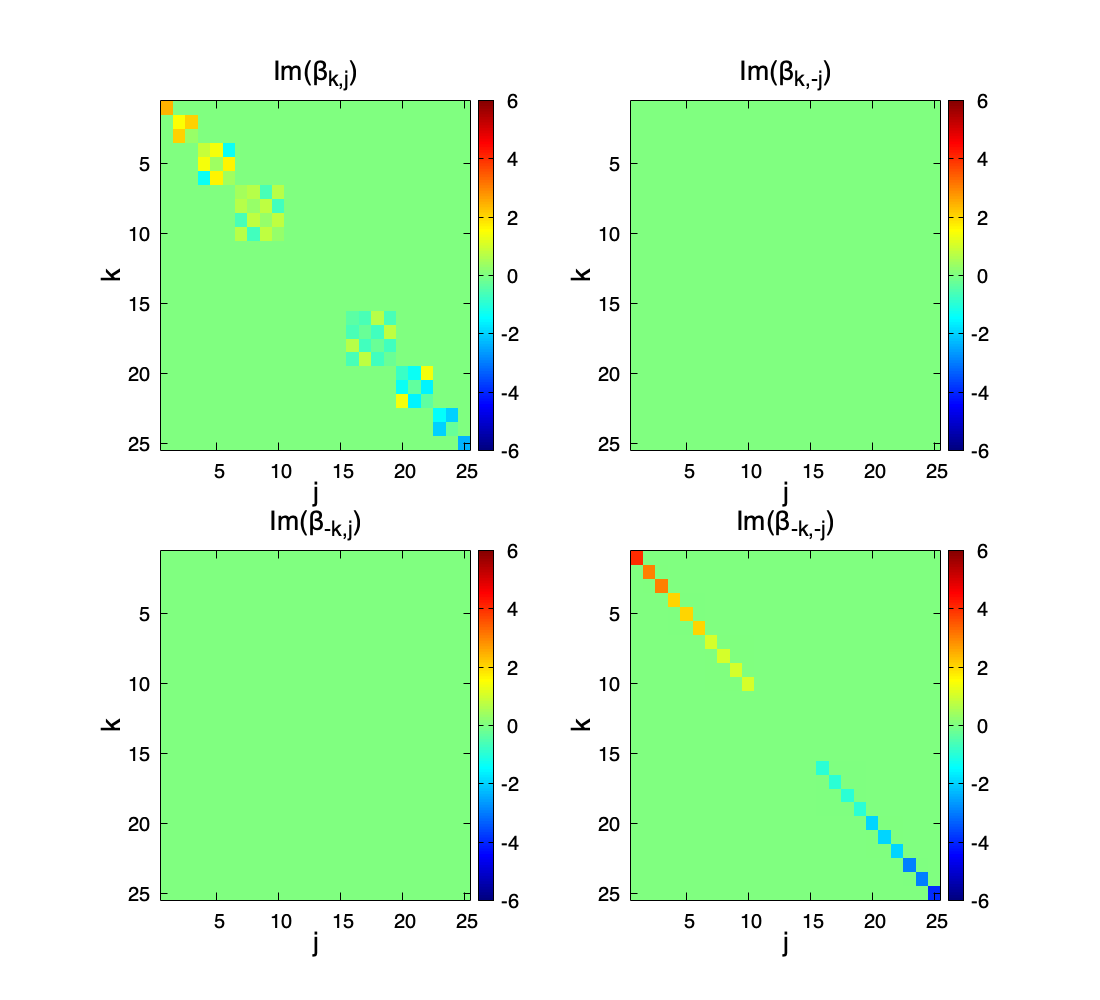}
  \includegraphics[height=\wpanel,trim = 21.05cm 18.4cm 5.45cm 3.3cm,clip]{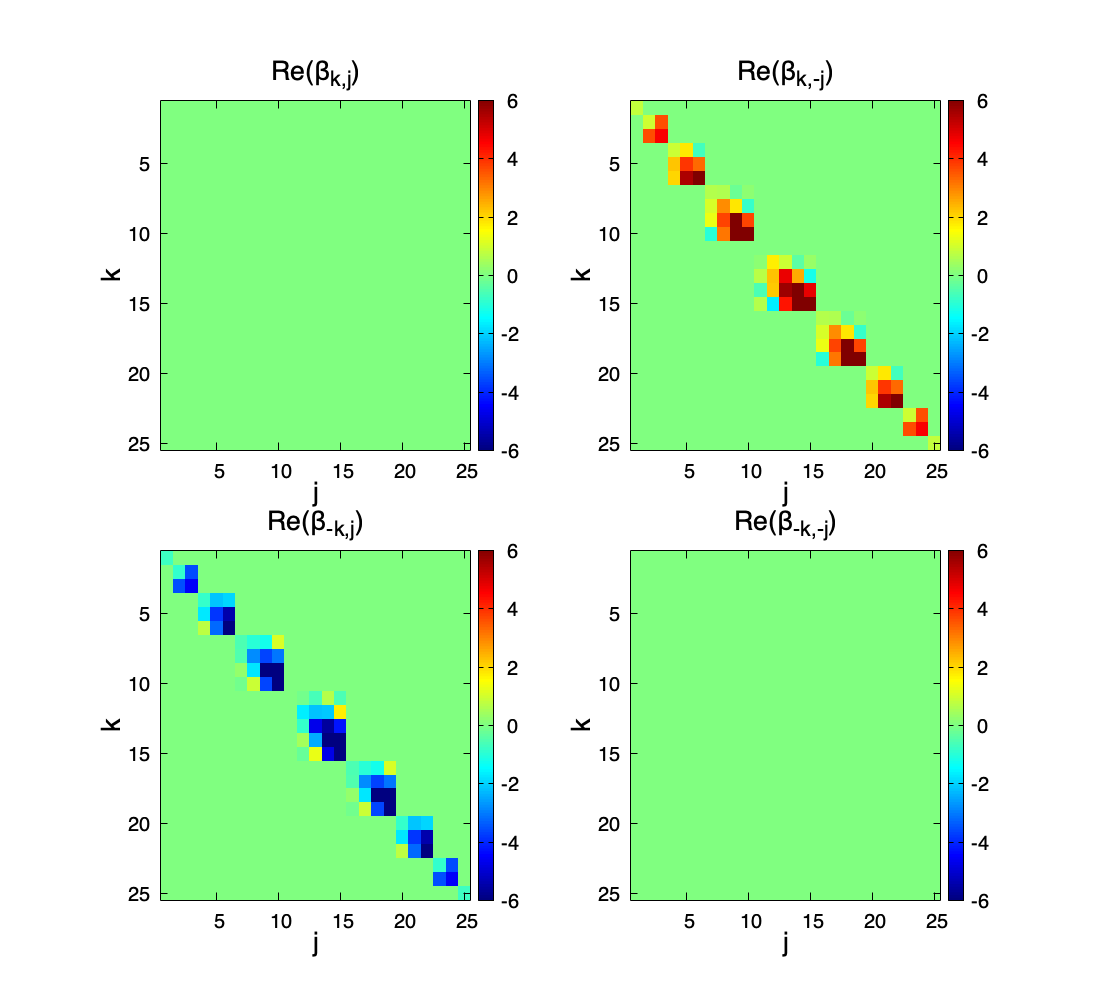}
   \includegraphics[height=\wpanel,trim = 4.5cm 2.5cm 22cm 19.2cm,clip]{auclair-desrotour_figG1b.png}
    \includegraphics[height=\wpanel,trim = 21.05cm 2.5cm 5.45cm 19.2cm,clip]{auclair-desrotour_figG1a.png}
    \includegraphics[height=\wpanel,trim = 33.45cm 18.4cm 3.9cm 3.3cm,clip]{auclair-desrotour_figG1b.png} \\
    \raisebox{\hraisebox}[1cm][0pt]{%
   \begin{minipage}{1.4cm}%
   $\coloc = 45^\degree$
\end{minipage}}
  \raisebox{\hraisek}[1cm][0pt]{$\jjb$}
   \includegraphics[height=\wpanel,trim = 4.5cm 18.4cm 22.cm 3.3cm,clip]{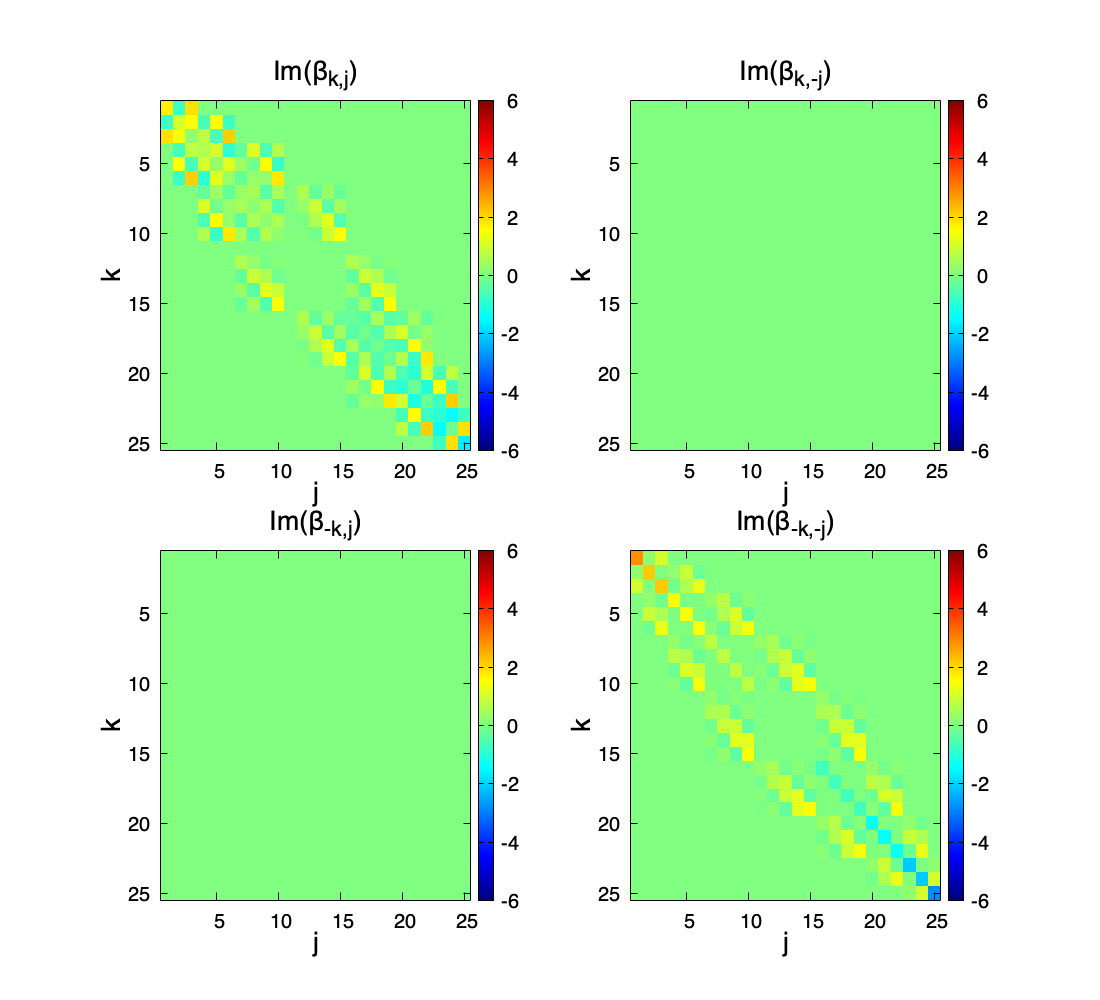}
  \includegraphics[height=\wpanel,trim = 21.05cm 18.4cm 5.45cm 3.3cm,clip]{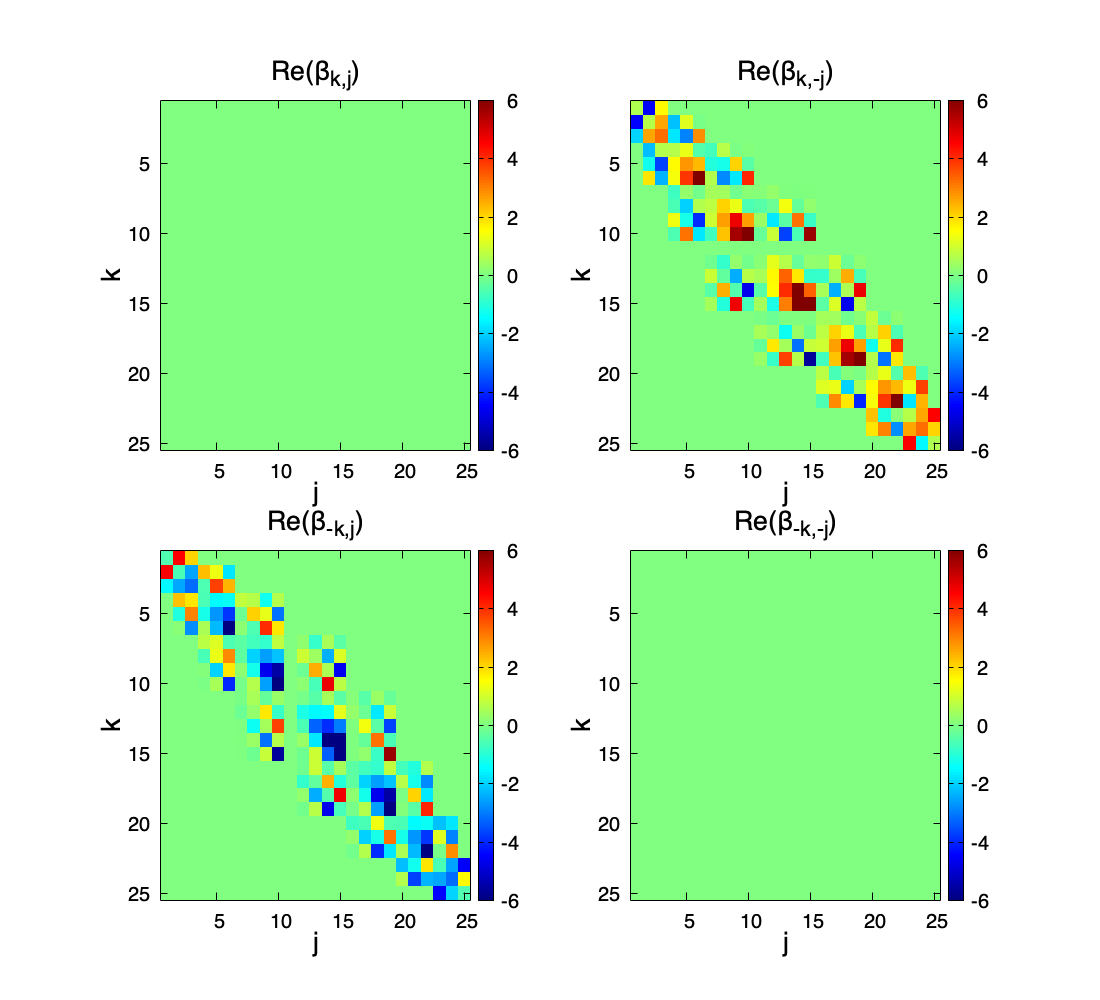}
   \includegraphics[height=\wpanel,trim = 4.5cm 2.5cm 22cm 19.2cm,clip]{auclair-desrotour_figG1d.png}
    \includegraphics[height=\wpanel,trim = 21.05cm 2.5cm 5.45cm 19.2cm,clip]{auclair-desrotour_figG1c.png} \hspace{0.4cm}~ \\
    \raisebox{\hraisebox}[1cm][0pt]{%
   \begin{minipage}{1.4cm}%
   $\coloc = 90^\degree$
\end{minipage}}
  \raisebox{\hraisek}[1cm][0pt]{$\jjb$}
   \includegraphics[height=\wpanel,trim = 4.5cm 18.4cm 22.cm 3.3cm,clip]{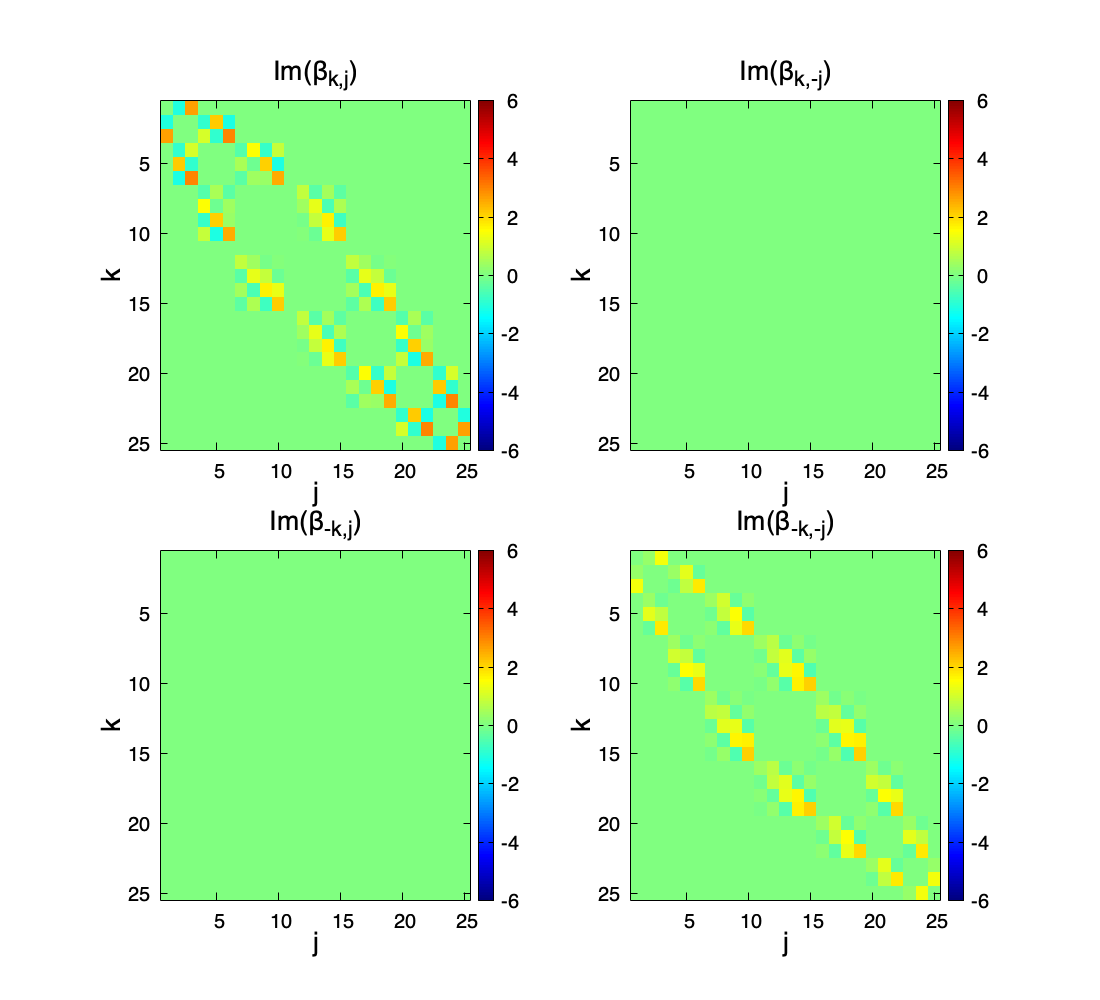}
  \includegraphics[height=\wpanel,trim = 21.05cm 18.4cm 5.45cm 3.3cm,clip]{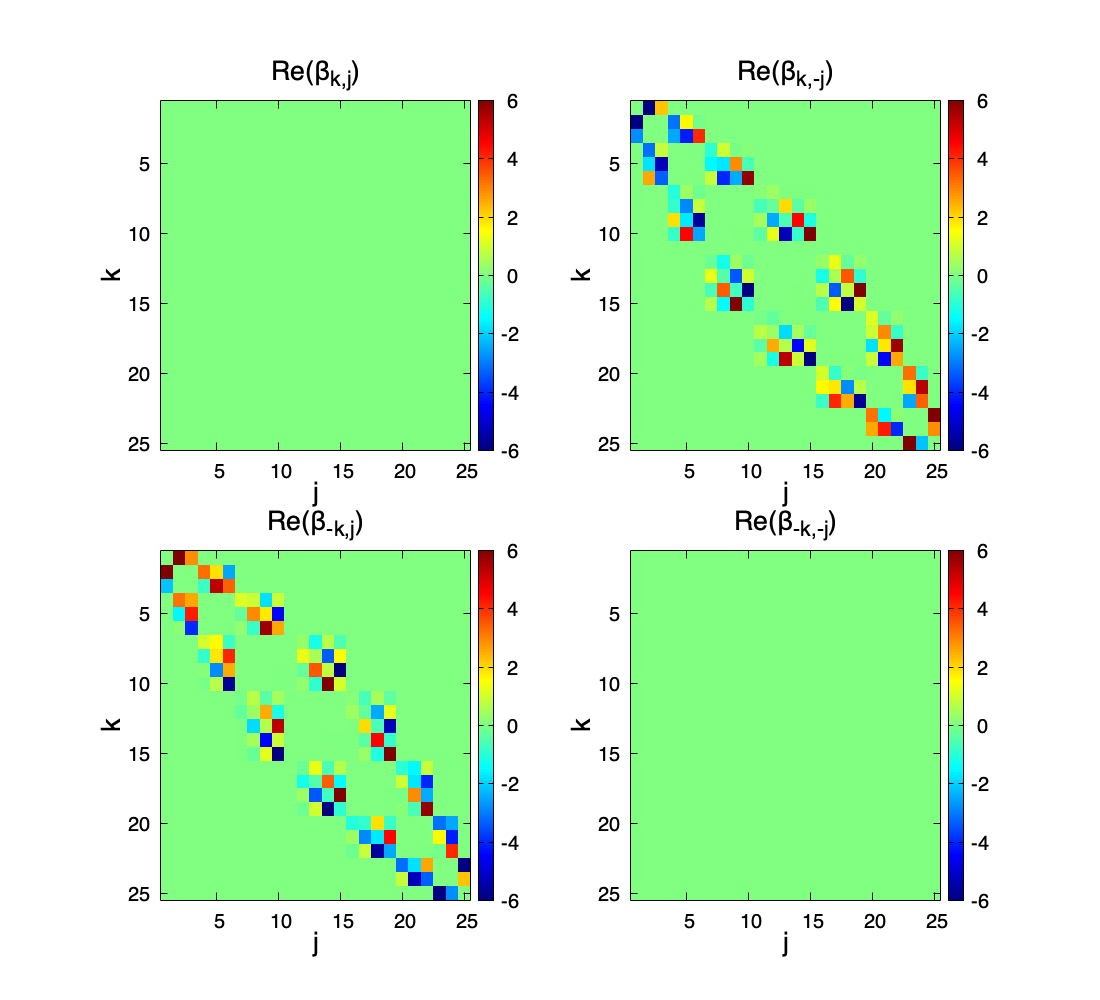}
   \includegraphics[height=\wpanel,trim = 4.5cm 2.5cm 22cm 19.2cm,clip]{auclair-desrotour_figG1f.png}
    \includegraphics[height=\wpanel,trim = 21.05cm 2.5cm 5.45cm 19.2cm,clip]{auclair-desrotour_figG1e.png} \hspace{0.4cm}~ \\
    \hspace{2.cm}  $\kkb$ \hspace{\jspace} $\kkb$ \hspace{\jspace} $\kkb$ \hspace{\jspace} $\kkb$ 
      \caption{Gyroscopic coefficients of an ocean basin of angular radius $\colbd=130^\degree$ (i.e. $\colcont = 50^\degree$) for various values of the colatitude of the ocean's centre. {\it Top:} $\coloc=0^\degree$ (polar ocean). {\it Middle:} $\coloc=45^\degree$ (mid-latitude ocean). {\it Bottom:} $\coloc = 90^\degree$ (equatorial ocean). The gyroscopic coefficients are computed from the expressions given by \eqsto{gyro1_exp}{gyro4_exp} with the truncation degree set to $\lmax=4$.  The basis functions of indices $\jjb$ and $\kkb$ are sorted in ascending order of orders ($\mm = -4,-3, \ldots,3,4$), with degrees ($\llat \geq \abs{\mm}$) grouped together. For convenience, the minimum and maximum of the colour bar are set to 6 in absolute value although the maximum values reached by gyroscopic coefficients are around ${\sim}$15 in this case.}
       \label{fig:betagyro}%
\end{figure*}

We emphasise that $\jjb$ and $\kkb$ are the indices of the functions in their respective sets. One shall be cautious not to mix up these indices with the orders and degrees of the functions. The third coefficient, given by \eq{gyro3}, is straightforwardly deduced from the second one, given by \eq{gyro2}, since $\gyroci{-\jjb}{\kkb} = - \conj{\gyroci{\kkb}{- \jjb}}$. In some particular cases, such as the hemispherical ocean configuration considered in previous studies \citep[e.g.][]{Webb1980,Farhat2022b}, the recurrence relations of the associated Legendre  functions can be used to obtain the analytical expressions of the gyroscopic coefficients \citep[see e.g.][Appendix~G]{Farhat2022b}. Unfortunately, these recurrence relations do not hold for non-integral degrees, which requires to evaluate the integrals of \eqsto{gyro1}{gyro4} numerically. However, it is still possible to write down these integrals as products of separated coordinates. 

Introducing the coordinates of the North pole in the ocean's coordinate system, $\left( \colnp , \lonnp \right) = \left( \coloc , \pi \right)$, and using trigonometric identities, we express $\cos \colpla$ as a function of the colatitude of the ocean's centre ($\coloc$) and the ocean's coordinate system,
\begin{align}
\cos \colpla & = \cos \colnp \cos \col + \sin \colnp \sin \col \cos \left( \lon - \lonnp \right) \\
 &  = \cos \coloc \cos \col - \sin \coloc \sin \col \cos \lon. 
\end{align}
Considering the definitions of the SPHs and SCHs given by \eqs{sphc}{schc}, the functions $\fpoti{\jjb}$ and $\fstreami{\jjb}$ are expressed in the general case as 
\begin{align}
\fpoti{\jjb} \left( \col , \lon \right) & = \fpotci{\jjb}  \phicolfi{\jjb} \left( \col \right) \expo{\inumber \mmi{\jjb} \lon} , \\
\fstreami{\jjb} \left( \col , \lon \right) & = \fstreamci{\jjb} \psicolfi{\jjb} \left( \col \right)  \expo{\inumber \mmi{\jjb} \lon},
\end{align}
where $ \phicolfi{\jjb} \left( \col \right) = \LegF{\llati{\jjb}}{\mmi{\jjb} } \left( \cos \col \right)$ designates the ALF of the $\jjb$-th function of the set of SCHNs (Neumann conditions) used to expand the potential function, and $ \psicolfi{\jjb} \left( \col \right) = \LegF{\llati{\jjb}}{\mmi{\jjb} } \left( \cos \col \right)$ the ALF of the $\jjb$-th function of the set of SCHDs (Dirichlet conditions) used to expand the stream function, as detailed in \append{app:sch}. The gyroscopic coefficients given by \eqsto{gyro1}{gyro4} are therefore expressed as 
\begin{align}
\gyroci{\jjb}{\kkb} = & - \inumber 2 \pi \fpotci{\jjb} \fpotci{\kkb}   \left\{  \cos \coloc \kron{\mmi{\jjb}}{\mmi{\kkb}} \mmi{\jjb}   \!\! \integ{ \! \! \! \! \cos \col \left[ \DD{\phicolfi{\jjb}}{\col} \phicolfi{\kkb} + \phicolfi{\jjb} \DD{\phicolfi{\kkb}}{\col}  \right] \!}{\col}{0}{\colbd}   \right. \nonumber \\
\label{gyro1_exp}
 & \left. - \frac{\sin \coloc}{2}  \kron{\abs{\mmi{\kkb} - \mmi{\jjb}}}{1} \!\! \integ{\! \! \! \! \sin \col \left[ \mmi{\kkb} \DD{\phicolfi{\jjb}}{\col} \phicolfi{\kkb} + \mmi{\jjb} \phicolfi{\jjb} \DD{\phicolfi{\kkb}}{\col}  \right] \! }{\col}{0}{\colbd}    \right\}, 
\end{align}
\begin{align}
\gyroci{\jjb}{-\kkb} = &  \ \pi \fpotci{\kkb} \fstreamci{\jjb} \left\{  \cos \coloc \kron{\mmi{\jjb}}{\mmi{\kkb}} \! \!   \integ{ \!\!\!\! \sin 2 \col  \left[ \DD{\phicolfi{\kkb}}{\col} \DD{\psicolfi{\jjb}}{\col} + \mmi{\jjb}^2 \frac{\phicolfi{\kkb} \psicolfi{\jjb}}{\sin^2 \col} \right] \! }{\col}{0}{\colbd}  \right. \nonumber \\ 
  & \left. - \sin \coloc  \kron{\abs{\mmi{\kkb} - \mmi{\jjb}}}{1} \!\! \integ{ \!\!\!\! \sin^2 \col  \left[ \DD{\phicolfi{\kkb}}{\col} \DD{\psicolfi{\jjb}}{\col} + \mmi{\jjb} \mmi{\kkb} \frac{\phicolfi{\kkb} \psicolfi{\jjb}}{\sin^2 \col} \right] \! }{\col}{0}{\colbd}  \! \right\}\! , 
\end{align}
\begin{equation}
\gyroci{-\jjb}{\kkb} = - \gyroci{\kkb}{-\jjb}, 
\end{equation}
\begin{align}
\gyroci{-\jjb}{-\kkb} = & - \inumber 2 \pi \fstreamci{\jjb} \fstreamci{\kkb} \!  \left\{ \!   \cos \coloc \kron{\mmi{\jjb}}{\mmi{\kkb}} \mmi{\jjb} \! \! \integ{ \! \! \! \! \! \cos \col \! \left[ \! \DD{\psicolfi{\jjb}}{\col} \psicolfi{\kkb} + \psicolfi{\jjb} \DD{\psicolfi{\kkb}}{\col} \! \right] \! }{\col}{0}{\colbd}   \right. \nonumber \\
\label{gyro4_exp}
 & \left. - \frac{\sin \coloc }{2} \kron{\abs{\mmi{\kkb} - \mmi{\jjb}}}{1} \! \! \integ{\! \! \! \!  \sin \col \left[ \mmi{\kkb} \DD{\psicolfi{\jjb}}{\col} \psicolfi{\kkb} + \mmi{\jjb} \psicolfi{\jjb} \DD{\psicolfi{\kkb}}{\col}  \right] \! }{\col}{0}{\colbd}    \right\} \!.
\end{align}

Interestingly, the coefficients $\gyroci{\pm \jjb}{\pm \kkb}$ are zero if $\abs{\mmi{\kkb}-\mmi{\jjb}} >1  $, meaning that Coriolis effects couple the eigenmodes having either the same order or neighbouring orders in the general case. If the ocean is polar ($\coloc=0,\pi$), the gyroscopic coefficients only mix up eigenmodes with the same order ($\mmi{\kkb}=\mmi{\jjb}$). In this configuration, $\gyroci{\pm\jjb}{\pm\kkb} \left( \coloc = \pi \right) = - \gyroci{\pm\jjb}{\pm\kkb} \left( \coloc = 0 \right)$. If the ocean is equatorial ($\coloc = \pi/2$), the gyroscopic coefficients only mix up eigenmodes of neighbouring orders ($\abs{\mmi{\kkb}-\mmi{\jjb}}=1$). Between these two extremal configurations, the gyroscopic coefficients are the sum of one component in $\cos \coloc$ associated with eigenmodes having the same order, and one component in $\sin \coloc$ associated with eigenmodes having neighbouring orders. The gyroscopic coefficients that do not mix up potential and stream functions ($\gyroci{\jjb}{-\kkb} $ and $\gyroci{-\jjb}{-\kkb} $) are pure imaginary numbers. Conversely, the coefficients that mix up the functions ($\gyroci{\jjb}{-\kkb}$ and $\gyroci{-\jjb}{\kkb} $) are pure real numbers. 

By comparing the expressions given by \eqs{gyro1_exp}{gyro4_exp}, we remark that the $\gyroci{\jjb}{\kkb} $ and $\gyroci{-\jjb}{-\kkb}  $ converge towards the same asymptotic values as $\colbd \rightarrow 180^\degree$ since  both the SCHNs and SCHDs converge towards the standard SPHs for the $\ltwonorm$-norm introduced in \eq{ltwonorm}, as illustrated by \fig{fig:degrees_size}. To our knowledge, no recurrence formula exist between the ALFs of real degrees of the SCHNs and SCHDs except in the specific case of the hemispherical ocean, where the degrees are integers \citep[see][and Table~\ref{tab:sch_degrees}]{Webb1980,Webb1982,Farhat2022b}. As a consequence, the integrals in \eqsto{gyro1_exp}{gyro4_exp} are evaluated numerically in practice.  

Figure~\ref{fig:betagyro} shows the gyroscopic coefficients characterising an ocean basin of angular radius $\colbd = 130^\degree$ for three typical configuration: $\coloc = 0^\degree$ (polar ocean), $\coloc = 45^\degree$ (mid-latitude ocean), and $\coloc = 90^\degree$ (equatorial ocean). The eigenmodes of indices $\jjb$ and $\kkb$ are sorted in ascending order of degrees and orders in the displayed plots, which emphasises the fact that gyroscopic coefficients only couple eigenmodes with the same or neighbouring orders. 

The coefficients $\cstreamsphi{\jjb}{\qq}$ introduced in \eq{cstreamkq} are evaluated similarly as the gyroscopic coefficients. These coefficients are defined as 
\begin{equation}
\cstreamsphi{\jjb}{\qq} \define \integ{ \! \! \er \! \dotp \! \left( \conj{\gradn \fstreami{\jjb}} \crossp \gradn \Ylmi{\qq}  \right) \!}{\surface}{\oceancap}{},
\end{equation}
where the SPHs $\Ylmi{\qq} $ are written as products of functions of separated variables,  
\begin{equation}
\Ylmi{\qq} \left( \col , \lon \right) = \sphci{\qq} \sphcolfi{\qq} \left( \col \right) \expo{\inumber \mmi{\qq} \lon }.
\end{equation}
After an integration by part, the coefficients are simply expressed as 
\begin{equation}
\cstreamsphi{\jjb}{\qq} = \inumber 2 \pi \fstreamci{\jjb} \sphci{\qq}  \mmi{\jjb}  \kron{\mmi{\jjb}}{\mmi{\qq}} \left[ \psicolfi{\jjb} \left( \colbd \right) \sphcolfi{\qq} \left( \colbd \right) - \psicolfi{\jjb} \left( 0 \right) \sphcolfi{\qq} \left( 0 \right) \right] .
\end{equation}
We note that the $\cstreamsphi{\jjb}{\qq}$ are zero if $\mmi{\jjb} = 0$. Besides, if $\abs{\mmi{\jjb}}>0$, the Dirichlet condition satisfied by the $\fstreami{\jjb}$ at $\colbd$ implies that $\psicolfi{\jjb} \left( \colbd \right) =0$, and the polar condition of $\Ylmi{\qq}$ implies that $\sphcolfi{\qq} \left( 0 \right) = 0$. As a consequence, $\cstreamsphi{\jjb}{\qq} = 0$ for all $\jjb$ and $\qq$. 


\section{Solid tidal Love numbers}
\label{app:solid_love_numbers}

The gravitational and load Love numbers introduced in \eqs{tiltd}{tiltg} describe the tidal response of the planet's solid part in the general case as far as the latter is spherically symmetric. In the present study, these parameters are formulated analytically from the prescriptions given by \cite{Bolmont2020} for solid Earth (see Table~2 of the article). This formulation is based on the tidal model described in \cite{Tobie2005}, which integrates the equations of the elasto-gravitational theory \citep[][]{TS1972} for realistic radial profiles of background quantities. As a first step, \cite{Bolmont2020} use this tidal model with the background profiles computed from the internal structure model of \cite{Sotin2007}. As a second step, they fit the parameters of simplified models to the obtained solutions. These simplified models describe the rheology of equivalent bodies with uniform interiors, for which the Love numbers can be explicitly formulated in terms of the degrees of the associated SPHs \citep[][]{MM1960}.

The degree-$\llat$ Love numbers of a homogeneous solid body of mass $\Mcore$, radius $\Rcore$, and uniform shear modulus $\mupla$, are expressed as \citep[e.g.][]{MM1960}
 \begin{equation}
\left\{ \kl , \hl , \kloadl, \hloadl \right\} \! = \! \frac{1}{1+\Cmul} \! \left\{ \! \frac{3}{2 \left( \llat - 1 \right)} , \frac{2 \llat +1}{2 \left( \llat - 1 \right)} , - 1 , -  \frac{2 \llat +1}{3} \!\right\} \! .
\label{love_solid}
\end{equation}
In the above equation, the degree-$\llat$ dimensionless shear modulus $\Cmul$ is defined as 
\begin{equation}
\Cmul  \define \Al \Cmu,
\label{Cmul}
\end{equation}
where $\Cmu$ designates the dimensionless visco-elastic rigidity accounting for the rheology of the material, and $\Al $ the dimensionless parameter given by \citep[][]{CR2011,Efroimsky2012}
\begin{equation}
\Al \define \frac{4 \left( 2 \llat^2 + 4 \llat + 3 \right) \pi \Rcore^4 \mupla}{3 \llat \Ggrav \Mcore^2}.
\label{Al_Cmu}
\end{equation}
Since the oceanic layer is thin in the present study, the mass and radius of the solid part can be approximated by the planet mass and radius, respectively, which yields $\Mcore \approx \Mpla$ and $\Rcore \approx \Rpla$.

The normalised rigidity $\Cmu$ is a complex parameter that describes the rheological behaviour of the solid part in the frequency domain. This parameter is determined by the adopted rheological model. For standard models such as the Maxwell visco-elastic rheology \citep[e.g.][]{Correia2014}, it is expressed as a simple function of the tidal frequency, $\ftide$. Following \cite{Bolmont2020}, we opt for an Andrade rheology. Similarly as the Maxwell rheology, the Andrade model describes the visco-elastic deformation of the material. However, it additionally accounts for the unrecoverable creep forming the anelastic component of the response, which affects the frequency-dependence of tidal dissipation in the high-frequency range  \citep[e.g.][]{Efroimsky2012}. In the Andrade rheology, $\Cmu$ is expressed as \citep[e.g.][]{CR2011}
\begin{equation}
\Cmu = \frac{1}{1 + \left( \inumber \ftide \tauA \right)^{- \alphaA} \GammaF \left( 1 + \alphaA \right) + \left( \inumber \ftide \tauM \right)^{-1}},
\label{Andrade_mu}
\end{equation}
where $\tauM$ designate the Maxwell time, $\tauA$ the Andrade time, $\GammaF$ the Gamma function introduced in \eq{GammaF}, and $\alphaA$ the parameter determining the duration of the transient response in the primary creep \citep[][]{Castelnau2008,CR2011}. 

\end{appendix}
\end{document}